%% file: main.tex
\documentclass[twoside,leqno,11pt]{article}
\usepackage[letterpaper, margin=1in]{geometry}
\usepackage[margin=1in]{geometry}
\usepackage{graphicx} 
\usepackage{hyperref}
\usepackage{amsmath, amssymb,amsthm}
\usepackage{xcolor}
\usepackage{enumitem}
\usepackage{comment}
\usepackage{cleveref}
\usepackage{thmtools,thm-restate}
\usepackage{algorithm}
\usepackage{algpseudocode}
\usepackage{mathtools} 

\usepackage{tikz}
\usetikzlibrary{positioning,fit,backgrounds,shapes.geometric,decorations.pathreplacing,arrows.meta}

\definecolor{ink}{HTML}{243647}
\definecolor{teal}{HTML}{087F8C}
\definecolor{lightteal}{HTML}{E2F4F2}
\definecolor{coreblue}{HTML}{E8EFF8}
\definecolor{removed}{HTML}{C34F47}
\definecolor{muted}{HTML}{84919C}
\definecolor{pale}{HTML}{F1F3F5}

\newtheorem{theorem}{Theorem}[section]
\newtheorem{lemma}[theorem]{Lemma}

\newtheorem{proposition}[theorem]{Proposition}
\newtheorem{definition}[theorem]{Definition}

\newtheorem{remark}[theorem]{Remark}
\newtheorem{fact}[theorem]{Fact}

\usepackage{lineno}

\DeclareMathOperator{\dist}{dist}
\DeclareMathOperator{\diam}{diam}

\newcommand{\G}{\mathcal{G}}
\newcommand{\Prob}{\mathbb{P}}
\newcommand{\E}{\mathbb{E}}
\newcommand{\ceil}[1]{\left\lceil#1\right\rceil}
\newcommand{\floor}[1]{\left\lfloor#1\right\rfloor}
\newcommand{\abs}[1]{\left\lvert#1\right\rvert}
\newcommand{\set}[1]{\left\{#1\right\}}

\newcommand{\hi}{\mathrm{hi}}
\newcommand{\lo}{\mathrm{lo}}
\newcommand{\AAA}{\mathcal A}
\newcommand{\D}{B_0}

\def\poly{\operatorname{poly}}
\def\polylog{\operatorname{polylog}}

\title{Smoothed Analysis for Distributed MIS}

\author{Dipan Dey\thanks{Department of Computer Science, University of Houston, Houston, TX 77204, USA.  \\ Email: \href{mailto:ddey@central.uh.edu}{\tt ddey@central.uh.edu} and \href{mailto:gopal@cs.uh.edu}{\tt gopal@cs.uh.edu}.  Research supported in 
part by ARO Grant W911NF-231-0191, NSF grant CCF-2402837, and NSF grant CCF-2552882.}
\and Fabien Dufoulon\thanks{Univ. Bordeaux, CNRS, Bordeaux INP, LaBRI, UMR 5800, F-33400 Talence, France.  \\ Email: \href{mailto: fabien.dufoulon@labri.fr}{\tt fabien.dufoulon@labri.fr}.}
\and  William K. Moses Jr.\thanks{\raggedright \mbox{Department of Computer Science, Durham University, Durham, UK.  Email: \nolinebreak \href{mailto: william.k.moses-jr@durham.ac.uk}{\tt william.k.moses-jr@durham.ac.uk}}.}
\and Gopal Pandurangan\footnotemark[1]}

\date{}

\begin{document}

\maketitle

\begin{abstract}
We study smoothed analysis for the distributed Maximal Independent Set (MIS) problem, a fundamental problem in distributed computing. For many decades, the best-known round complexity of distributed MIS was $O(\log n)$, achieved by the seminal Luby's algorithm [STOC 1985, SICOMP 1986], and a long-standing open problem was whether a significantly faster algorithm exists. The best lower bound known for a while was  $\Omega(\sqrt{\log{n}/\log\log{n}})$ [Kuhn, Moscibroda,  Wattenhofer, JACM 2016], which was recently improved to $\Omega(\sqrt{\log{n}})$ [Khoury and Schild, FOCS 2025]. Very recently, Khoury and Schild [Arxiv, 2026] made a major breakthrough, showing that $\Omega(\log n/\log\log n)$ is a lower bound for any MIS algorithm (even) in the
LOCAL model; this means that Luby's algorithm is essentially optimal (up to a $\log \log n$ factor). The lower bound shown is \emph{existential}, i.e., there exist graphs where this lower bound holds for any MIS algorithm.

The above lower bound result for MIS raises an important question: \emph{how robust (or fragile)} is the MIS problem in relation to obtaining significantly faster algorithms? Motivated by this question, we study smoothed analysis of distributed MIS. In smoothed analysis, given an arbitrary (input or \emph{base}) graph $G$, we smooth (perturb) it by adding/deleting a small amount of \emph{random} edges. In particular, if $G$ is a graph of $n$ nodes, we consider the smoothed graph $\G_s = G \triangle G(n,p)$, where $G(n,p)$
is an Erdős-Rényi random graph ($\triangle$ denotes the symmetric difference). The goal is to show whether one can obtain significantly faster algorithms for as small a $p$ as possible, and if so, to quantify the required $p$ (which captures the amount of randomness needed to smooth).

 Our main results are almost \emph{tight bounds} on the round complexity of smoothed MIS vis-\`{a}-vis the smoothing parameter $p$: 
\begin{enumerate}[nosep]
\item For every base graph $G$, smoothing with parameter $p=\beta\ln n/n$ (for some sufficiently large constant $\beta$), there is a randomized LOCAL algorithm that, \emph{given only} the smoothed graph $\G_s = G \triangle G(n,p)$, computes an MIS of $\G_s$ in $O(\sqrt{\log n})$ rounds with high probability.\footnote{Throughout, this means with probability at least $1-n^{-c}$, where $c>0$ is an arbitrary constant fixed in advance.} This is (almost) quadratically faster than the worst-case bound of $\Omega(\log n/\log\log n)$ [Khoury and Schild, 2026] as well as the best-known MIS algorithm of Luby [STOC 1985].
\item  We show that the above smoothing parameter is optimal up to a constant factor for achieving such a significantly faster runtime. More precisely, if $p\le(1/2-\delta)\ln n/n$ for a constant $\delta>0$, then there is some base graph $G$ on which every randomized LOCAL MIS algorithm needs $\Omega(\log n/\log\log n)$ rounds on the smoothed graph $\G_s$; in other words, there is no advantage in smoothing.
\end{enumerate}

Our results are the first of their kind for distributed graph algorithms for local problems, and they show that smoothed analysis provides a way to overcome lower-bound barriers for well-studied problems in distributed graph algorithms.
\end{abstract}

\thispagestyle{empty}
\newpage

\tableofcontents
\thispagestyle{empty}
\newpage

\input{intro}

\input{prelims}

\input{structural}

\input{upper_bound}

\section{Lower Bound: Less Smoothing Gives No Advantage}
\label{sec:lower-bound-less-smoothening-bad}
\input{lower-bound-less-smoothening}

\bibliographystyle{plainurl}
\bibliography{references,refs-mst-smooth,refs1}

\input{appendix}

\end{document}

%% file: intro.tex
\newpage
\setcounter{page}{1}
\section{Introduction}

\subsection{Smoothed Analysis}
\label{sec:smoothed}
In this work, we study {\em smoothed analysis} of a fundamental problem in distributed computing, namely the Maximal Independent Set (MIS) problem.

Spielman and Teng's seminal work \cite{Spielman_2004}  introduced 
\emph{smoothed analysis} of algorithms to explain why the popular simplex algorithm for linear programming does very well in practice, despite having a worst-case exponential run time. 
At a high level, smoothed analysis involves perturbing a given input instance (which can be a ``bad'' instance) with \emph{random} noise and then analyzing the algorithm's (expected) performance on the \emph{perturbed} input (where the expectation is taken with respect to the random noise and possibly the random choices of the algorithm, if it is randomized). 
In particular, smoothed analysis of the simplex algorithm 
involves perturbing the input data with a \emph{small} amount of  Gaussian noise (with mean zero, parameterized by the variance of the noise) and then showing that the perturbed input can be solved efficiently in (expected) polynomial time by the simplex algorithm. In fact, Spielman and Teng show that the run time depends on the perturbation; the more the perturbation, i.e., the larger the variance of the added noise, the faster the run time.

Smoothed analysis has since become an important paradigm for going \emph{beyond worst-case analysis} of algorithms.
Smoothed analysis differs from {\em worst-case} analysis, which is determined by the algorithm's performance on the worst input. It may be that only a few inputs yield poor performance, while on the vast majority of inputs performance is significantly better. Thus, while the worst-case  \emph{lower bounds} the performance of the algorithm on \emph{any} input, such worst-case inputs can be few and often specially constructed (e.g., the input that forces simplex algorithm to take exponential time). Hence, such lower bounds can be called {\em fragile}.

Smoothed analysis also differs from {\em average-case analysis}, another well-studied paradigm for addressing the ``fragility'' of worst-case analysis. In average-case analysis, we assume a probability distribution on the set of all possible inputs and analyze the expected performance of the algorithm.\footnote{Again, the expectation is with respect to the input distribution as well as possibly the random choices made by the algorithm, in case it is randomized.} While average-case analysis may not be determined only by performance on a small number of inputs, its weakness is that the chosen input distribution may not reflect real-world inputs, and its performance may depend mainly on the chosen distribution (which is typically taken to be uniform over all inputs). 

Smoothed analysis can be thought of as a hybrid, combining the good aspects of both worst-case and average-case analysis
--- it considers any input (including worst-case inputs) and then randomly perturbs it. If even small perturbations (by adding random noise) lead to significantly better performance, then the worst case is quite sensitive to the input parameters. In practice, there is usually noise, so the algorithm is likely to avoid worst-case behavior.

\subsection{Smoothed Analysis of Graph Isomorphism}
\label{sec:iso}

To illustrate a recent ``success story'' for smoothed analysis in the centralized setting, we consider a well-known graph-theoretic problem: {\em graph isomorphism}. The best-known algorithm for graph isomorphism takes quasi-polynomial time~\cite{babai1,babai2}, and it is not known whether the problem admits a polynomial-time algorithm.
On the other hand, an average-case analysis has been very well studied. This is known as {\em random graph isomorphism} where one compares a $G(n,p)$ (Erdős-Rényi) random graph with an {\em arbitrary} graph. In particular, an early result by Babai, Erdős, and Selkow~\cite{selkow} showed that random graph isomorphism can be solved efficiently (in linear expected time) for $G(n,1/2)$, which models a graph drawn randomly from the space of all graphs on $n$ nodes. Later, these results were extended to a much sparser regime of $p$ \cite{bollobas,cp08,linial}. These results crucially exploit properties of random graphs. However, they do not address the original problem where both graphs can be arbitrary.

In graphs, smoothed analysis typically involves adding (or deleting) a small amount of {\em random edges} to a given arbitrary input graph $G$ --- we call this the {\em base} graph (cf. Section \ref{sec:model}).
The recent works of \cite{AKM25,sridhar} study \emph{smoothed analysis of graph isomorphism}, where we compare two arbitrary graphs, but one of them is smoothed by adding a small amount of \emph{random edges} ---  by adding a $G(n,p)$ (Erdős-Rényi) random graph with $p = O(\log n/n)$. More precisely, given $G$, they consider the symmetric difference graph --- $G \triangle G(n,p)$ --- where each existing edge is also deleted with probability $p$ (our smoothed model is also the same).
They show that graph isomorphism between a smoothed graph and an arbitrary graph can be solved in polynomial time. In fact, they show a similar polynomial time solvability (via a more complicated algorithm) for even smaller $p = O(1/n)$. However,
if $p=o(1/n)$, they show that smoothing is not very helpful as it essentially reduces to solving isomorphism on a hard instance of the \emph{base graph} $G$ (see \cite{AKM25} for more details). 

\subsection{The Maximal Independent Set (MIS) Problem}
\label{sec:MIS}
The \emph{maximal independent set} (\emph{MIS}) problem is one of the most fundamental and well-studied problems in distributed graph algorithms.  Given a graph with $n$ nodes, each node must decide whether it belongs to a subset $M \subseteq V$ (called the MIS) or not such that (i) every node is either in $M$ or has a neighbor in $M$ and (ii) no two nodes in $M$ are adjacent to each other.

Because of the importance of MIS, distributed algorithms for MIS  have been studied extensively for
the last four decades mainly with a focus on improving the  time complexity (i.e.,\ the number of \emph{rounds}).
In 1986, Alon, Babai, and Itai \cite{Alon_1986} and Luby \cite{Luby_1986} presented a randomized distributed MIS algorithm that runs in $O(\log{n})$ rounds (where $n$ is the number of nodes in the graph) with high probability. Since these seminal results, significant progress has been made in designing progressively faster distributed MIS algorithms \cite{Barenboim_2016, Ghaffari16,Rozhon2020}.

The currently best-known algorithm in the LOCAL model takes $O(\log{\Delta}  + \poly(\log{\log{n}}))$ rounds \cite{Rozhon2020,Ghaffari26}.\footnote{The currently
best-known randomized algorithm in the CONGEST model takes
$O(\log{\Delta}\log\log{n} + \poly(\log{\log{n}}))$ rounds~\cite{GGR20}. See Section \ref{sec:model} for definitions of LOCAL and CONGEST models --- two standard models in distributed computing.}
Thus, the currently known
best algorithms of MIS (\cite{Rozhon2020, GGR20, Ghaffari16}) are dependent on
$\Delta$ (the maximum degree), and hence still take $O(\log n)$ rounds (even in the LOCAL model) for graphs with $O(n^{\Omega(1)})$
degree. As far as deterministic algorithms are concerned, the best known algorithms take $O(\poly(\log n))$ rounds in the LOCAL and CONGEST models \cite{Rozhon2020,GGR20}. 

There are faster distributed algorithms known for special classes of
graphs such as trees~\cite{Ghaffari16,lenzen} and Erdős-Rényi random graphs~\cite{Ghaffari16,KR15}, but they
still take $O(\sqrt{\log{n}})$ rounds.\footnote{The algorithm in \cite{KR15} for $G(n,p)$ random graphs actually takes $O(\sqrt{\log n \log \log n})$ rounds, though it is claimed it takes $O(\sqrt{\log n/\log \log n})$ (due to a calculation error).}
There is also a fast deterministic distributed algorithm running in $\mathcal{O}(\log^*{n})$ for growth-bounded graphs (which includes unit-disk graphs)~\cite{logstar}.
There are also MIS
algorithms that run faster on graphs with low arboricity, but they nevertheless
take $O(\log n)$ rounds on high arboricity graphs~\cite{Ghaffari16,barenboim}.

While the above results make significant progress on the round complexity of the MIS problem for some specific graphs, in general graphs the best known running time is still $O(\log{n})$ (even for randomized algorithms in the LOCAL model). Till recently,  the best-known lower bound for MIS was
    $\Omega\left(\min\left\{\frac{\log \Delta}{\log \log \Delta}, \sqrt{\frac{\log n}{\log \log n}}\right\}\right)$ rounds due to Kuhn, Moscibroda, and Wattenhofer~\cite{Kuhn_2016} that also applies to randomized algorithms and holds even in the LOCAL model.  This lower bound was later improved  to $\Omega(\sqrt{\log n})$ by Khoury and Schild in 2025 (in the LOCAL model) \cite{khoury-focs25}. Thus till very recently (see next), there existed
essentially a \emph{quadratic} gap between the known upper and lower bounds
for the round complexity of MIS.

Very recently, in October 2026, Khoury and Schild~\cite{khoury2026}, in a major breakthrough, announced that $\Omega(\log n/\log\log n)$ is a lower bound for any MIS algorithm (even) in the
LOCAL model; this means that Luby's algorithm is essentially optimal (up to a $\log \log n$ factor).
This lower bound is existential; it applies to specially constructed graphs. Even prior to this result, the status of the MIS problem gave  a good motivation for studying smoothed analysis, as no algorithm with $o(\log n)$
rounds was known; but the new result motivates studying smoothed complexity of MIS even more, as it rules out getting any algorithm substantially better (up to a $\log \log n$ factor) than $\Theta(\log n)$ in the worst case.

\subsection{Smoothed Analysis of Distributed MIS}
In this paper, we initiate a study of smoothed analysis for distributed graph algorithms for \emph{local problems} by focusing on the fundamental MIS problem. Local problems (not to be confused with the LOCAL model --- cf. Section \ref{sec:model}) are problems that admit algorithms that are significantly faster than the \emph{diameter} of the graph, typically running in $\polylog{n}$ rounds.\footnote{On the other hand, global problems need at least the graph diameter number of rounds, since they involve traversing the whole network. They include fundamental problems such as broadcast, leader election, and minimum spanning tree (MST) construction. Smoothed analysis in the context of distributed algorithms has been applied earlier for the MST problem \cite{CPP20} as well as for dynamic networks --- cf. Section \ref{sec:related} for details.} They include fundamental problems such as MIS, maximal matching, and coloring.

Our motivation is simple: Can we use smoothed analysis to overcome the (essentially) $\Theta(\log n)$ lower bound barrier of distributed MIS? As mentioned earlier, a natural smoothing model for studying graph problems in prior work (see e.g., \cite{AKM25,sridhar,Dinitz_2018,newport}) is adding (and deleting) a small number of random edges to the given graph $G$ (called the base graph). More precisely, this is typically done by adding a $G(n,p)$ random graph to the base graph. In other words, for each pair of nodes, an edge is added with probability $p$ independently of the others (if an edge already existed between the node pair, then this does nothing). One may also delete an existing edge with probability $p$. We will mainly focus on adding random edges, i.e., the smoothed graph $\G=G\cup G(n,p)$, since both our upper and lower bounds also apply when edges are deleted (see the end of this subsection, \Cref{sec:symdiff,subsec:lower-bound-edge-deletions}).

Why should smoothing give substantially better bounds than without? One main intuition is that (this is conceptually analogous to the situation in the graph isomorphism problem)  
the \emph{average case analysis} of distributed MIS, i.e., on random graphs,  gives a substantially better bound. As mentioned earlier,
distributed MIS on $G(n,p)$ random graphs (for every $p$) can be solved
in $O(\sqrt{\log n})$ rounds \cite{Ghaffari16,KR15} --- this is the best known bound if $p$ is arbitrary. 
As discussed in Section \ref{sec:MIS}, this is almost quadratically faster than the recent lower bound of 
$\Omega(\log n/\log \log n)$ as well as the best-known general upper bound of $O(\log n)$ (e.g.,  Luby's algorithm \cite{Luby_1986}) that has stood for over four decades. 
So a key question is: can we obtain a substantially faster algorithm --- in particular an $O(\sqrt{\log n})$-round MIS algorithm --- via smoothed analysis for \emph{any arbitrary (base) graph} $G$? And if so, what is the smallest $p$ (the smoothing parameter) needed? We want to keep the number of added random edges as small as possible.

Our main result answers the above question in the affirmative: one can obtain a $O(\sqrt{\log n})$-round MIS algorithm (cf. Section \ref{sec:contributions} for a precise statement and Section \ref{sec:upper} for the algorithm) for $p = O(\log n/n)$, i.e., by adding $O(n \log n)$ random edges to $G$. We also show that this amount of smoothing is \emph{the best possible} for obtaining an $O(\sqrt{\log n})$-round algorithm, in the sense that $\Omega(n \log n)$ random edges need to be added for \emph{some base graph} $G$ (this also illustrates why smoothed analysis is a hybrid of both worst-case and average-case analysis: it is worst case in the base graph $G$ and average-case in the amount of random edges added).

Our results seem quite robust with respect to both the runtime obtained and the amount of smoothing.
We note that, in general, it is unclear how to obtain better than $\sqrt{\log n}$ rounds via smoothing, i.e., even if we use a larger $p$.
As mentioned earlier, if the base graph itself is taken to be a $G(n,p)$ random graph, in particular,  say if $p= 2^{\sqrt{\log n}}/n$, then smoothing adds/deletes only a small number of random edges and thus boils down to finding an efficient distributed algorithm for such a $G(n,p)$ graph. Currently, neither a faster than $O(\sqrt{\log n})$ round algorithm nor a matching lower bound is known for $G(n,p)$ random graphs, and it seems hard to significantly improve this bound by more than a $\poly(\log \log n)$ factor.\footnote{In fact, ChatGPT  claims an $O(\sqrt{\log n/\log \log n})$ bound for $G(n,p)$ random graphs, but nothing better. Also see footnote 4.} 

We also point out that our results (both upper and lower bounds) also apply if we smooth by \emph{also deleting} edges, i.e., say we delete each existing edge (independently) with probability  $p$, i.e., the smoothed graph is the symmetric difference between the base graph $G$ and $G(n,p)$.
The $O(\sqrt{\log n})$ upper bound holds for 
$p=O(\log n/n)$; informally, this follows by writing $G\triangle G(n,p)$ as the smoothing of a random subgraph of $G$ (\Cref{sec:symdiff}).\footnote{One can argue that the probability that a deleted edge is added is negligible; for a more precise analysis see Section \ref{sec:symdiff}.}
Again, as mentioned earlier, improving on the $O(\sqrt{\log n})$ runtime seems hard,
since the base itself can be taken to be a $G(n,p)$ with
$p = 2^{\sqrt{\log n}}/n$ and deleting a few random edges does not make a difference. Also, our lower bound holds in both models: if $p\le(1/2-\delta)\ln n/n$ for a constant $\delta>0$, that is, if fewer than $\frac14 n\ln n$ edges are added or deleted in expectation, then there is no advantage in smoothing --- there is a base graph that requires $\Omega(\log n/\log\log n)$ rounds (cf.\ Section~\ref{sec:lower-bound-less-smoothening-bad}).

Conceptually, our MIS results mirror those obtained for the graph isomorphism, albeit under very different settings. For MIS, ``hard" means breaking the $\Theta(\log n)$  barrier for arbitrary graphs (as opposed to the quasi-polynomial barrier for graph isomorphism for arbitrary graphs) using smoothing and showing that the amount of smoothing needed is essentially tight, i.e., $\Theta(n \log n)$ random edges added/deleted are required for obtaining an $O(\sqrt{\log n})$-round algorithm (for
graph isomorphism it is adding/deleting $\Theta(n)$ edges for polynomial time algorithms). 
Thus, smoothed analysis seems to be a promising ``beyond worst-case analysis'' technique even for distributed graph problems such as MIS.

\subsection{Model}
\label{sec:model}

\paragraph{Smoothing Model.}

We are given an arbitrary graph $G$, called the \emph{base graph}. We smooth $G$ as follows.  
Given a \emph{smoothing parameter} $p \in [0,1]$, for each pair of nodes of $G$, flip it with probability $p$, i.e., each non-edge becomes an edge with probability $p$ and each edge becomes a non-edge with probability $p$. Let the resulting graph be $\G_s$. In other words, $\G_s = G \triangle G(n,p)$, where $G(n,p)$
is the (standard) \emph{Erdős-Rényi random graph} and $\triangle$ denotes the symmetric difference between the two graphs (the graph with the same set of nodes and the set of edges that are either only present in $G$ or only present in $G(n,p)$ but not both). As mentioned in Section \ref{sec:iso}, this model is quite natural for smoothing in graphs (this is the same model used in \cite{AKM25,sridhar}), and this model and similar ones have  been used in several prior works, including in distributed algorithms \cite{Dinitz_2018,CPP20,newport}.
In our analysis, we also use $\G$, which denotes the smoothed graph under just additions, i.e., $\G = G \cup G(n,p)$. It turns out that it is simpler to analyze $\G$ rather than $\G_s$; it is easy to extend the results obtained for $\G$ to $\G_s$ (cf. Section \ref{sec:symdiff}).\footnote{We use $\G_s$ to denote the smoothed graph under the symmetric difference, to distinguish it from $\G=G\cup G(n,p)$, which is smoothed by additions only.}

\paragraph{Distributed Computing Model.}
 We have a distributed network of $n$ nodes, modeled as an undirected graph. Each node hosts a processor with limited initial knowledge. We assume that each node has a distinct ID of size $O(\log n)$ bits.
 We assume that each node has ports (each port having a unique port number); each incident edge is connected to one distinct port. 
  We assume that each node knows $n$.

The smoothed graph $\G_s$ (or $\G$) serves as input to the algorithm. 
We do not assume that the algorithm knows which edges were added or deleted. 
 For our upper bound (i.e., distributed algorithm --- cf. Section \ref{sec:upper}), we assume that nodes do not know the smoothing parameter $p$ nor can they distinguish which incident edges are random and which belonged to the base graph $G$ (or got deleted).  On the other hand, our lower bound (cf. Section \ref{sec:lower-bound-less-smoothening-bad}) holds even if nodes know $p$ and even if they know which edges are random and which belong(ed) to $G$ (including knowing which edges were deleted).

Nodes are allowed to communicate through the edges of the smoothed graph, and it is assumed that communication is synchronous and occurs in rounds. We identify each node of the network with a vertex of the smoothed graph, and use the two terms interchangeably. In particular, we assume that each node knows the current round number (starting from round 0). In each round, each node can perform some local computation (which finishes in the same round), including accessing a private source of randomness.
We assume the LOCAL model, a standard model~\cite{Peleg_2000_Book} with no restriction on the size of messages sent per edge per 
round.\footnote{The other standard model is the CONGEST model, where only $O(\log n)$ bits can be exchanged per edge per round.}

\subsection{Our Contributions}
\label{sec:contributions}

Our main conceptual contribution is positing a smoothed analysis of distributed MIS, motivated by the desire to go beyond its worst-case runtime bound of $\Omega(\log n/\log \log n)$ \cite{khoury2026} in arbitrary graphs. Even before this lower bound was established, the best-known runtime was $O(\log n)$, which the classical Luby's algorithm~\cite{Luby_1986} achieves.

Our main technical results are almost \emph{tight bounds} on the round complexity of smoothed MIS vis-\`{a}-vis the smoothing parameter $p$: 
\begin{enumerate}
\item (cf. Theorem \ref{thm:upperfinal}) For every base graph $G$ and smoothing with parameter $p=\beta\ln n/n$ (for some sufficiently large constant $\beta$), there is a randomized LOCAL algorithm that, \emph{given only} the smoothed graph $\G_s = G \triangle G(n,p)$, computes an MIS of $\G_s$ in $O(\sqrt{\log n})$ rounds with high probability.\footnote{Throughout, this means with probability at least $1-n^{-c}$, where $c>0$ is an arbitrary constant fixed in advance.} This is almost quadratically faster than the worst-case bound of $\Omega(\log n/\log\log n)  $\cite{khoury2026} and the best-known (in terms of the parameter $n$) runtime of $O(\log n)$ \cite{Luby_1986}.

\item (cf. \Cref{cor:lower-bound-edge-deletions}) We show that the above smoothing parameter is optimal up to a constant factor for achieving such a significantly faster runtime. More precisely, if $p\le(1/2-\delta)\ln n/n$ for a constant $\delta\in(0,1/2]$, then there is a base graph $G$ such that every randomized LOCAL algorithm that computes an MIS of the smoothed graph $\G_s$ with constant success probability takes $\Omega(\log n/\log\log n)$ rounds; in other words, this amount of smoothing gives no advantage.
\end{enumerate}

Our results are the first of their kind for distributed graph algorithms for local problems such as MIS, and they show that smoothed analysis provides a way to overcome lower-bound barriers for well-studied problems in distributed graph algorithms.

\subsection{High-level Overview of our Results \& Technical Challenges}
\label{sec:high-level}
\paragraph{Smoothed Algorithm and its Analysis.}
Our algorithmic result (\Cref{thm:upperfinal}) shows that smoothing with $p=\beta\ln n/n$, for a sufficiently large constant $\beta$, makes MIS solvable in $O(\sqrt{\log n})$ LOCAL rounds with high probability, for \emph{every} base graph $G$. The algorithm sees only the smoothed graph $\G_s$: it cannot tell base edges from random ones. The random edges are both part of the communication network and part of the input, since the output must be an MIS of $\G_s$. Our  analysis focuses on \emph{adding} random edges only, i.e., the smoothed graph $\G = G \cup G(n,p)$; as argued in Section \ref{sec:symdiff}, it can be extended to $\G_s$, i.e., under
both additions and deletions.

The basic high-level intuition comes from the analysis of distributed MIS in $G(n,p)$ random graphs \cite{KR15,Ghaffari16}. If $p = O(2^{\sqrt{\log n}}/n)$, then
the degree of a node is $O(2^{\sqrt{\log n}})$ with high probability and hence one can simply use the degree-sensitive algorithm of \Cref{prop_Ghaffari} to obtain a $O(\sqrt{\log n})$-round algorithm. On the other hand, if $p = \Omega(2^{\sqrt{\log n}}/n)$, then the \emph{diameter} of the $G(n,p)$  graph is $O(\sqrt{\log n})$ (this follows from
the fact that the diameter $D$ of a $G(n,p)$ graph is $O(\log n/\log(np))$ with high probability --- cf. Lemma \ref{lem:er-diam}); hence one can simply collect all the topology information in a single node in the LOCAL model in $O(D)$ rounds and compute the MIS locally in the node and distribute the answers to all nodes in another $O(D)$ rounds.

In smoothed analysis, when the base graph $G$ is arbitrary, there are key technical challenges in implementing the above intuition. Call a node \emph{low-degree} if its degree in $\G$ is at most a threshold $B=e^{\sqrt{\ln n}}$, and \emph{high-degree} otherwise.   Of course, if all nodes in $G$ have low degree (note that the difference in degree of a node in $G$ versus $\G$ is $O(\log n)$, which is quite small if the node's degree is high), then one can immediately obtain the result by applying the degree-sensitive algorithm. However, some (or even all) nodes can have high degree; unlike the $G(n,p)$
case, this does not imply $O(\sqrt{\log n})$ diameter of the smoothed graph, since the added edges form a $G(n,p)$ graph with $p = O(\log n/n)$, which is very small.

Next, consider the other extreme case where all nodes have high degree. In this case, one can show that the vertices can be partitioned into $n'=O(n/B)$ parts of size $\Theta(B)$, such that any two vertices of a part are within constant distance in $G$ (\Cref{lem:cells}). Two parts span $\Theta(B^2)$ vertex pairs, so a random edge joins them with probability $p'=\Omega(pB^2)$, which is much larger than $p$. Viewing each part as a single node, the parts therefore behave like a $G(n',p')$ random graph, whose diameter is $O(\sqrt{\log n})$ (cf. \Cref{lem:er-diam}). Hence, the diameter of the smoothed graph is also $O(\sqrt{\log n})$ (\Cref{lem:seeded}), and we can again gather the topology, using the primitive of \Cref{lem:gather}.

 The bulk of the technical complication arises when $G$ (and hence $\G$) contains both high-degree and low-degree nodes. 
First consider the subgraph induced by the low-degree nodes.
The degree-sensitive algorithm of \Cref{prop_Ghaffari} handles these nodes in $O(\sqrt{\log n})$ rounds. The difficulty lies in the high-degree vertices, which our algorithm removes first, in two steps (\Cref{sec:alg}). In the first step, every high-degree vertex marks each edge to a low-degree neighbor with probability about $1/2$.  Every low-degree vertex with a marked edge and no (low-degree) neighbor that also has a marked edge joins the MIS. This low-degree MIS node serves as a \emph{helper}, dominating the high-degree vertices next to it. In the second step, the remaining high-degree vertices learn their components by gathering information for $O(\sqrt{\log n})$ rounds, using \emph{all} nodes as relays, and solve MIS locally on the components. A completeness check ensures that this step never outputs two adjacent vertices; only maximality depends on the random edges.

The main challenge is to show that, with high probability, either no high-degree vertex remains after the first step, or all remaining ones are pairwise within distance $O(\sqrt{\log n})$ in $\G$ (\Cref{sec:alg-analysis}). Although the algorithm never sees $G$, the analysis distinguishes cases using sets defined solely from $G$. In \Cref{lem:seeded}, we show that if many vertices are within two hops of vertices of large degree in $G$, then the random edges shrink the diameter of this region to $O(\sqrt{\log n})$. Otherwise, the helpers dominate every high-degree vertex. 

The technical difficulty is dealing with the  
low-degree helper nodes in the first step. 
The potential helpers of a high-degree vertex are the low-degree neighbors it reaches by a marked edge; such a vertex becomes a helper \emph{only if none of its neighbors has a marked edge}. Two potential helpers of the same high-degree vertex therefore have dependent outcomes if they are adjacent or have a common neighbor 
(in this case, if the common neighbor is also a potential helper, then neither of them becomes a helper vertex). Moreover, other high-degree vertices may prevent a potential helper from joining by marking edges to its neighbors. A counting argument in \Cref{lem:sep} shows that all but a constant number of the potential helpers are pairwise at distance at least three, which makes their outcomes independent. We then show that each such potential helper is, with constant probability, not blocked by marked edges by any vertex of large degree in $G$.  The crucial intuition behind this analysis is to exploit the additional \emph{random} edges introduced by smoothing. Note that a high-degree node $v$ in the base graph $G$ might have low-degree nodes with an arbitrary connectivity structure among them; e.g., they can form a clique. In this case, with high probability no low-degree neighbor of $v$ in $G$ will join the MIS, since it is likely that at least one of its neighbors (which is also a neighbor of $v)$ is also marked by $v$. However, if there are lots of low-degree \emph{non-neighbors} of $v$ in $G$, then it is likely that one of them (say $u$) will become a low-degree neighbor of $v$ in $\G$ due to the addition of a random edge between $u$ and $v$.
Since the number of added random edges per node is small, it is unlikely that another low-degree neighbor of $u$ \emph{also} got a random edge added between itself and $v$.
Hence, it is likely that $u$ is isolated and becomes a helper node which dominates the high-degree node. 

On the other hand, if there are not too many low-degree vertices, and hence not too many helpers,  then the analysis shows that most of the remaining vertices are high-degree and can be partitioned into components of $O(\sqrt{\log n})$ diameter (which can be used to solve MIS by gathering).

\paragraph{Lower Bound.}
Our lower bound (Theorem~\ref{the:lower-bound-less-smoothing})
shows that, when $p\leq(1/2-\delta)\ln n/n$ for a fixed
$\delta\in(0,1/2]$, any randomized LOCAL algorithm
that computes an MIS of $\G=G\cup G(n,p)$ with
constant positive success probability for every
$n$-node graph $G$ requires
$\Omega(\ln n/\ln\ln n)$ rounds. 
This holds even if the algorithm can distinguish
original edges from added edges.
We prove this by reducing MIS on an arbitrary input graph
$H$ with $h$ nodes to MIS on $\G$, where $n$ is polynomially related to $h$. This allows us to leverage lower bounds that apply to MIS in hard (non-smoothed) instances (such as that of \cite{khoury2026}) to derive lower bounds for MIS in the smoothed setting. 

The construction of $\G$ from $H$ proceeds as follows. 
We replace each node $u\in H$ with a clique $C_u$
of size $k$, called its replacement clique. For every edge $(u,v)\in E(H)$, we add
edges that form a complete bipartite graph between
$C_u$ and $C_v$. We then add isolated padding nodes
to obtain $G$, and smooth $G$ to obtain
$\G=G\cup G(n,p)$.
In order to leverage the reduction to prove our lower bound, we must ensure that the reduction (i) preserves the validity of
the projected solution (the MIS on $\G$ projected back to the input graph $H$) and (ii) the running time.

The first challenge is preserving maximality (for the projection).
Given an MIS $I$ of $\G$, we select $u\in H$ exactly
when $I$ contains a node of $C_u$.
Independence follows from the edges between
replacement cliques. However, selected padding nodes
may dominate clique nodes without providing selected
neighbors in $H$, so maximality does not follow
automatically.
We overcome this using \emph{protected nodes}, whose
neighborhoods are unchanged by smoothing.
Choosing $k=\Theta(e^{np}\ln h)$ with a sufficiently large constant factor ensures that every
replacement clique contains a protected node with
probability $1-o(1)$
(Lemma~\ref{lb:protection}).
If an unselected node of $H$ had no selected neighbor,
the protected node in its clique would neither be
selected nor have a selected neighbor in $\G$,
contradicting maximality of $I$.
Thus every MIS of $\G$ projects to an MIS of $H$
on this event (Lemma~\ref{lb:projection}).

The second challenge is preserving the running time.
Each node $u\in H$ can simulate its clique $C_u$
locally, and communication between adjacent cliques
can use the corresponding edge of $H$.
Random edges can disrupt this simulation by connecting
virtual nodes whose simulators are not adjacent in $H$.
We first rule out added edges between clique nodes
with probability $1-o(1)$
(Lemma~\ref{lb:core-edges}).
This alone is insufficient: paths through padding
nodes can also create communication between different
local simulations, even when these paths do not
shorten distances between the original nodes.
We therefore analyze the radius-$(T+2)$ neighborhoods
of all clique nodes after removing all edges between
clique nodes, where $T$ is the running time of the
algorithm being simulated.

Each node of $H$ independently generates virtual
trees for its clique nodes using a random process. The main difficulty is ensuring that these
locally generated trees all represent
the relevant neighborhoods of a genuine smoothed graph.
For $T<\varepsilon\ln n/\ln\ln n$, where
$\varepsilon>0$ is sufficiently small, the total
number of virtual nodes is at most a cutoff $s$
with probability $1-o(1)$, where $ps^2=o(1)$
(Lemma~\ref{lb:population}).
In the coupling argument described below, this cutoff lets us bound
the probability of repeated IDs or unwanted edges
before the cutoff by $O(ps^2)=o(1)$.
We then construct a coupling under which, with
probability $1-o(1)$, all virtual trees are
 identical to their corresponding
neighborhoods, including node IDs.
These neighborhoods are pairwise disjoint and have
no edges between them in the graph obtained after
removing all edges between clique nodes
(Lemma~\ref{lb:coupling}).
There is a tension between the two parts of the proof:
larger cliques make the existence of protected nodes more likely, but also
increase the number of trees and the chance of
interactions between them. 
The slack provided by $\delta$ allows both requirements
to hold simultaneously. Still, it should be noted that choosing the correct parameters to satisfy both properties is non-trivial.

Consequently, each node of $H$ can locally simulate
its clique and attached trees, while combining all
messages between adjacent cliques into one LOCAL
message over the corresponding edge of $H$.
A $T$-round algorithm on $\G$ with success probability
$\sigma$ therefore gives a $T+O(1)$-round algorithm
on $H$ with success probability $\sigma-o(1)$
(Lemma~\ref{lb:simulation}).
Since
$\ln h/\ln\ln h=\Theta(\ln n/\ln\ln n)$,
the lower bound of Khoury and Schild~\cite{khoury2026}
yields the claimed bound.
The same construction also handles random deletions:
all original edges lie among the clique nodes,
and thus the probability that any of them is deleted is $o(1)$
(\Cref{cor:lower-bound-edge-deletions}).

\subsection{Additional Related Work}
\label{sec:related}
Smoothed analysis was first used in the context of the simplex algorithm \cite{Spielman_2004}.   Subsequently, it has been applied to various algorithmic problems in the sequential setting (see, e.g., \cite{Spielman_2009} for a survey and further references). Recent examples of smoothed analysis include online algorithms \cite{roughgarden} and graph isomorphism \cite{sridhar,AKM25}.
Here we focus on prior work on smoothed analysis for distributed algorithms.

To the best of our knowledge, the two main works on smoothed analysis of distributed algorithms are due to
 Dinitz, Fineman, Gilbert, and Newport~\cite{Dinitz_2018,newport} on smoothed analysis of {\em dynamic networks} and Chatterjee, Pandurangan, and Pham~\cite{CPP20} on smoothed analysis of the distributed minimum spanning tree (MST) problem. 

The works of \cite{Dinitz_2018,newport} consider a dynamic graph $\mathcal{H} = G_1, G_2, \dots $ that describes an evolving network topology, where $G_i$ is the graph at round $i$.  Graphs in $\mathcal{H}$ share the same node set and are connected, but the edges can change arbitrarily from one round to the next.
They define a smoothing model for a dynamic graph parameterized by a smoothing factor $k \in \{1,2,\dots, \binom{n}{2}\}$. To $k$-smooth a dynamic graph $\mathcal{H}$ is to replace each static graph $G_i$ in $\mathcal{H}$ with a smoothed graph $G'_i$ sampled uniformly from the space of connected graphs that are within {\em edit distance} $k$ of $G_i$. 
They study three fundamental problems with strong lower bounds in dynamic network models: \emph{flooding}, \emph{random walks}, and \emph{aggregation}. For each problem, they study the robustness/fragility of the existing bounds and show how they improve under increasing amounts of smoothing.

 The work of \cite{CPP20} introduced smoothed analysis in the context of distributed graph algorithms. Their smoothing model is somewhat different from what is considered here and in other papers such as \cite{AKM25}. It must be noted that MST is a weighted graph problem and it can be shown that smoothing the weights by adding a small amount of random noise (à la Spielman and Teng \cite{Spielman_2004}) does not overcome the worst-case bound of $\tilde{\Omega}(D+\sqrt{n})$ rounds for distributed MST. Instead, they consider a smoothing model where, in each round of the distributed algorithm, every node can add an edge to another random node (chosen uniformly at random from $V$) with probability $p(n)$ (smoothing parameter). Thus, a key difference in this model is that each node \emph{knows} the random edges added, which is crucial for the analysis. 
The added random edges (with weight $\infty$), once they are added, can be used (only) for communication and do not affect the MST of the base graph $G$. They
show upper and lower bounds on the time complexity of distributed MST in the above smoothing model that shows how the worst-case barrier of $D+\sqrt{n}$ can be overcome. In particular, they show that MST can be solved in $\tilde{O}(\min\{\frac{1}{\sqrt{p(n)}} 2^{O(\sqrt{\log n})}, D + \sqrt{n}\})$ rounds. They also show a lower bound of $\tilde{\Omega}(\min\{\frac{1}{\sqrt{p(n)}}, D+\sqrt{n}\})$.

%% file: prelims.tex
\section{Preliminaries}\label{sec:prelims}

The \emph{base graph} $G$ is a fixed simple graph on a vertex set $V$ with $\abs V=n$. Every pair of vertices that is not an edge of $G$ is added to a set $R$ independently with probability
\[
  p=\frac\lambda n,\qquad\text{where}\qquad \lambda=\beta\ln n
\]
for a constant $\beta\ge1$; the value of $\beta$ is fixed in \Cref{sec:upper}. The \emph{smoothed graph} is $\G=G\cup R$. The algorithm sees only $\G$: it knows neither $G$ nor which edges of $\G$ are random. We call the edges of $G$ \emph{base edges} and the edges of $R$ \emph{random edges}. The vertices communicate in the LOCAL model over $\G$, and the goal is to compute a maximal independent set (MIS) of $\G$. A set $I\subseteq V$ is an MIS of $\G$ if no two vertices of $I$ are adjacent in $\G$ and every vertex of $V$ lies in the closed neighborhood $N_\G[I]$. We also use the degree scale
\[
  \D=\floor{e^{\sqrt{\ln n}}}. 
\]
An event holds \emph{with high probability (w.h.p.)} if it holds with probability at least $1-n^{-c}$, where $c>0$ is an arbitrary fixed constant.

\paragraph{Notation.}
All logarithms are natural. For a graph $F$, by abuse of notation, we use $u \in F$ to refer to $u$ being a node of $F$ and $(u,v) \in E(F)$ to refer to the edge $(u,v)$ present in the graph $F$. Furthermore, for a graph $F$, a vertex $v$ and a vertex set $Q$ of $F$, we write $\deg_F(v)$ for the number of neighbors of $v$ in $F$, $N_F(v)$ for the set of these neighbors, which we also call the $F$-neighbors of $v$, and $N_F[v]=N_F(v)\cup\set v$. For a set $Q$, we write $N_F[Q]=\bigcup_{v\in Q}N_F[v]$. For vertices $u,v$ of $F$, $\dist_F(u,v)$ is the length of a shortest path from $u$ to $v$ in $F$ ($\infty$ if there is none), and $\dist_F(u,Q)=\min_{y\in Q}\dist_F(u,y)$. Finally,
\[
  N_F^2[Q]=\set{v\in V(F):\ \dist_F(v,Q)\le2}
\]
is the set of vertices within distance two of $Q$ in $F$.

\begin{fact}[Chernoff Bounds \cite{MUBook}]\label[fact]{fact:tails}
Let $X$ be a sum of independent Bernoulli variables.
\begin{enumerate}
  \item[(F1)] If $\E[X]\le\mu$, then $\Prob[X>2\mu]\le e^{-\mu/3}$.
  \item[(F2)] If $\E[X]\ge\mu$ and $0<\delta<1$, then $\Prob[X\le(1-\delta)\mu]\le e^{-\delta^2\mu/2}$; in particular $\Prob[X<\mu/2]\le e^{-\mu/8}$.
\end{enumerate}
\end{fact}

\paragraph{Diameter of random graphs.}
For an integer $N\ge1$ and $q\in[0,1]$, let $G(N,q)$ denote the random graph on $N$ vertices in which every pair of vertices is an edge independently with probability $q$. Chung and Lu~\cite{ChungLu01} showed that if $Nq\to\infty$, then $\diam(G(N,q))=(1+o(1))\ln N/\ln(Nq)$ with probability $1-o(1)$. For our analysis, we need an upper bound of the same order whose failure probability is explicit and exponentially small in $Nq$. The following lemma provides it; its proof is the standard expansion argument, which we include in Appendix \ref{app:er-diam}.

\begin{lemma}\label{lem:er-diam}
There is an absolute constant $c'\ge1$ such that the following holds. Let $0<q\le1/2$ and $\nu=Nq\ge c'(\ln N+1)$. Then $X\sim G(N,q)$ satisfies
\[
  \Prob\bigl[\diam(X)>2r+1\bigr]\le3e^{-\nu/c'},\qquad\text{where } r=\ceil{\frac{\ln N}{\ln(\nu/8)}}+1.
\]
\end{lemma}

\paragraph{MIS on graphs of bounded degree.}We use the following MIS algorithm as a black box in our result. This result was reproved in Theorem $7$ of \cite{Ghaffari26} and also appeared previously in \cite{Ghaffari24} and \cite{Ghaffari16}.

\begin{proposition}[Theorem 7 of \cite{Ghaffari26}]\label[proposition]{prop_Ghaffari}
For any $n$-vertex graph $H$ with maximum degree $\Delta$, there exists a randomized LOCAL algorithm that computes an MIS with high probability in $O(\log \Delta) +\widetilde{O}(\log^3 \log n)$ rounds.
\end{proposition}

\paragraph{A Gathering Primitive.} Let $H$ be a graph whose vertices have distinct identifiers and communicate in the LOCAL model over $H$. Let $A\subseteq V(H)$ be a set of vertices such that every vertex knows whether it lies in $A$, and for $v\in A$, let $C_v$ denote the component of $H[A]$ containing $v$. We use a deterministic primitive in which every $v\in A$ learns $C_v$ whenever every vertex of $C_v$ is within distance $r$ of $v$ in $H$, within $r+1$ rounds. The algorithm relays messages through all vertices and edges of $H$, including those outside $A$. The algorithm also uses the fact that LOCAL messages have unbounded size. It must also be safe when this distance condition fails: $v$ checks that what it has learned is complete and, if not, reports failure, so it never outputs an incorrect graph.

\begin{lemma}[Gathering]\label{lem:gather}
Let $H$ be a graph and $A\subseteq V(H)$ be a set of vertices such that every vertex knows whether it lies in $A$, and let $r\ge1$ be an integer. The procedure $\textsc{Gather}(H,A,r)$ below is deterministic and takes $r+1$ rounds on $H$. At its end, every $v\in A$ outputs either a graph or the symbol $\bot$, and:
\begin{enumerate}[label=(\roman*)]
  \item if $v$ outputs a graph, this graph is $C_v$;
  \item if every vertex of $C_v$ is within distance $r$ of $v$ in $H$, then $v$ outputs $C_v$.
\end{enumerate}
\end{lemma}

\paragraph{The procedure $\textsc{Gather}(H,A,r)$.}
In one round, every vertex $v\in A$ learns which of its neighbors lie in $A$, and forms its \emph{adjacency list}
\[
  \mathrm{adj}_A(v)=\bigl(\mathrm{ID}(v),\ \set{\mathrm{ID}(u):u\in N_H(v)\cap A}\bigr).
\]
\begin{enumerate}[label=(\arabic*)]
  \item \emph{Collect.} Every $v\in A$ sends $\mathrm{adj}_A(v)$. For $r$ rounds, every vertex of $H$, in $A$ or not, forwards all lists it has received.
  \item \emph{Check.} For $v\in A$, let $M_v$ be the graph whose vertices are the identifiers with a list received by $v$, including its own, and whose edges are the pairs reported by these lists. Let $K_v$ be the component of $M_v$ containing $v$, and let $\mathcal L_v$ be the set of lists of the vertices of $K_v$. Note that the graph $M_v$ may contain vertices of other components of $H[A]$, since lists travel through all of $H$; but no list reports an edge of $H[A]$ between different components, so these vertices are not in $K_v$. Every identifier in the lists belongs to a vertex of $A$, since a list reports only neighbors in $A$. The set $\mathcal L_v$ is \emph{closed} if every vertex that appears as a neighbor in a list of $\mathcal L_v$ also has its own list in $\mathcal L_v$.
  \item \emph{Output.} If $\mathcal L_v$ is closed, $v$ outputs $K_v$. Otherwise, $v$ outputs $\bot$.
\end{enumerate}

\begin{proof}[Proof of \Cref{lem:gather}]
The procedure is deterministic and requires $r+1$ rounds: one round to create the lists, followed by $r$ rounds of Collect. The Check and Output steps are local, and no step involves randomness.

\emph{(i)} Let $v\in A$ be such that $\mathcal L_v$ is closed. Every list is the true adjacency list in $H[A]$ of a vertex of $A$, so $K_v$ is a connected subgraph of $H[A]$ containing $v$, and hence a subgraph of $C_v$. By definition of $\mathcal L_v$ being closed, every neighbor in $H[A]$ of a vertex of $K_v$ has its list in $\mathcal L_v$, so it is a vertex of $K_v$, and the edge between them is reported. Following edges of $H[A]$ from $v$ therefore never leaves $K_v$, so $K_v=C_v$.

\emph{(ii)} Assume that every vertex of $C_v$ is within distance $r$ of $v$ in $H$. A list travels one edge per round, so $v$ receives the list of every vertex of $C_v$ during the Collect step. These lists report all edges of $C_v$, so $C_v$ is a connected subgraph of $M_v$ containing $v$. No list reports an edge from $C_v$ to a vertex outside $C_v$, so $K_v=C_v$. The lists of $\mathcal L_v$ name only vertices of $C_v$, all of which have their lists in $\mathcal L_v$. So $\mathcal L_v$ is closed, and $v$ outputs $C_v$.
\end{proof}

%% file: structural.tex
\section{Structural Properties of Smoothed Graphs}\label{sec:structural}

This section proves two properties (that do not depend on any algorithm) of the smoothed graph $\G$. The first (\Cref{sec:seeded}) shows that if the two-hop neighborhood $Z$ of a set $Q$ of vertices of large degree in a subgraph of the base graph $G$ has many vertices, then its diameter in $\G$, that is $\diam(\G [Z])$, is $O(\sqrt{\ln n})$, with very high probability.  The second (\Cref{sec:separation}) shows that, in a fixed graph of maximum degree $e^{O(\sqrt{\ln n})}$, all but a constant number of the vertices of a small random sample are pairwise at distance greater than two, this makes their outcomes in the algorithm independent.
\Cref{sec:upper} uses the first property to show that the high-degree vertices that the algorithm does not handle in its first step are pairwise close in $\G$, and the second to make the success events of different helpers independent.

\subsection{Smoothed Diameter of Large Two-Hop Neighborhoods}\label{sec:seeded}

Let $F$ be a fixed subgraph of the base graph and let $Q$ be a set of vertices of large degree in $F$, called \emph{seeds}. The two-hop neighborhood $Z=N_F^2[Q]$ may have large diameter in $F$, or even be disconnected, since different seeds can be far apart or even lie in different components of $F$. We show that if $Z$ has many vertices, then the random edges reduce its diameter in $\G$ to $O(\sqrt{\ln n})$. All sets in this subsection are fixed before the random edges are drawn, so the only randomness is $R$.

\begin{lemma}[Smoothed diameter of a large two-hop neighborhood]\label{lem:seeded}
Let $F$ be a fixed subgraph of $G$, let $s$ be an integer with $e^{0.49\sqrt{\ln n}}\le s\le \D$, and let $Q\subseteq V(F)$ be a fixed set of seeds with $\deg_F(v)>s$ for every $v\in Q$. If $Z=N_F^2[Q]$ satisfies $\abs Z\ge n/\sqrt s$, then, for all sufficiently large $n$ (depending only on $\beta$),
\[
  \Prob\Bigl[\diam(\G[Z])>200\sqrt{\ln n}\Bigr]\le3\exp\Bigl(-\frac{\lambda\sqrt s}{4c'}\Bigr),
\]
where $c'$ is the constant of \Cref{lem:er-diam}. In particular, the right-hand side is smaller than every fixed inverse polynomial in $n$.
\end{lemma}

The idea is to partition $Z$ into parts of between $s$ and $2s$ vertices, such that any two vertices of the same part are joined by a path of length at most eight in $F[Z]$ (\Cref{lem:cells}). We use these parts in the proof of \Cref{lem:seeded} to create a \emph{quotient graph} $\Gamma$: 
its vertices are the parts, and two vertices (representing two parts) are adjacent if $\G$ has an edge between them. Two parts span at least $s^2$ vertex pairs, so they are adjacent in $\Gamma$ with probability at least $\eta_0=\lambda s^2/(2n)$. This probability is small, but there are $m\ge |Z|/(2s)$ parts, so $\Gamma$ contains an Erd\H{o}s--R\'enyi graph $G(m,\eta_0)$ for which $m\eta_0\ge\lambda\sqrt s/4=e^{\Omega(\sqrt{\ln n})}$. By \Cref{lem:er-diam}, its diameter is $O(\ln m/\ln(m\eta_0))=O(\sqrt{\ln n})$. Finally, a short path in $\Gamma$ lifts to a short path in $\G[Z]$, which implies the diameter upper bound on $\G[Z]$. Throughout the following, $z=\abs Z$.

\begin{lemma}[A balanced partition]\label{lem:cells}
Let $F$ be a graph, let $s\ge1$ be an integer, and let $Q\subseteq V(F)$ be a nonempty set of vertices called \emph{seeds} with $\deg_F(v)>s$ for every $v\in Q$. Let $Z=N_F^2[Q]$ and $z=\abs Z$. Then $Z$ can be partitioned into $m$ sets $W_1,\dots,W_m$, called \emph{parts}, such that
\begin{enumerate}[label=(\roman*)]
  \item $s\le\abs{W_i}\le2s$ for every $i$, and hence $z/(2s)\le m\le z/s$; and
  \item any two vertices of the same part are joined by a path of length at most eight in $F[Z]$.
\end{enumerate}
\end{lemma}

\begin{proof}

\Cref{fig:partition} illustrates the construction. Using the seed vertices, let us build \emph{cores}, each consisting of a seed, called its \emph{center}, and $s$ of its $F$-neighbors. Fix an arbitrary order $c_1,\dots,c_{\abs Q}$ of the seeds, and process them in this order. When $c_i$ is processed, if $c_i$ lies in no core built so far and at least $s$ of its $F$-neighbors lie in no core built so far, we build a core consisting of $c_i$ and $s$ such neighbors; otherwise we build nothing. By construction, the cores are pairwise disjoint. Let $U$ be the set of all vertices of the cores. At least one core is built, since $c_1$ has more than $s$ $F$-neighbors and no core exists before it is processed. Moreover, $U\subseteq Z$, since every vertex of a core is within distance one of its center.

\emph{Now, we show that every vertex of $Z$ is within distance three of $U$ in $F[Z]$.}
Let $y$ be a seed with $y\notin U$. In this case, either $y$ is connected directly to the seed vertex of an already created core (this contradicts $y \notin U$), or $y$ has fewer than $s$ neighbors in $F \setminus U$ while it was processed. As $y$ has more than $s$ $F$-neighbors, at least one of them lies in $U$. Now, as $y \in Z$, we can say that $y$ is at distance one from $U$ in $F[Z]$. This implies that every seed vertex of $Q$ is at distance at most one from $U$ in $F[Z]$. Now let $x \in Z$. Then, by definition of $Z$, $x$ is connected to at least one seed vertex within distance two in $F[Z]$. So, it has distance at most three from $U$ in $F[Z]$.

\emph{Clusters.} Next, we form disjoint clusters of the vertices of $Z$, by assigning a center to every vertex of $Z$. If $x\in U$, then $x$ belongs to exactly one core, and we assign to $x$ the center of that core. Otherwise, $x$ is joined to $U$ by a path of length at most three in $F[Z]$. Let $u\in U$ be a vertex closest to $x$ in $F[Z]$, breaking ties arbitrarily, and assign to $x$ the center of $u$. For each center $c$, the vertices assigned to $c$ form the \emph{cluster} of $c$. The subgraph of $F$ induced by a cluster need not be connected, since a shortest path from $x$ to $u$ may pass through vertices of other clusters. We only need the following properties. Every vertex of $Z$ is assigned exactly one center, so the clusters are pairwise disjoint and cover $Z$. Every vertex $x$ of a cluster is joined to its center by a path of length at most four in $F[Z]$: a path of length at most three from $x$ to $u$, followed by the edge from $u$ to its center if $u$ is not the center itself. Finally, each cluster contains its core, and hence at least $s+1$ vertices.

\emph{Finally, we split the clusters into parts.} Split a cluster with $w\ge s+1$ vertices into $g=\floor{w/s}\ge1$ parts whose sizes differ by at most one, so each size is $\floor{w/g}$ or $\ceil{w/g}$. Since $g\le w/s$, each size is at least $s$. Since $w<(g+1)s\le2gs$, each size is at most $2s$. This proves (i); the bounds on $m$ follow since the part sizes sum to $z$. Two vertices of one part lie in the same cluster, and joining their paths to its center gives a path of length at most eight in $F[Z]$. This proves (ii).
\end{proof}

\input{fig_partition}
Like the clusters, a part need not be connected; (ii) only says that its vertices are close in $F[Z]$. The parts depend only on $F$ and are used only in the analysis. We are now ready to prove \Cref{lem:seeded}.

\begin{proof}[Proof of \Cref{lem:seeded}]
Since $z\ge n/\sqrt s>0$ and $Z=N^2_F[Q]$, we have $Q\ne\emptyset$. Fix the parts $W_1,\dots,W_m$ of \Cref{lem:cells}, and let $\Gamma$ be their \emph{quotient graph}: its vertices are $1,\dots,m$, and $i\ne j$ are adjacent if $\G$ has an edge between $W_i$ and $W_j$.

\emph{The quotient graph contains $G(m,\eta_0)$.} Since $s\le \D$, we have $ps^2\le\lambda \D^2/n \to 0$; so for large $n$, $\xi:=ps^2\le1$. Let $\eta_0=\xi/2=\lambda s^2/(2n)$. If $G$ has an edge between $W_i$ and $W_j$, the parts are adjacent. Otherwise, all $\abs{W_i}\abs{W_j}\ge s^2$ pairs between them are missing from $G$ and enter $R$ independently, so the parts are adjacent with probability at least $1-(1-p)^{s^2}\ge1-e^{-\xi}\ge \xi/2=\eta_0$, using $e^{-\xi}\le1-\xi/2$ for $0\le \xi\le1$. Different pairs of parts use disjoint sets of vertex pairs, so these adjacencies are independent. 
Hence we may realize $\Gamma$ using independent random numbers $\upsilon_{ij}$, each uniform on $[0,1]$, by declaring $i,j$ adjacent when $\upsilon_{ij}\le\Prob[i,j\text{ adjacent}]$. Let $X$ be the graph on $\set{1,\dots,m}$ in which $i,j$ are adjacent when $\upsilon_{ij}\le\eta_0$. Then $X\sim G(m,\eta_0)$, and $X\subseteq\Gamma$ since $\eta_0\le\Prob[i,j\text{ adjacent}]$. Hence $\diam(\Gamma)\le\diam(X)$.

\emph{Bounding the diameter of $X$.} We apply \Cref{lem:er-diam} with $N=m$ and $q=\eta_0\le1/2$. Using $m\ge z/(2s)$ and $z\ge n/\sqrt s$,
\[
  \nu:=m\eta_0\ \ge\ \frac z{2s}\cdot\frac{\lambda s^2}{2n}\ \ge\ \frac{\lambda\sqrt s}4\ \ge\ \frac{\ln n}4\,e^{0.245\sqrt{\ln n}},
\]
since $\lambda\ge\ln n$. This is at least $c'(\ln n+1)\ge c'(\ln m+1)$ for large $n$. So, with $r=\ceil{\ln m/\ln(\nu/8)}+1$, we get $\diam(X)\le2r+1$ except with probability $3e^{-\nu/c'}\le3\exp(-\lambda\sqrt s/(4c'))$.
Now since $s\ge e^{0.49\sqrt{\ln n}}$ and $\lambda\ge\ln n$,
\[
  \ln\frac\nu8\ \ge\ \tfrac12\ln s+\ln\lambda-\ln32\ \ge\ 0.245\sqrt{\ln n}+\ln\ln n-\ln32\ \ge\ 0.24\sqrt{\ln n}
\]
for large $n$. With $\ln m\le\ln n$, this gives $r\le\sqrt{\ln n}/0.24+2$, so $2r+1\le8.4\sqrt{\ln n}+5$.

\emph{Putting everything together.} We now bound the diameter of $\G[Z]$. Suppose that $\diam(\Gamma)\le2r+1$, and let $u,v\in Z$. Take a shortest path $i_0,\dots,i_\tau$ in $\Gamma$ from the part of $u$ to the part of $v$, so $\tau\le2r+1$, and for each $j$ an edge $x_jy_j$ of $\G$ with $x_j\in W_{i_{j-1}}$ and $y_j\in W_{i_j}$. With $y_0=u$ and $x_{\tau+1}=v$, the vertices $y_j$ and $x_{j+1}$ lie in the same part, so by \Cref{lem:cells}(ii) they are joined by a path of length at most eight in $F[Z]$. Concatenating these paths with the edges $x_jy_j$ gives a walk from $u$ to $v$ in $\G[Z]$ of length at most
\[
  8(\tau+1)+\tau\ \le\ 9(2r+1)+8\ \le\ 75.6\sqrt{\ln n}+53\ <\ 200\sqrt{\ln n}.
\]
So $\diam(\Gamma)\le2r+1$ implies $\diam(\G[Z])\le200\sqrt{\ln n}$. Since $X\subseteq\Gamma$, the event $\diam(\G[Z])>200\sqrt{\ln n}$ therefore implies $\diam(X)>2r+1$. By \Cref{lem:er-diam}, applied to $X\sim G(m,\eta_0)$ in the previous paragraph, this has probability at most $3e^{-\nu/c'}\le3\exp(-\lambda\sqrt s/(4c'))$. All conditions ``for large $n$'' above involve only $n$ and $\beta$. Finally, $\lambda\sqrt s\ge\ln n\cdot e^{0.245\sqrt{\ln n}}$, so this bound is smaller than every fixed inverse polynomial in $n$.

\end{proof}

\begin{remark}\label{rem:ahk}
Partitioning into clusters around high-degree vertices and then analyzing a random quotient graph is adapted from the diameter arguments of Aigner-Horev, Hefetz and Krivelevich~\cite{AHK25}. \Cref{lem:seeded} extends the idea to two-hop neighborhoods of high-degree seeds in an arbitrary fixed subgraph, with explicit failure probabilities.
\end{remark}

\subsection{Most Randomly Sampled Vertices Are Isolated}\label{sec:separation}
Let $F_0$ be a fixed graph with $V(F_0)\subseteq V$, so that $F_0$ has at most $n$ vertices, and with maximum degree at most $\Delta\ge1$, and let $V_0\subseteq V(F_0)$ be fixed.  We show that, for a fixed constant $q$, fewer than $q$ sampled vertices have another sampled vertex within distance two in $F_0$, except with probability $n^{-q/2+o(1)}$ (\Cref{lem:sep}). 

\begin{definition}[Squares, isolated vertices, witnesses]\label{def:witness}
The \emph{square} $F_0^2$ of $F_0$ is the graph on $V(F_0)$ in which two distinct vertices $u,v$ are adjacent if $\dist_{F_0}(u,v)\le2$. For $P\subseteq V(F_0)$, a vertex $v\in P$ is \emph{isolated} in $F_0^2[P]$ if $\dist_{F_0}(u,v)>2$ for every $u\in P\setminus\set v$, and \emph{non-isolated} otherwise. A component of a graph is \emph{trivial} if it has exactly one vertex. A set $W\subseteq V(F_0)$ is a \emph{witness} if no component of $F_0^2[W]$ is trivial.
\end{definition}

By definition, a vertex of $P$ is isolated in $F_0^2[P]$ exactly when it forms a trivial component of $F_0^2[P]$. So the non-isolated vertices of $F_0^2[P]$ are exactly the vertices of the nontrivial components of $F_0^2[P]$. The following lemma is the main result of this subsection.

\begin{lemma}[Few sampled vertices are close]\label[lemma]{lem:sep}
Let $F_0$ be a fixed graph with at most $n$ vertices and maximum degree at most $\Delta\ge1$, and let $V_0\subseteq V(F_0)$ be fixed. Let $1\le k\le n$, and let $\mathcal B$ be the random set that contains each vertex of $V_0$ independently with probability $k/n$. Then, for every integer $q\ge2$,
\[
  \Prob\Bigl[\text{at least $q$ vertices of $\mathcal B$ are non-isolated in }F_0^2[\mathcal B]\Bigr]
  \ \le\ \sum_{j=q}^{2q-1}(j+1)^j\Bigl(\frac{\Delta^4k^2}{n}\Bigr)^{j/2}.
\]
In particular, if $\ln\Delta=O(\sqrt{\ln n})$, $k=O(\ln n)$ and $q$ is fixed, then the right-hand side is at most $n^{-q/2+o(1)}$, where the $o(1)$ term depends only on $q$ and the implied constants, and not on $F_0$ or $V_0$.
\end{lemma}

We prove \Cref{lem:sep} by a union bound over small sets that certify the bad event. \Cref{lem:witness} shows that whenever at least $q$ sampled vertices are non-isolated, the sample contains a witness of size between $q$ and $2q-1$. \Cref{lem:witness-count} bounds the number of such witnesses. Finally, a fixed witness of size $j$ lies entirely in the sample with probability at most $(k/n)^j$. A union bound over all witnesses of size between $q$ and $2q-1$ then bounds the probability that at least $q$ sampled vertices are non-isolated by the sum in \Cref{lem:sep}.

\begin{lemma}[A bad sample contains a small witness]\label[lemma]{lem:witness}
Let $P\subseteq V(F_0)$, and let $q\ge2$ be an integer. If at least $q$ vertices of $P$ are non-isolated in $F_0^2[P]$, then $P$ contains a witness $W$ with $q\le\abs W\le2q-1$.
\end{lemma}

\begin{proof}
The non-isolated vertices of $F_0^2[P]$ are the vertices of its nontrivial components, and there are at least $q$ of them. We distinguish two cases.

\emph{Some nontrivial component $K$ of $F_0^2[P]$ has at least $q$ vertices.} Run a breadth-first search in $K$ from any vertex, and let $W$ be the set of the first $q$ vertices it visits. Every vertex of $W$ other than the first is adjacent in $F_0^2$ to its BFS parent, which was visited earlier and hence lies in $W$. So $F_0^2[W]$ is connected. Since $\abs W=q\ge2$, its only component is nontrivial, and $W$ is a witness with $\abs W=q$.

\emph{Every nontrivial component of $F_0^2[P]$ has fewer than $q$ vertices.} Starting from $W=\emptyset$, add whole nontrivial components of $F_0^2[P]$ to $W$, one at a time, and stop as soon as $\abs W\ge q$. This happens, since these components together contain at least $q$ vertices. Just before the last addition, $\abs W\le q-1$, and the last component added has at most $q-1$ vertices, so $\abs W\le2q-2$. There are no edges of $F_0^2$ between different components of $F_0^2[P]$, so the components of $F_0^2[W]$ are exactly the components added. Each of them is nontrivial, so $W$ is a witness.

In both cases $W\subseteq P$ and $q\le\abs W\le2q-1$.
\end{proof}

\begin{lemma}[Counting witnesses]\label[lemma]{lem:witness-count}
Let $F_0$ be a graph with at most $n$ vertices and maximum degree at most $\Delta\ge1$, and let $j\ge1$. The number of witnesses $W\subseteq V(F_0)$ with $\abs W=j$ is at most
\[
  (j+1)^j\,n^{j/2}\,\Delta^{2j}.
\]
\end{lemma}

\begin{proof}
A witness is defined through $F_0^2$: every vertex of a witness $W$ has another vertex of $W$ within distance two in $F_0$, but not necessarily adjacent to it in $F_0$. So, once one vertex of $W$ is known, each vertex of $W$ linked to it lies among its neighbors in $F_0^2$, not only among its neighbors in $F_0$.

A vertex of $F_0$ has at most $\Delta$ neighbors, and at most $\Delta(\Delta-1)$ vertices at distance exactly two, each reached through one of these neighbors. So every vertex has at most $\Delta+\Delta(\Delta-1)=\Delta^2$ neighbors in $F_0^2$.

Let us now fix a witness $W$ with $\abs W=j$. In each component of $F_0^2[W]$, choose a spanning tree and a root. These trees form a spanning forest of $F_0^2[W]$. Each tree has at least two vertices, since $W$ is a witness, so the number $\rho$ of trees satisfies $\rho\le j/2$.

\emph{Labels.} Now, let us list the trees one after another, and list the vertices of each tree in breadth-first order from its root. This gives labels $1,\dots,j$ to the vertices of $W$ such that the parent of every non-root vertex has a smaller label. Record the forest by the \emph{parent map}
\[
  \sigma\colon\set{1,\dots,j}\to\set{0,1,\dots,j},
\]
where $\sigma(i)=0$ if vertex $i$ is a root, and $\sigma(i)$ is the label of its parent otherwise. Every parent map obtained in this way satisfies $\sigma(i)<i$ whenever $\sigma(i)\ne0$, and $\sigma(i)=0$ for at most $j/2$ labels $i$. We call a map with these two properties \emph{admissible}.

\emph{Encoding $W$.} The witness $W$ is determined by its admissible parent map $\sigma$ together with the sequence $w_1,\dots,w_j$ of its vertices in label order. There are at most $(j+1)^j$ admissible maps $\sigma$, since each of the $j$ labels is sent to one of $j+1$ values. Fix an admissible $\sigma$, and let $\rho\le j/2$ be its number of zeros. Choose $w_1,\dots,w_j$ in this order:
\begin{itemize}
  \item if $\sigma(i)=0$, then $w_i$ is any vertex of $F_0$, so there are at most $n$ choices;
  \item otherwise, $w_i$ is a neighbor in $F_0^2$ of $w_{\sigma(i)}$, which is already chosen since $\sigma(i)<i$; so there are at most $\Delta^2$ choices.
\end{itemize}
This gives at most $n^{\rho}\Delta^{2(j-\rho)}\le n^{j/2}\Delta^{2j}$ sequences, using $\rho\le j/2$, $n\ge1$ and $\Delta\ge1$.

\emph{Conclusion.} Every witness of size $j$ arises from at least one pair consisting of an admissible map $\sigma$ and a sequence $w_1,\dots,w_j$. Some pairs may give no witness or the same witness twice; this only increases the count. Hence there are at most $(j+1)^j\,n^{j/2}\,\Delta^{2j}$ witnesses of size $j$.
\end{proof}

We are now ready to prove the main result of this subsection.
\begin{proof}[Proof of \Cref{lem:sep}] Suppose that at least $q$ vertices of $\mathcal B$ are non-isolated in $F_0^2[\mathcal B]$. By \Cref{lem:witness} with $P=\mathcal B$, the set $\mathcal B$ contains a witness $W$ with $q\le\abs W\le2q-1$. So it suffices to bound the probability that $\mathcal B$ contains such a witness.

Now, fix a witness $W$ with $\abs W=j$. If $W\not\subseteq V_0$, then $W\not\subseteq\mathcal B$, since $\mathcal B\subseteq V_0$. Otherwise, the $j$ vertices of $W$ lie in $\mathcal B$ independently, each with probability $k/n$, so
\[
  \Prob[W\subseteq\mathcal B]\le\Bigl(\frac kn\Bigr)^j .
\]

 By \Cref{lem:witness-count}, for each $j$ there are at most $(j+1)^jn^{j/2}\Delta^{2j}$ witnesses of size $j$. Summing over $j=q,\dots,2q-1$,
\[
  \Prob[\text{at least $q$ non-isolated}]
  \ \le\ \sum_{j=q}^{2q-1}(j+1)^j\,n^{j/2}\,\Delta^{2j}\Bigl(\frac kn\Bigr)^j
  \ =\ \sum_{j=q}^{2q-1}(j+1)^j\Bigl(\frac{\Delta^4k^2}{n}\Bigr)^{j/2},
\]
where the last step uses $n^{j/2}\Delta^{2j}k^jn^{-j}=(\Delta^4k^2/n)^{j/2}$.

Now let $\ln\Delta=O(\sqrt{\ln n})$, $k=O(\ln n)$ and $q$ be fixed, and put $y=\Delta^4k^2/n$. Then
\[
  \ln\bigl(\Delta^4k^2\bigr)\ =\ 4\ln\Delta+2\ln k\ =\ O\bigl(\sqrt{\ln n}\bigr)+O(\ln\ln n)\ =\ o(\ln n),
\]
and hence $y\le n^{-1+o(1)}$. In particular, $y\le1$ for all sufficiently large $n$, so $y^{j/2}\le y^{q/2}$ for every $j\ge q$. Moreover, $(j+1)^j\le(2q)^{2q}$ for $j\le2q-1$, and the sum has $q$ terms. Since $q$ is fixed,
\[
  \sum_{j=q}^{2q-1}(j+1)^j\,y^{j/2}\ \le\ q\,(2q)^{2q}\,y^{q/2}\ \le\ n^{-q/2+o(1)} .
\]
The $o(1)$ terms depend only on $q$ and the implied constants, and not on $F_0$ or $V_0$.
\end{proof}

%% file: fig_partition.tex
\begin{figure}[hpt!]
\centering
\begin{tikzpicture}[
  seed/.style={circle,fill=black,inner sep=0pt,minimum size=6pt},
  vtx/.style={circle,draw=black,fill=white,inner sep=0pt,minimum size=6pt},
  pv/.style={circle,draw=black,inner sep=0pt,minimum size=7pt,line width=0.6pt},
  core/.style={line width=1.2pt},
  par/.style={->,>=stealth,line width=0.8pt,blue!60!black},
  faint/.style={line width=0.5pt,gray!50},
  clu/.style={draw=gray!60,fill=gray!12,rounded corners=8pt,inner sep=7pt},
  lbl/.style={font=\small},
  ttl/.style={font=\small\bfseries,anchor=west},
  capt/.style={font=\small,align=center,text width=6.4cm}
]

\def\coords{
  \coordinate (c1) at (0,0);      \coordinate (l11) at (-0.8,0.7);  \coordinate (l12) at (-0.8,-0.7);
  \coordinate (c2) at (3.3,0);    \coordinate (l21) at (4.1,0.7);   \coordinate (l22) at (4.1,-0.7);
  \coordinate (y) at (1.35,0.75); \coordinate (x1) at (1.35,1.75);  \coordinate (x2) at (-1.75,0.0);
  \coordinate (x3) at (5.0,0.0);  \coordinate (x4) at (2.55,-0.9);
}
\begin{scope}
  \coords
  \begin{scope}[on background layer]
    \node[clu,fit=(c1)(l11)(l12)(y)(x1)(x2)] {};
    \node[clu,fit=(c2)(l21)(l22)(x3)(x4)] {};
  \end{scope}
  \draw[faint] (y)--(c2) (x2)--(l12) (l21)--(x3);
  \draw[core] (c1)--(l11) (c1)--(l12) (c2)--(l21) (c2)--(l22);
  \draw[par,shorten >=3pt,shorten <=3pt] (y)--(c1);
  \draw[par,shorten >=3pt,shorten <=3pt] (x1)--(y);
  \draw[par,shorten >=3pt,shorten <=3pt] (x2)--(l11);
  \draw[par,shorten >=3pt,shorten <=3pt] (x3)--(l22);
  \draw[par,shorten >=3pt,shorten <=3pt] (x4)--(c2);
  \foreach \v in {c1,c2,y} \node[seed] at (\v) {};
  \foreach \v in {l11,l12,l21,l22,x1,x2,x3,x4} \node[vtx] at (\v) {};
  \node[lbl,below=3pt] at (c1) {$c$};
  \node[lbl,below=4pt] at (c2) {$c'$};
  \node[lbl,right=3pt] at (y) {$y$};
  \node[ttl] at (-2.0,2.55) {(a) cores and clusters ($s=2$)};
  \node[capt,anchor=north] at (1.6,-1.5) { Seeds are filled, and the cores, with centers $c$ and $c'$, are drawn thick. Arrows show shortest paths to the cores, and the two clusters are shaded.};
\end{scope}

\begin{scope}[xshift=8.3cm]
  \coords
  \begin{scope}[on background layer]
    \node[clu,fill=white,fit=(c1)(l11)(l12)(y)(x1)(x2)] {};
    \node[clu,fill=white,fit=(c2)(l21)(l22)(x3)(x4)] {};
  \end{scope}
  \draw[faint] (y)--(c2) (x2)--(l12) (l21)--(x3);
  \draw[faint] (c1)--(l11) (c1)--(l12) (c2)--(l21) (c2)--(l22) (y)--(c1) (x1)--(y) (x2)--(l11) (x3)--(l22) (x4)--(c2);
 
  \foreach \v in {c1,l11} \node[pv,fill=blue!35] at (\v) {};
  \foreach \v in {l12,x2} \node[pv,fill=orange!55] at (\v) {};
  \foreach \v in {y,x1}   \node[pv,fill=green!45!black!35] at (\v) {};
 
  \foreach \v in {c2,l21}    \node[pv,fill=red!40] at (\v) {};
  \foreach \v in {l22,x3,x4} \node[pv,fill=yellow!70!orange!60] at (\v) {};
  \node[lbl,below=4pt] at (c1) {$c$};
  \node[lbl,below=4pt] at (c2) {$c'$};
  \node[ttl] at (-2.0,2.55) {(b) splitting the clusters into parts};
  \node[capt,anchor=north] at (1.6,-1.5) {Colors show the parts. A cluster of $w$ vertices is split into $\floor{w/s}$ parts, each of size between $s$ and $2s$.};
\end{scope}
\end{tikzpicture}
\caption{The construction in the proof of \Cref{lem:cells}, with $s=2$. (a)~Cores are built greedily, i.e., each core consists of a seed, called its center, together with $s$ of its $F$-neighbors. The seed $y$ lies in no core, but it has an $F$-neighbor in a core. Every other vertex of $Z$ is assigned the center of a closest core vertex, and the arrows show shortest paths to the cores. The vertices assigned to the same center form a cluster, and each of them is within distance four of its center in $F[Z]$. (b)~Each cluster is split into parts of size between $s$ and $2s$. The left cluster has $6$ vertices and is split into $3$ parts of size $2$; the right cluster has $5$ vertices and is split into parts of sizes $2$ and $3$. A part need not be connected, but any two of its vertices are joined, through the center of their cluster, by a path of length at most eight in $F[Z]$.}
\label{fig:partition}
\end{figure}

%% file: upper_bound.tex
\section{An $O(\sqrt {\log n})$-round Smoothed MIS Algorithm}\label{sec:upper}

This section proves that, for $\lambda=\beta\ln n$ with a suitable constant $\beta$, an MIS of the smoothed graph $\G$ can be computed in $O(\sqrt{\log n})$ rounds. The algorithm is given only $\G$, the number $n$ and distinct identifiers; it knows neither which edges of $\G$ are random nor the degrees in the base graph $G$.

\begin{restatable}{theorem}{upperthm}\label{thm:upper}
For every constant $c>0$, there is a constant $\beta\ge1$, depending only on $c$, such that the following holds for all sufficiently large $n$. Let $G$ be any graph on $n$ vertices, and let $\G$ be its smoothing with $\lambda=\beta\ln n$. There is a randomized LOCAL algorithm that, given $\G$, the number $n$, and distinct identifiers of $O(\log n)$ bits, computes an MIS of $\G$ in $O(\sqrt{\log n})$ rounds with probability at least $1-n^{-c}$. The probability is over the random edges and the random bits of the algorithm.
\end{restatable}

The constant $\beta$ grows with $c$, i.e., a smaller failure probability requires more random edges.

We describe the algorithm in \Cref{sec:alg} and prove \Cref{thm:upper} in \Cref{sec:alg-analysis}. In its second step, the algorithm uses the gathering primitive of \Cref{lem:gather}.

\subsection{Algorithm Description}\label{sec:alg}

  The algorithm is given $\G$, the number $n$, and distinct identifiers. The constant $\beta$ of \Cref{thm:upper} is part of the algorithm: with $\lambda=\beta\ln n$, every vertex computes the following three values:
\[
  B=\D+2\lambda,\qquad \theta=\frac{\ceil{\lambda/2}}{\lambda},\qquad \ell=1000\ceil{\sqrt{\ln n}}.
\]
Here $B$ is a degree threshold, $\theta$ is a marking probability, and $\ell$ is a round budget. For $\lambda>2$ we have $\ceil{\lambda/2}<\lambda$, so $\theta<1$ is a valid probability. 

\begin{definition}[High and low vertices, $V_{\hi}$ and $V_{lo}$]
    A vertex $v$ is \emph{high} if $\deg_\G(v)>B$, and \emph{low} otherwise. We write $V_{\hi}$ to denote the set of high vertices and $V_{\lo}$ to denote the set of low vertices.
\end{definition}

Given a graph $\G$, the output of the algorithm should be a maximal independent set, $I$, of $\G$. Since $\log B=O(\sqrt{\ln n})$, \Cref{prop_Ghaffari} computes an MIS of a graph of maximum degree at most $B$ in $O(\sqrt{\ln n})$ rounds. So the algorithm first computes an independent set that, w.h.p., dominates all high vertices. The low vertices that this set does not dominate induce a subgraph of $\G$ of maximum degree at most $B$, and the algorithm runs \Cref{prop_Ghaffari} on it.

\paragraph{Description.}
\Cref{alg:main} gives the algorithm, and \Cref{fig:steps} illustrates it. After a set-up round, the algorithm has three steps, and each step adds an independent set to the output: $S$, $J$ and $K$, respectively. The output is $I=S\cup J\cup K$. 

\emph{Set-up.} Every vertex tells its neighbors whether it is high or low.

\emph{Step 1 (helpers).} Every high vertex marks each of its edges to a low neighbor independently with probability $\theta$, and tells that neighbor. A low vertex with at least one marked edge is a \emph{candidate}. A candidate joins the set $S$ if none of its neighbors is a candidate. The vertices of $S$ are called \emph{helpers}.

\emph{Step 2 (gathering).} A high vertex with no neighbor in $S$ is \emph{active}, and $\AAA=V_{\mathrm{hi}}\setminus N_\G[S]$ is the set of active vertices. All vertices of $\G$ run $\textsc{Gather}(\G,\AAA,\ell)$ (\Cref{lem:gather}), so that every active vertex $v$ either learns its component $C_v$ of $\G[\AAA]$ or outputs $\bot$. Fix a deterministic rule that maps every graph $C$ with vertex identifiers to an MIS $J_C$ of $C$, for example the greedy MIS in increasing order of identifier. If $v$ learns $C_v$, it computes $J_{C_v}$ and joins $J$ if $v\in J_{C_v}$. If $v$ outputs $\bot$, it does not join $J$.

\emph{Step 3 (completion).} Let $V_{\mathrm{rest}}=V_{\mathrm{lo}}\setminus N_\G[S\cup J]$ be the set of low vertices with no neighbor in $S\cup J$. Every low vertex has degree at most $B$ in $\G$, so $\G[V_{\mathrm{rest}}]$ has maximum degree at most $B$. The algorithm of \Cref{prop_Ghaffari} is run on $\G[V_{\mathrm{rest}}]$ with $\Delta=B$, and its output is $K$.

\begin{remark}[The round budget $\ell$]
All vertices run \textsc{Gather} for the same number $\ell+1$ of rounds, fixed in advance, so that all of them start Step~3 together. \Cref{lem:gather-ok} shows that, w.h.p., either $\AAA=\emptyset$ or all active vertices are pairwise within distance $\ell$ in $\G$, so that \Cref{lem:gather}(ii) applies to every active vertex. The value $\ell=1000\ceil{\sqrt{\ln n}}$ is chosen so that it exceeds the distance $200\sqrt{\ln n}+4$ that arises in the proofs of \Cref{lem:case1,lem:case2a}.
\end{remark}

\begin{algorithm}[hpt!]
\caption{Smoothed MIS}\label{alg:main}
\begin{algorithmic}[1]
\Require the graph $\G$, the number $n$, distinct identifiers.
\Statex \textbf{Set-up: classify by degree}
\State $V_{\mathrm{hi}}\gets\set{v:\deg_\G(v)>B}$, $V_{\mathrm{lo}}\gets V\setminus V_{\mathrm{hi}}$; every vertex tells its neighbors its class.
\Statex \textbf{Step 1: helpers}
\For{all edges $uv$ of $\G$ with $u\in V_{\mathrm{hi}}$, $v\in V_{\mathrm{lo}}$}
  \State $u$ marks $uv$ independently with probability $\theta$ and tells $v$.
\EndFor
\State $Y\gets\set{v\in V_{\mathrm{lo}}: v\text{ has a marked incident edge}}$ \Comment{candidates}
\State $S\gets\set{v\in Y: v\text{ has no neighbor in }Y}$ \Comment{helpers}
\Statex \textbf{Step 2: gather and solve the remaining high vertices}
\State $\AAA\gets V_{\mathrm{hi}}\setminus N_\G[S]$ \Comment{active vertices}
\State Run $\textsc{Gather}(\G,\AAA,\ell)$ \Comment{\Cref{lem:gather}}
\State $J\gets\set{v\in\AAA: v\text{ outputs a graph }C\text{ and }v\in J_C}$ \Comment{fixed MIS rule on each component}
\Statex \textbf{Step 3: finish the low vertices}
\State $V_{\mathrm{rest}}\gets V_{\mathrm{lo}}\setminus N_\G[S\cup J]$
\State Run the algorithm of \Cref{prop_Ghaffari} with $\Delta=B$ on $\G[V_{\mathrm{rest}}]$; let $K$ be its output.
\State \Return $I=S\cup J\cup K$
\end{algorithmic}
\end{algorithm}

\input{fig_overview}

\subsection{Algorithm Analysis}\label{sec:alg-analysis}

In this subsection, we prove \Cref{thm:upper}. Throughout, the base graph $G$ is fixed, and the randomness comes from the random edges $R$ and the random bits of the algorithm. If every high vertex is dominated by $S\cup J$, then Step~3 completes $S\cup J$ to an MIS of $\G$, using the algorithm of \Cref{prop_Ghaffari} on the remaining low vertices. So the main question is whether $S\cup J$ dominates all high vertices.

We first show, in \Cref{lem:J} below, that the set $J$ computed in Step~2 is independent, and that it dominates every active vertex whenever the active vertices are pairwise within distance $\ell$ in $\G$. This reduces the analysis to proving that the following condition holds on the active vertices:
\begin{equation}\label{eq:gather-ok}
  \AAA=\emptyset,\quad\text{or all vertices of $\AAA$ are pairwise within distance $\ell$ in $\G$.}
\end{equation}

\Cref{lem:gather-ok} states that condition~\eqref{eq:gather-ok} holds w.h.p., for every base graph, and \Cref{sec:analysis} proves it. This is the most involved part of the analysis. It splits the base graphs into three cases, using sets defined from $G$ alone. In two of the cases, \Cref{lem:seeded} shows that all active vertices are close together. In the third case, we use the random choice of the helpers in Step~1, together with \Cref{lem:sep}, to show that the helpers dominate every high vertex, so that no vertex is active. Finally, \Cref{sec:proof-upper} combines these results and proves \Cref{thm:upper}.

\paragraph{Parameters of the analysis.}
Fix $c>0$. Let $a\ge 48(c+3)$ be a constant and $q\ge 2c+7$ an integer, also put $\beta=2a$, so that $\lambda=2a\ln n$. Let
\[
  k=\theta\lambda=\ceil{\lambda/2}=\ceil{a\ln n}.
\]

\begin{lemma}\label{lem:J}
The set $J$ computed in Step~2 is independent in $\G$. If condition~\eqref{eq:gather-ok} holds, then $\AAA\subseteq N_\G[J]$.
\end{lemma}

\begin{proof}
We apply \Cref{lem:gather} with $H=\G$, $A=\AAA$ and $r=\ell$. For $v\in\AAA$, let $C_v$ be the component of $\G[\AAA]$ containing $v$.

\emph{Independence.} Let $u,w\in J$ be adjacent in $\G$. Then $u,w\in\AAA$ lie in the same component $C$ of $\G[\AAA]$. Both output a graph (i.e., they do not fail), so by \Cref{lem:gather}(i) both output $C$. Both apply the same deterministic rule to $C$, so they compute the same MIS $J_C$ of $C$, and both lie in $J_C$. This contradicts the independence of $J_C$.

\emph{Domination.} If $\AAA=\emptyset$, there is nothing to show. Otherwise, let $v\in\AAA$. Since condition~\eqref{eq:gather-ok} holds and $C_v\subseteq\AAA$, every vertex of $C_v$ is within distance $\ell$ of $v$ in $\G$. By \Cref{lem:gather}(ii), $v$ outputs $C_v$, and it joins $J$ if and only if $v\in J_{C_v}$. Hence $J\cap C=J_C$ for every component $C$ of $\G[\AAA]$. Since $J_C$ is an MIS of $C$, every vertex of $C$ lies in $N_\G[J]$. So $\AAA\subseteq N_\G[J]$.
\end{proof}

By \Cref{lem:J}, the set $J$ dominates $\AAA$ whenever condition~\eqref{eq:gather-ok} holds. The next lemma shows that this happens w.h.p.; we prove it in \Cref{sec:analysis}.

\begin{restatable}{lemma}{gathering}\label{lem:gather-ok}
For every base graph $G$ and all sufficiently large $n$, condition~\eqref{eq:gather-ok} holds with probability at least $1-n^{-c-1}$.
\end{restatable}

\subsubsection{Proof of \Cref{lem:gather-ok}}\label{sec:analysis}

The analysis splits the base graphs into cases, using sets defined from $G$ alone. These sets are fixed before the random edges are drawn, and the algorithm never computes them. Let
\[
  H_0=\set{v:\deg_G(v)>\D},\qquad L_0=V\setminus H_0,\qquad M=\abs{H_0},
\]
\[
  X_0=N_G(H_0)\cap L_0,\qquad Z_H=N_G^2[H_0],\qquad z_H=\abs{Z_H}.
\]
We call the vertices of $H_0$ and $L_0$ \emph{base-high} and \emph{base-low}. Thus $X_0$ is the set of base-low vertices with a base edge to $H_0$, and $H_0\cup X_0\subseteq Z_H$. In contrast, the sets $V_{hi}$ and $V_{lo}$ of high and low vertices (\Cref{sec:alg}) are defined by the degrees in $\G$, and they depend on the random edges. The next lemma relates the two classifications.

\begin{lemma}[Few random edges per vertex]\label{lem:degree-event}
Let $\mathcal E$ be the event that every vertex has at most $2\lambda$ incident edges in $R$. Then $\mathcal E$ fails with probability at most $ne^{-\lambda/3}=n^{1-\beta/3}$. Moreover, if $\mathcal E$ holds, we have $L_0\subseteq V_{\mathrm{lo}}$, and hence $\AAA\subseteq V_{\mathrm{hi}}\subseteq H_0$.
\end{lemma}

\begin{proof}
\emph{The failure probability.} Fix a vertex $v$. The number of edges of $R$ incident to $v$ is a sum of at most $n-1$ independent indicators, one for each vertex $u\ne v$ with $uv\notin E(G)$, each equal to $1$ with probability $p=\lambda/n$. So its mean is at most $(n-1)\lambda/n\le\lambda$. By \Cref{fact:tails}(F1) with $\mu=\lambda$, it exceeds $2\lambda$ with probability at most $e^{-\lambda/3}$. The event $\mathcal E$ fails only if this happens for some vertex, so a union bound over the $n$ vertices shows that $\mathcal E$ fails with probability at most
\[
  ne^{-\lambda/3}\ =\ n\cdot n^{-\beta/3}\ =\ n^{1-\beta/3},
\]
using $\lambda=\beta\ln n$.

\emph{The inclusions when $\mathcal E$ holds.} Assume that $\mathcal E$ holds, and let $v\in L_0$. Then $v$ has at most $\D$ incident base edges, since $v\notin H_0$, and at most $2\lambda$ incident random edges, by $\mathcal E$. So $\deg_\G(v)\le \D+2\lambda=B$, and $v\in V_{\mathrm{lo}}$. This shows $L_0\subseteq V_{\mathrm{lo}}$. Since $V_{\mathrm{hi}}=V\setminus V_{\mathrm{lo}}$ and $H_0=V\setminus L_0$, it follows that $V_{\mathrm{hi}}\subseteq H_0$. Finally, $\AAA\subseteq V_{\mathrm{hi}}$ by the definition of $\AAA$.
\end{proof}

If $M=0$, then $\AAA=\emptyset$ whenever $\mathcal E$ holds; we treat this case in the proof of \Cref{lem:gather-ok} at the end of this subsection. Otherwise $z_H\ge\D+1$, since a vertex of $H_0$ has more than $\D$ neighbors in $G$. We distinguish three cases, determined by $G$ alone. Case~(1) is $z_H\ge n/\sqrt\D$. Otherwise, we define below a degree threshold $t$ and a set $Z_Q$ of vertices in $G[L_0]$, and let $z_Q=\abs{Z_Q}$; Case~(2a) is $z_Q\ge n/\sqrt t$, and Case~(2b) is $z_Q<n/\sqrt t$. In Cases~(1) and~(2a), we show that all vertices of $H_0$, and hence, if $\mathcal E$ holds, all active vertices, are pairwise close in $\G$. In Case~(2b), we show that the helpers of Step~1 dominate every high vertex, so $\AAA=\emptyset$. \Cref{fig:cases} summarizes the cases.

We use the inclusions of \Cref{lem:degree-event} in the following way only. In the proof of \Cref{lem:gather-ok}, each case bounds the probability that condition~\eqref{eq:gather-ok} fails by a union bound in which $\Prob[\mathcal E^c]$ is one of the terms; the other terms are probabilities of events that do not involve $\mathcal E$. We never condition on $\mathcal E$, so the edges of $R$ remain independent in all later calculations.

\input{fig_case}

\paragraph{Case 1: $z_H\ge n/\sqrt \D$.}

\begin{lemma}[Case (1)]\label{lem:case1}
If $M>0$ and $z_H\ge n/\sqrt \D$, then, for all sufficiently large $n$, condition~\eqref{eq:gather-ok} holds with probability at least $1-n^{-c-1}$.
\end{lemma}

\begin{proof}
\emph{All vertices in $Z_H$ are pairwise within distance $\ell$ in $\G$.} We apply \Cref{lem:seeded} with $F=G$, $Q=H_0$ and $s=\D$. Its hypotheses hold: $e^{0.49\sqrt{\ln n}}\le e^{\sqrt{\ln n}}-1\le \D$ for $n\ge3$, every $v\in H_0$ has $\deg_G(v)>\D$, and $\abs{N_G^2[H_0]}=z_H\ge n/\sqrt \D$. So, except with probability $3e^{-\lambda\sqrt \D/(4c')}$, the diameter of $\G[Z_H]$ is at most $200\sqrt{\ln n}$. On this event, all vertices of $Z_H$ are pairwise within distance $200\sqrt{\ln n}\le\ell$ in $\G[Z_H]$, and hence in $\G$, since $\G[Z_H]$ is a subgraph of $\G$.

\emph{Now, the active vertices lie in $Z_H$.} If $\mathcal E$ and the event in the previous paragraph hold, we have $\AAA\subseteq H_0$ by \Cref{lem:degree-event}, and $H_0\subseteq Z_H$ by the definition of $Z_H$. So, if $\mathcal E$ holds and the event of the previous paragraph, all vertices of $\AAA$ are pairwise within distance $\ell$ in $\G$, and \eqref{eq:gather-ok} holds.

\emph{Let us now argue about the failure probability.}  Condition~\eqref{eq:gather-ok} can fail only if $\mathcal E$ fails, with probability at most $ne^{-\lambda/3}$ by \Cref{lem:degree-event}, or $\G[Z_H]$ has large diameter, with probability at most $3e^{-\lambda\sqrt \D/(4c')}$ by \Cref{lem:seeded}. So it fails with probability at most
\[
  ne^{-\lambda/3}+3e^{-\lambda\sqrt \D/(4c')}.
\]
The first term equals $n\cdot e^{-\beta\ln n/3}=n^{1-\beta/3}$, since $\lambda=\beta\ln n$; this is at most $n^{-c-2}$ once $\beta=2a\ge3(c+3)$, which holds since $a$ is large enough depending on $c$. For the second term, $\lambda=\beta\ln n\ge1$ for $n\ge3$, so $3e^{-\lambda\sqrt \D/(4c')}\le3e^{-\sqrt \D/(4c')}$. Since $\sqrt \D$ grows faster than every multiple of $\ln n$, we have $\sqrt \D/(4c')\ge(c+2)\ln n+\ln3$ for all sufficiently large $n$, and then the second term is at most $n^{-c-2}$. Hence \eqref{eq:gather-ok} fails with probability at most $2n^{-c-2}\le n^{-c-1}$.
\end{proof}

\paragraph{Case (2): $z_H<n/\sqrt \D$.}

In this case, $H_0\subseteq Z_H$ implies $M<n/\sqrt \D$. We now define a degree threshold $t$ for the base-low part of $G$. We need $e^{0.49\sqrt{\ln n}}\le t\le \D$, so that \Cref{lem:seeded} applies with $s=t$ in Case~(2a), and $t\le n/(16Mk)$, which is used in Case~(2b) to bound the failure probability of a test vertex (\Cref{lem:test-succeed}). Since $M<n/\sqrt \D$, all these constraints can be met. Define
\[
  t=\min\set{\D,\ \floor{\frac{n}{16Mk}}},\qquad Q=\set{v\in L_0:\ \deg_{G[L_0]}(v)>t},
\]

and let $Z_Q=N^2_{G[L_0]}[Q]$ and $z_Q=\abs{Z_Q}$. Since $M<n/\sqrt \D$, we have $n/(16Mk)>\sqrt \D/(16k)$, which is at least $1$ for all sufficiently large $n$ (depending on the constant $a$). Using $\floor x\ge x/2$ for $x\ge1$, we get $\floor{n/(16Mk)}\ge\sqrt \D/(32k)$. Together with $\D\ge e^{\sqrt{\ln n}}/2$ and $k=O(\ln n)$, this gives, for all sufficiently large $n$,
\begin{equation}\label{eq:t-range}
  \frac{\sqrt \D}{32k}\le t\le \D,\qquad \ln t\ \ge\ \tfrac12\ln \D-\ln(32k)\ \ge\ \tfrac12\sqrt{\ln n}-O(\ln\ln n)\ \ge\ 0.49\sqrt{\ln n}.
\end{equation}
So $s=t$ is a valid choice in \Cref{lem:seeded}. Case~(2a) is $z_Q\ge n/\sqrt t$, and Case~(2b) is $z_Q<n/\sqrt t$.

\paragraph{Case (2a): $z_Q\ge n/\sqrt t$.}

\begin{lemma}[Case (2a)]\label{lem:case2a}
If $M>0$, $z_H<n/\sqrt \D$ and $z_Q\ge n/\sqrt t$, then, for all sufficiently large $n$, condition~\eqref{eq:gather-ok} holds with probability at least $1-n^{-c-1}$.
\end{lemma}

\begin{proof}
\emph{$Z_Q$ has small diameter.} We apply \Cref{lem:seeded} with $F=G[L_0]$, seeds $Q$ and $s=t$. Its hypotheses hold: $G[L_0]$ is a fixed subgraph of $G$, $e^{0.49\sqrt{\ln n}}\le t\le \D$ by~\eqref{eq:t-range}, every $v\in Q$ has $\deg_{G[L_0]}(v)>t$, and $\abs{N^2_{G[L_0]}[Q]}=z_Q\ge n/\sqrt t$. So, except with probability $3e^{-\lambda\sqrt t/(4c')}$, all vertices of $Z_Q$ are pairwise within distance $200\sqrt{\ln n}$ in $\G[Z_Q]$, and hence in $\G$.

\emph{Every base-high vertex is close to $Z_Q$.} We claim that each $h\in H_0$ is within distance two of $Z_Q$ in $\G$, except with probability $e^{-\lambda\sqrt \D}$. This is clear if a $G$-neighbor of $h$, or a $G$-neighbor of such a vertex, lies in $Z_Q$. Otherwise, fix $\D$ distinct $G$-neighbors $y_1,\dots,y_{\D}$ of $h$. None of them lies in $Z_Q$ or has a $G$-neighbor in $Z_Q$, so the $\D z_Q$ pairs $\set{y_i,z}$ with $z\in Z_Q$ are distinct and are not edges of $G$. Each of them lies in $R$ independently with probability $p$, and if one does, then $h$ is within distance two of $Z_Q$. All of them are missing with probability
\[
  (1-p)^{\D z_Q}\ \le\ e^{-\lambda \D z_Q/n}\ \le\ e^{-\lambda \D/\sqrt t}\ \le\ e^{-\lambda\sqrt \D},
\]
using $1-p\le e^{-p}$, $z_Q\ge n/\sqrt t$ and $t\le \D$. A union bound over the at most $n$ vertices of $H_0$ proves the claim for all $h\in H_0$ at once, except with probability $ne^{-\lambda\sqrt \D}$.

\emph{The active vertices are close.} Suppose that $\mathcal E$ holds, that $Z_Q$ has small diameter, and that every base-high vertex is within distance two of $Z_Q$. By \Cref{lem:degree-event}, $\AAA\subseteq H_0$. So any $h,h'\in\AAA$ are within distance two of vertices $z,z'\in Z_Q$, respectively, and $\dist_\G(z,z')\le200\sqrt{\ln n}$. Hence $\dist_\G(h,h')\le2+200\sqrt{\ln n}+2\le\ell$, and \eqref{eq:gather-ok} holds.

\emph{The failure probability.} Condition~\eqref{eq:gather-ok} can fail only if one of three events fails: $\mathcal E$, which fails with probability at most $ne^{-\lambda/3}$; the event that $Z_Q$ has small diameter, which fails with probability at most $3e^{-\lambda\sqrt t/(4c')}$ by \Cref{lem:seeded}; and the event that every base-high vertex is within distance two of $Z_Q$, which fails with probability at most $ne^{-\lambda\sqrt \D}$ by the union bound above. So, by a union bound, \eqref{eq:gather-ok} fails with probability at most
\[
  ne^{-\lambda/3}+3e^{-\lambda\sqrt t/(4c')}+ne^{-\lambda\sqrt \D}.
\]
We bound each term by $n^{-c-2}$, for all sufficiently large $n$. The first term equals $n\cdot e^{-\beta\ln n/3}=n^{1-\beta/3}$, since $\lambda=\beta\ln n$; this is at most $n^{-c-2}$ once $\beta=2a\ge3(c+3)$, which holds since $a$ is large enough depending on $c$. For the other two terms, $\lambda\ge1$ for $n\ge3$. By~\eqref{eq:t-range}, $\sqrt t\ge \D^{1/4}/\sqrt{32k}$, and both $\sqrt t$ and $\sqrt \D$ grow faster than every multiple of $\ln n$, since $k=O(\ln n)$ and $\D\ge e^{\sqrt{\ln n}}/2$. So, for all sufficiently large $n$,
\[
  3e^{-\lambda\sqrt t/(4c')}\ \le\ 3e^{-\sqrt t/(4c')}\ \le\ n^{-c-2},\qquad ne^{-\lambda\sqrt \D}\ \le\ ne^{-\sqrt \D}\ \le\ n^{-c-2}.
\]
Hence \eqref{eq:gather-ok} fails with probability at most $3n^{-c-2}\le n^{-c-1}$, for $n\ge3$.
\end{proof}

\paragraph{Case (2b): $z_Q<n/\sqrt t$.}
In this case we show that $\AAA=\emptyset$, that is, $S$ dominates every high vertex. Fix $h\in H_0$, and suppose that $h$ is high. A base-low vertex $v$ dominates $h$ if (i)~$hv$ is a random edge that $h$ marks in Step~1, so that $v$ is a candidate, and (ii)~no neighbor of $v$ is a candidate, so that $v\in S$. If $v$ is not a base neighbor of $h$, then $v$ satisfies~(i) with probability $p\theta=k/n$. We proceed in two steps.
\begin{itemize}
     \item In \Cref{lem:helpers}, we show that, with probability at least $1-2n^{-c-3}$, at least $\floor{k/3}$ vertices satisfy (i) and are pairwise at distance at least three in the base-low graph $F_0$ defined below. We call them the \emph{test vertices} of $h$. Since they are far apart, the events that they satisfy (ii) are independent.
    \item In \Cref{lem:test-succeed}, we show that each test vertex fails (ii) with probability at most $1/4$, so all of them fail with probability at most $4^{-\floor{k/3}}$.

\end{itemize}
If some test vertex satisfies~(ii), it joins $S$ and dominates $h$ (\Cref{lem:robust-dom}). So $h$ is dominated with high probability.

\emph{The base-low graph.} Let $F_0=\G[L_0]$ be the subgraph of $\G$ induced on the base-low vertices. It contains both the base edges and the random edges inside $L_0$; the former are fixed by $G$, so $F_0$ is determined by $G$ and the random edges inside $L_0$. Let $X_0^+=N_{F_0}[X_0]$ be the set of base-low vertices that lie in $X_0$ or are adjacent in $F_0$ to a vertex of $X_0$. We use only helpers that are joined to $h$ by a random edge and lie outside $X_0^+$. The reason is that base edges are always present: if a vertex $w$ has a base edge to a high vertex $h'$, then $h'$ marks this edge with probability $\theta\ge1/2$, so $w$ is a candidate with constant probability and may block its neighbors. A pair $h'w$ that is not a base edge, in contrast, is present and marked only with probability $p\theta=k/n$.

\begin{lemma}[The base-low graph is well behaved]\label{lem:F0-good}
Assume $M>0$ and $z_H<n/\sqrt{\D}$. Let $\mathcal E_0$ be the event that the following two properties hold:
\begin{enumerate}[label=(\alph*)]
  \item every base-low vertex has at most $2\lambda$ random neighbors in $L_0$; in particular, $F_0$ has maximum degree at most $\D+2\lambda=B$;
  \item $\abs{X_0^+}\le n/4$.
\end{enumerate}
Then $\mathcal E_0$ depends only on the random edges inside $L_0$, and, for all sufficiently large $n$, $\mathcal E_0$ fails with probability at most $2n^{-c-2}$.
\end{lemma}

\begin{proof}
\emph{Property (a).} This is proved similarly to \Cref{lem:degree-event}, counting only random edges inside $L_0$: by \Cref{fact:tails}(F1) and a union bound over at most $n$ vertices, (a) fails with probability at most $ne^{-\lambda/3}=n^{1-\beta/3}$, which is at most $n^{-c-2}$ since $\beta=2a\ge3(c+3)$. A base-low vertex has at most $\D$ base neighbors, so if (a) holds, it has degree at most $\D+2\lambda$ in $F_0$.

\emph{Property (b).} A vertex of $X_0^+$ lies in $X_0$, or is a base neighbor of a vertex of $X_0$, or is joined to a vertex of $X_0$ by a random edge inside $L_0$. A vertex of the first two kinds is within distance two of $H_0$ in $G$, so it lies in $Z_H$; there are at most $z_H$ of them. Let $Y$ be the number of random edges inside $L_0$ with at least one endpoint in $X_0$. Each vertex of the third kind is the endpoint outside $X_0$ of such an edge, and distinct vertices of the third kind give distinct edges; so there are at most $Y$ of them. Hence
\[
  \abs{X_0^+}\ \le\ z_H+Y.
\]

\emph{If $Y\le2\lambda z_H$, then (b) holds.} In this case,
\[
  \abs{X_0^+}\ \le\ z_H+2\lambda z_H\ <\ \frac{(1+2\lambda)\,n}{\sqrt \D}\ \le\ \frac n4 .
\]
Here the second inequality uses $z_H<n/\sqrt \D$. The last one is equivalent to $\sqrt \D \ge4+8\beta\ln n$, which holds for all sufficiently large $n$ depending only on $c$, since $\sqrt \D\ge e^{\sqrt{\ln n}/2}/\sqrt2$ grows faster than every multiple of $\ln n$.

\emph{Let us now calculate the probability that $Y>2\lambda z_H$.} The variable $Y$ is a sum of independent indicators, one for each pair of vertices of $L_0$ that has an end in $X_0$ and is not a base edge, each equal to $1$ with probability $p$. There are at most $\abs{X_0}\cdot n$ such pairs, so
\[
  \E[Y]\ \le\ \abs{X_0}\cdot n\cdot\frac\lambda n\ \le\ \lambda z_H,
\]
since $X_0\subseteq Z_H$. By \Cref{fact:tails}(F1) with $\mu=\lambda z_H$, we get $\Prob[Y>2\lambda z_H]\le e^{-\lambda z_H/3}$. Since $M>0$, there is a vertex $h\in H_0$ with more than $\D$ neighbors in $G$, and $h$ together with its neighbors lies in $Z_H$, so $z_H\ge \D+1$. Together with $\lambda\ge1$, this gives $e^{-\lambda z_H/3}\le e^{-\D/3}$. Moreover, $\D=\floor{e^{\sqrt{\ln n}}}\ge e^{\sqrt{\ln n}}/2$ for $n\ge2$. Since $e^{\sqrt{\ln n}}$ grows faster than every multiple of $\ln n$, we have $e^{\sqrt{\ln n}}/6\ge(c+2)\ln n$ for all sufficiently large $n$ depending only on $c$. Hence $\D/3\ge(c+2)\ln n$, and
\[
  \Prob[Y>2\lambda z_H]\ \le\ e^{-\D/3}\ \le\ n^{-c-2}.
\]
By the previous paragraph, (b) can fail only if $Y>2\lambda z_H$, so (b) fails with probability at most $n^{-c-2}$.

\emph{Conclusion.} By a union bound over the failures of (a) and (b), $\mathcal{E}$ fails with probability at most $n^{-c-2}+n^{-c-2}=2n^{-c-2}$. Both properties depend only on the random edges inside $L_0$, and hence so does $\mathcal E_0$.
\end{proof}

\emph{The pool.} Fix an outcome of $F_0$ that satisfies $\mathcal E_0$ (mentioned in \Cref{lem:F0-good}), and define the \emph{pool}
\[
  \mathcal U=L_0\setminus\bigl(X_0^+\cup Q\bigr).
\]
The pool consists of the base-low vertices that we use as potential helpers: neither they nor their neighbors in $F_0$ have base edges to $H_0$, and they have few neighbors in $F_0$.

We exclude $Q$ because a vertex with many base-low neighbors is easily blocked by a candidate neighbor. Every $v\in\mathcal U$ has at most $t$ base neighbors in $L_0$, since $v\notin Q$, and at most $2\lambda$ random neighbors in $L_0$, since the fixed outcome of $F_0$ satisfies \Cref{lem:F0-good}(a). The pool is large, since $\abs{L_0}=n-M>n-n/\sqrt \D$ and $\abs Q\le z_Q<n/\sqrt t$. So
\begin{equation}\label{eq:pool}
  \deg_{F_0}(v)\le t+2\lambda\quad(v\in\mathcal U),\qquad \abs{\mathcal U}\ \ge\ n-\frac n{\sqrt \D}-\frac n4-\frac n{\sqrt t}\ \ge\ \frac{2n}3,
\end{equation}
where the last inequality holds for all sufficiently large $n$, because $\D$ and $t$ tend to infinity.

\emph{The sample of a base-high vertex.} Fix $h\in H_0$. The potential helpers of $h$ are the base-low vertices joined to $h$ by a random edge that $h$ marks in Step~1; we call the set of these vertices the \emph{sample} of $h$ and denote it by $\mathcal B_h$. To define it formally, give every pair $\set{u,v}$ a coin $\kappa_{uv}$ equal to $1$ with probability $\theta$, independently of each other and of all random edges. Step~1 marks a present edge $uv$ with $u\in V_{\mathrm{hi}}$ and $v\in V_{\mathrm{lo}}$ exactly when $\kappa_{uv}=1$; this has the same distribution as the algorithm, since every such edge uses its own coin. For $v\in L_0\setminus X_0$, the pair $hv$ is not a base edge, by the definition of $X_0$. Let
\[
  I_{hv}=\mathbf 1\bigl[hv\in R\text{ and }\kappa_{hv}=1\bigr],\qquad \mathcal B_h=\set{v\in L_0\setminus X_0:\ I_{hv}=1}.
\]
Then $\Prob[I_{hv}=1]=p\theta=k/n$, and these indicators are independent over $v$. They depend only on the random edges and coins on pairs $\set{h,v}$, which are not inside $L_0$, so $\mathcal B_h$ is independent of $F_0$. If $h\in V_{\mathrm{hi}}$, every vertex of $\mathcal B_h\cap V_{\mathrm{lo}}$ is a candidate.

\begin{lemma}[Many test vertices]\label{lem:helpers}
Fix an outcome of $F_0$ that satisfies $\mathcal E_0$, and let $h\in H_0$. Then, for all sufficiently large $n$, with probability at least $1-2n^{-c-3}$, at least $\floor{k/3}$ vertices of $\mathcal B_h\cap\mathcal U$ are isolated in $F_0^2[\mathcal B_h]$.
\end{lemma}

\begin{proof}
\emph{Many sampled vertices in the pool.} Let $Y_h=\abs{\mathcal B_h\cap\mathcal U}$ be the number of sampled vertices of $h$ that lie in the pool. Since $\mathcal U\subseteq L_0\setminus X_0$, the variable $Y_h$ is a sum of $\abs{\mathcal U}$ independent indicators $I_{hv}$, each equal to $1$ with probability $k/n$. Its mean is $\mu=\abs{\mathcal U}k/n\ge2k/3$, by~\eqref{eq:pool}. Since $\frac34\mu\ge\frac34\cdot\frac{2k}3=\frac k2$, we have $Y_h<k/2$ only if $Y_h\le\frac34\mu$. By \Cref{fact:tails}(F2) with $\delta=1/4$, this has probability at most
\[
  e^{-\mu/32}\ \le\ e^{-k/48}\ \le\ e^{-a\ln n/48}\ =\ n^{-a/48}\ \le\ n^{-c-3},
\]
using $k\ge a\ln n$ and $a\ge48(c+3)$. Hence $Y_h\ge k/2$, except with probability at most $n^{-c-3}$.

\emph{Few non-isolated sampled vertices.} We apply \Cref{lem:sep} to the fixed graph $F_0$, with $V_0=L_0\setminus X_0$ and $\Delta=B$. The graph $F_0$ has at most $n$ vertices and, since the fixed outcome of $F_0$ satisfies $\mathcal E_0$, $F_0$ has maximum degree at most $B$ by \Cref{lem:F0-good}(a). The sample $\mathcal B_h$ contains each vertex of $V_0$ independently with probability $k/n$, and it is independent of $F_0$. Moreover, $\ln B\le\sqrt{\ln n}+1$ and $k=O(\ln n)$. So, by \Cref{lem:sep}, fewer than $q$ vertices of $\mathcal B_h$ are non-isolated in $F_0^2[\mathcal B_h]$, except with probability $n^{-q/2+o(1)}$. Since $q\ge2c+7$, this is at most $n^{-c-3}$ for all sufficiently large $n$.

\emph{Conclusion.} Suppose that $Y_h\ge k/2$ and that fewer than $q$ vertices of $\mathcal B_h$ are non-isolated in $F_0^2[\mathcal B_h]$. Then, among the $Y_h$ vertices of $\mathcal B_h\cap\mathcal U$, at least $Y_h-q\ge k/2-q$ are isolated in $F_0^2[\mathcal B_h]$. Since $q$ is fixed and $k\ge a\ln n$, we have $k/2-q\ge k/3\ge\floor{k/3}$ for all sufficiently large $n$. By a union bound over the two failure events, this holds with probability at least $1-2n^{-c-3}$.
\end{proof}

\emph{Test vertices.} Let $T_h$ be the set of vertices of $\mathcal B_h\cap\mathcal U$ that are isolated in $F_0^2[\mathcal B_h]$; we call them the \emph{test vertices} of $h$. On the event of \Cref{lem:helpers}, we have $\abs{T_h}\ge\floor{k/3}$.

The set $T_h$ is determined by $F_0$ and $\mathcal B_h$, and every $v\in T_h$ satisfies:
\begin{enumerate}[label=(T\arabic*)]
  \item\label{it:T1} $\deg_{F_0}(v)\le t+2\lambda$ and $N_{F_0}[v]\cap X_0=\emptyset$, since $v\in\mathcal U$;
  \item\label{it:T2} $N_{F_0}(v)\cap\mathcal B_h=\emptyset$, since $v$ is isolated in $F_0^2[\mathcal B_h]$;
  \item\label{it:T3} distinct $u,v\in T_h$ have $\dist_{F_0}(u,v)\ge3$, so $N_{F_0}[u]\cap N_{F_0}[v]=\emptyset$.
\end{enumerate}

If $h$ is high, every test vertex $v$ is a candidate, since $h$ marks the random edge $hv$. So $v$ joins $S$ unless some neighbor of $v$ is a candidate. By~\ref{it:T2}, no neighbor of $v$ is a candidate because of a mark from $h$. A neighbor of $v$ can still be a candidate because of another base-high vertex $h'$, in two ways: $h'$ itself may be a neighbor of $v$ (through a random edge $h'v$), or $h'$ may mark a random edge $h'w$ to a neighbor $w$ of $v$. We do not know which base-high vertices are high, so we exclude both possibilities for all of $H_0\setminus\set h$.

\emph{Robust success.} Call $v\in T_h$ \emph{robustly successful} if
\begin{enumerate}[label=(S\arabic*)]
  \item\label{it:S1} $h'v\notin R$ for every $h'\in H_0\setminus\set h$, and
  \item\label{it:S2} $I_{h'w}=0$ for every $w\in N_{F_0}(v)$ and every $h'\in H_0\setminus\set h$.
\end{enumerate}
By~\ref{it:T1}, all pairs involved are non-edges of $G$. \Cref{fig:case2b} illustrates the test vertices and the two ways in which they can fail. The next lemma shows that robust success is enough: a robustly successful test vertex joins $S$ and dominates $h$.

\input{fig_2b}

\begin{lemma}[Robust success implies domination]\label{lem:robust-dom}
If $\mathcal E$ holds, $h\in V_{\mathrm{hi}}$, and some $v\in T_h$ is robustly successful, then $v\in S$; hence $h\notin\AAA$.
\end{lemma}

\begin{proof}
Since $\mathcal E$ holds, we have $v\in L_0\subseteq V_{\mathrm{lo}}$ and $V_{\mathrm{hi}}\subseteq H_0$ (\Cref{lem:degree-event}). Since $v\in\mathcal B_h$, the edge $hv$ is a random edge that $h$ marks, so $v$ is a candidate. It remains to show that no neighbor $w$ of $v$ is a candidate. The neighbor $h$ is high, so it is not a candidate. Since $v\notin X_0$, no base edge joins $v$ to $H_0$, and by~\ref{it:S1} no random edge joins $v$ to $H_0\setminus\set h$; so every other neighbor $w$ lies in $L_0$, that is, $w\in N_{F_0}(v)$. By~\ref{it:T1}, $w\notin X_0$, so a mark on an edge $h'w$ with $h'\in V_{\mathrm{hi}}\subseteq H_0$ would mean $I_{h'w}=1$. This is excluded by~\ref{it:T2} for $h'=h$, and by~\ref{it:S2} for $h'\ne h$. So $v$ is a candidate with no candidate neighbor, hence $v\in S$, and $h\in N_\G[S]$.
\end{proof}

So it remains to show that, with high probability, some test vertex of $h$ is robustly successful.

\begin{lemma}[Some test vertex succeeds]\label{lem:test-succeed}
Fix an outcome of $F_0$ that satisfies $\mathcal E_0$, and an outcome of $\mathcal B_h$ for which the event of \Cref{lem:helpers} holds. Then, for all sufficiently large $n$, the conditional probability that no vertex of $T_h$ is robustly successful is at most $n^{-c-3}$.
\end{lemma}

\begin{proof}
We say that a test vertex \emph{fails} if it is not robustly successful.

\emph{The conditioning does not matter.} Conditions~\ref{it:S1} and~\ref{it:S2} depend only on the random edges and coins on pairs between $H_0\setminus\set h$ and $L_0$. The graph $F_0$ and the sample $\mathcal B_h$ depend on other pairs, those inside $L_0$ and those containing $h$. So, even after fixing $F_0$ and $\mathcal B_h$, each pair $\set{h',v}$ with $h'\in H_0\setminus\set h$ and $v\in L_0$ lies in $R$ with probability $p$, and each $I_{h'w}$ with $w\in L_0\setminus X_0$ equals $1$ with probability $k/n$, all independently.

\emph{One test vertex fails with probability at most $1/4$.} Fix $v\in T_h$. Condition~\ref{it:S1} fails only if one of the at most $M$ pairs $\set{h',v}$ with $h'\in H_0\setminus\set h$ lies in $R$; each does so with probability $p=\lambda/n$. Condition~\ref{it:S2} fails only if one of the at most $\deg_{F_0}(v)\cdot M$ indicators $I_{h'w}$ with $w\in N_{F_0}(v)$ and $h'\in H_0\setminus\set h$ equals $1$; each does so with probability $k/n$. By a union bound and~\ref{it:T1}, the probability that $v$ fails is at most
\[
  Mp+(t+2\lambda)M\frac kn\ =\ \frac{M\lambda}n+\frac{tMk}n+\frac{2\lambda Mk}n .
\]
We bound the three terms. Since $M<n/\sqrt \D$, the first is at most $\lambda/\sqrt \D$ and the third is at most $2\lambda k/\sqrt \D$. Since $t\le n/(16Mk)$, the second is at most $1/16$. As $\lambda$ and $k$ are $O(\ln n)$ while $\sqrt \D$ grows faster than every power of $\ln n$, the sum is at most $1/4$ for all sufficiently large $n$.

\emph{Different test vertices are independent.} The event that $v$ fails is determined by the variables on the pairs $\set{h',u}$ with $h'\in H_0\setminus\set h$ and $u\in N_{F_0}[v]$. By~\ref{it:T3}, these sets of pairs are disjoint for different test vertices, so the events are independent. Hence all test vertices fail with probability at most $4^{-\abs{T_h}}$.

\emph{Conclusion.} On the event of \Cref{lem:helpers}, we have $\abs{T_h}\ge\floor{k/3}\ge k/3-1$. So the probability that all test vertices fail is at most
\[
  4^{-\floor{k/3}}\ \le\ 4\cdot4^{-k/3}\ =\ 4e^{-(\ln4)k/3}\ \le\ 4n^{-(\ln4)a/3}\ \le\ n^{-c-3},
\]
using $k\ge a\ln n$, and $(\ln4)a/3\ge c+4$ for $a\ge48(c+3)$, and $n\ge4$.
\end{proof}

\begin{lemma}[Case (2b)]\label{lem:case2b}
If $M>0$, $z_H<n/\sqrt \D$ and $z_Q<n/\sqrt t$, then, for all sufficiently large $n$, $\AAA=\emptyset$ with probability at least $1-n^{-c-1}$.
\end{lemma}

\begin{proof}

\emph{Bad base-high vertices.} Call $h\in H_0$ \emph{bad} if the event of \Cref{lem:helpers} fails, or no vertex of $T_h$ is robustly successful. Fix an outcome of $F_0$ that satisfies $\mathcal E_0$, and fix $h\in H_0$. Conditional on this outcome, the event of \Cref{lem:helpers} fails with probability at most $2n^{-c-3}$, by that lemma. For every outcome of $\mathcal B_h$ for which this event holds, \Cref{lem:test-succeed} shows that, conditional on $F_0$ and $\mathcal B_h$, no vertex of $T_h$ is robustly successful with probability at most $n^{-c-3}$. Hence, conditional on $F_0$, the vertex $h$ is bad with probability at most $2n^{-c-3}+n^{-c-3}=3n^{-c-3}$. We need no independence between different base-high vertices: a union bound over the at most $n$ vertices of $H_0$ shows that, conditional on $F_0$, some vertex of $H_0$ is bad with probability at most $3n^{-c-2}$. Since this holds for every outcome of $F_0$ that satisfies $\mathcal E_0$, the probability that $\mathcal E_0$ holds and some vertex of $H_0$ is bad is at most $3n^{-c-2}$.

\emph{No active vertex.} Suppose that $\mathcal E$ and $\mathcal E_0$ hold and no vertex of $H_0$ is bad. Let $h\in V_{\mathrm{hi}}$. Then $h\in H_0$ by \Cref{lem:degree-event}, and since $h$ is not bad, some $v\in T_h$ is robustly successful. By \Cref{lem:robust-dom}, $h\notin\AAA$. Since $\AAA\subseteq V_{\mathrm{hi}}$, we get $\AAA=\emptyset$.

\emph{The failure probability.} By the previous paragraph, $\AAA\ne\emptyset$ only if $\mathcal E$ fails, $\mathcal E_0$ fails, or $\mathcal E_0$ holds and some vertex of $H_0$ is bad. By \Cref{lem:degree-event}, $\mathcal E$ fails with probability at most $n^{1-\beta/3}\le n^{-c-2}$, since $\beta=2a\ge3(c+3)$. By \Cref{lem:F0-good}, $\mathcal E_0$ fails with probability at most $2n^{-c-2}$. By the first paragraph, the event $\mathcal E_0$ holds and some vertex of $H_0$ is bad with probability at most $3n^{-c-2}$. By a union bound, $\AAA\ne\emptyset$ with probability at most
\[
  n^{-c-2}+2n^{-c-2}+3n^{-c-2}\ =\ 6n^{-c-2}\ \le\ n^{-c-1},
\]
for $n\ge6$.
\end{proof}

We are now ready to prove \Cref{lem:gather-ok}. Let us restate it here for convenience.

\gathering*

\begin{proof}[Proof of \Cref{lem:gather-ok}]
The quantities $M$, $z_H$ and $z_Q$ depend only on $G$, so for a given $G$ exactly one of the following applies: $M=0$, Case~(1), Case~(2a), or Case~(2b). If $M=0$, then $H_0=\emptyset$. Let $\mathcal E$ be the event of \Cref{lem:degree-event}. If $\mathcal E$ holds, then $\AAA\subseteq H_0=\emptyset$ by that lemma, so condition~\eqref{eq:gather-ok} holds. Hence it fails with probability at most $\Prob[\mathcal E^c]\le n^{1-\beta/3}\le n^{-c-1}$, since $\beta=2a\ge3(c+3)$. In Cases~(1) and~(2a), \Cref{lem:case1,lem:case2a} show that condition~\eqref{eq:gather-ok} holds with probability at least $1-n^{-c-1}$. In Case~(2b), \Cref{lem:case2b} shows that $\AAA=\emptyset$, and hence that condition~\eqref{eq:gather-ok} holds, with probability at least $1-n^{-c-1}$.
\end{proof}

\subsubsection{Proof of \Cref{thm:upper}}\label{sec:proof-upper}
We can now show the desired $O(\sqrt{ \log n})$ round complexity for solving MIS in the smoothed setting.
Let us restate \Cref{thm:upper} here for convenience.

\upperthm*

\begin{proof}
Recall that the algorithm outputs $I=S\cup J\cup K$, where $S$ is the set of helpers of Step~1, $J$ is the set computed in Step~2, and $K$ is the output of the algorithm of \Cref{prop_Ghaffari} on $\G[V_{\mathrm{rest}}]$ in Step~3. We say that Step~3 \emph{succeeds} if $K$ is an MIS of $\G[V_{\mathrm{rest}}]$. We show that $I$ is an MIS of $\G$ whenever condition~\eqref{eq:gather-ok} holds and Step~3 succeeds, that both happen with probability at least $1-n^{-c}$, and that the algorithm takes $O(\sqrt{\ln n})$ rounds.

\emph{Independence.} The sets $S$ and $J$ are independent in $\G$: a vertex joins $S$ only if it is a candidate with no candidate neighbor, so no two vertices of $S$ are adjacent, and $J$ is independent by \Cref{lem:J}. If Step~3 succeeds, then $K$ is independent as well. Moreover, no edge joins any two of the three sets: $J\subseteq\AAA=V_{\mathrm{hi}}\setminus N_\G[S]$ has no neighbor in $S$, and $K\subseteq V_{\mathrm{rest}}=V_{\mathrm{lo}}\setminus N_\G[S\cup J]$ has no neighbor in $S\cup J$. Hence, if Step~3 succeeds, $I$ is independent in $\G$.

\emph{Maximality.} Assume that condition~\eqref{eq:gather-ok} holds and Step~3 succeeds, and let $v\notin N_\G[S\cup J]$. If $v$ were high, it would be active, since $v\notin N_\G[S]$; but \Cref{lem:J} gives $\AAA\subseteq N_\G[J]$, a contradiction. So $v$ is low, and hence $v\in V_{\mathrm{rest}}$. Since $K$ is an MIS of $\G[V_{\mathrm{rest}}]$, we get $v\in N_\G[K]$. Hence every vertex lies in $N_\G[I]$.

\emph{Condition~\eqref{eq:gather-ok} holds and Step~3 succeeds with probability at least $1-n^{-c}$.} By \Cref{lem:gather-ok}, condition~\eqref{eq:gather-ok} fails with probability at most $n^{-c-1}$, for every base graph $G$ and all sufficiently large $n$. Next, every low vertex has degree at most $B$ in $\G$, so $\G[V_{\mathrm{rest}}]$ has at most $n$ vertices and maximum degree at most $B$. Step~3 runs the algorithm of \Cref{prop_Ghaffari} with $\Delta=B$, using fresh random bits, so for every outcome of the earlier steps it fails with probability at most $n^{-c-1}$. By a union bound, condition~\eqref{eq:gather-ok} holds and Step~3 succeeds with probability at least $1-2n^{-c-1}\ge1-n^{-c}$.

\emph{Running time.} The set-up takes one round. Step~1 takes $O(1)$ rounds, since the marks, the candidate status and the membership in $S$ are each sent to the neighbors in one round. One more round tells every vertex whether it is active, and \textsc{Gather} then takes $\ell+1=O(\sqrt{\ln n})$ rounds by \Cref{lem:gather}. One more round tells every vertex whether it has a neighbor in $S\cup J$, and hence whether it lies in $V_{\mathrm{rest}}$. Finally, $B=\D+2\lambda\le2e^{\sqrt{\ln n}}$ for large $n$, so $\log B=O(\sqrt{\ln n})$, and Step~3 takes $O(\log B)+\widetilde O(\log^3\log n)=O(\sqrt{\ln n})$ rounds by \Cref{prop_Ghaffari}. Each step runs for a number of rounds that depends only on $n$ and $\beta$, so the total is $O(\sqrt{\ln n})$.
\end{proof}

\subsection{Extension to the Symmetric Difference Model}
\label{sec:symdiff}

\Cref{thm:upper} also holds for the symmetric-difference model $\G_s=G\triangle G(n,p)$ of \Cref{sec:model}, with the same algorithm. Let $p\le1/2$, and let $G'\subseteq G$ keep each edge of $G$ independently with probability $\gamma=(1-2p)/(1-p)$. Let $\G'$ be the smoothing of $G'$, that is, $G'$ together with each non-edge of $G'$ added independently with probability $p$. A non-edge of $G$ is present in $\G'$ with probability $p$, and an edge of $G$ is present with probability $\gamma+(1-\gamma)p=\gamma(1-p)+p=1-p$, independently over all pairs; so $\G'$ has the same distribution as $G\triangle G(n,p)$. \Cref{thm:upper} holds for every base graph, in particular for every outcome of $G'$, and the algorithm does not use the base graph. Averaging over $G'$, the algorithm computes an MIS of $G\triangle G(n,p)$ in $O(\sqrt{\log n})$ rounds with probability at least $1-n^{-c}$.

\begin{theorem}\label{thm:upperfinal}
For every constant $c>0$, there is a constant $\beta\ge1$, depending only on $c$, such that the following holds for all sufficiently large $n$. Let $G$ be any graph on $n$ vertices, and let $\G_s = G \triangle G(n,p)$ be the smoothed graph with $p=\beta\ln n/n$. There is a randomized LOCAL algorithm that, given $\G_s$, the number $n$, and distinct identifiers of $O(\log n)$ bits, computes an MIS of $\G_s$ in $O(\sqrt{\log n})$ rounds with probability at least $1-n^{-c}$. The probability is over the random edges and the algorithm's random bits.
\end{theorem}

%% file: fig_overview.tex
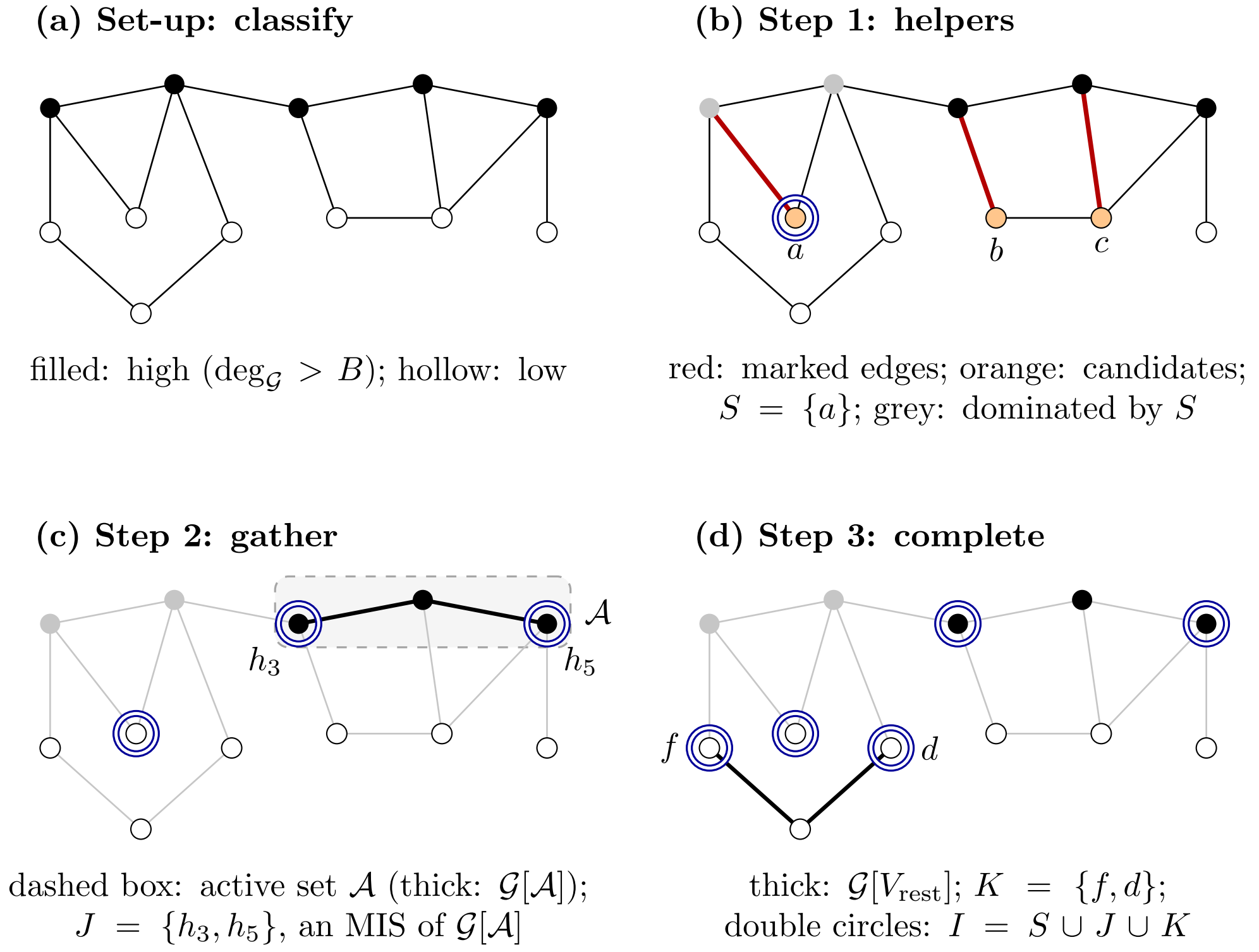
\begin{figure}[hpt!]
\centering
\begin{tikzpicture}[
  scale=1.0,
  hi/.style={circle,fill=black,inner sep=0pt,minimum size=6pt},
  dom/.style={circle,fill=gray!45,inner sep=0pt,minimum size=6pt},
  low/.style={circle,draw=black,fill=white,inner sep=0pt,minimum size=6pt},
  cand/.style={circle,draw=black,fill=orange!45,inner sep=0pt,minimum size=6pt},
  outp/.style={circle,draw=blue!60!black,double=white,double distance=0.9pt,line width=0.6pt,inner sep=0pt,minimum size=12pt},
  e/.style={line width=0.5pt},
  ae/.style={line width=1.2pt},
  faint/.style={line width=0.5pt,gray!45},
  mrk/.style={line width=1.4pt,red!70!black},
  actbox/.style={draw=gray!70,dashed,fill=gray!8,rounded corners=4pt,line width=0.6pt,inner sep=7pt},
  lbl/.style={font=\small},
  ttl/.style={font=\small\bfseries,anchor=west},
  capt/.style={font=\small,align=center,text width=6.2cm}
]
\def\coords{
  \coordinate (h1) at (0.0,2.0);  \coordinate (h2) at (1.3,2.25);
  \coordinate (h3) at (2.6,2.0);  \coordinate (h4) at (3.9,2.25);
  \coordinate (h5) at (5.2,2.0);
  \coordinate (f) at (0.0,0.7);   \coordinate (a) at (0.9,0.85);
  \coordinate (d) at (1.9,0.7);   \coordinate (b) at (3.0,0.85);
  \coordinate (c) at (4.1,0.85);  \coordinate (ee) at (5.2,0.7);
  \coordinate (g) at (0.95,-0.15);
}
\begin{scope}
  \coords
  \draw[e] (h1)--(h2)--(h3)--(h4)--(h5);
  \draw[e] (h1)--(f) (h1)--(a) (h2)--(a) (h2)--(d) (h3)--(b) (h4)--(c) (h5)--(c) (h5)--(ee);
  \draw[e] (b)--(c) (f)--(g) (d)--(g);
  \foreach \v in {h1,h2,h3,h4,h5} \node[hi] at (\v) {};
  \foreach \v in {f,a,d,b,c,ee,g} \node[low] at (\v) {};
  \node[ttl] at (-0.3,2.9) {(a) Set-up: classify};
  \node[capt,anchor=north] at (2.6,-0.45) {filled: high ($\deg_\G>B$);\ hollow: low};
\end{scope}
\begin{scope}[xshift=6.9cm]
  \coords
  \draw[e] (h1)--(h2)--(h3)--(h4)--(h5);
  \draw[e] (h1)--(f) (h2)--(a) (h2)--(d) (h5)--(c) (h5)--(ee);
  \draw[e] (b)--(c) (f)--(g) (d)--(g);
  \draw[mrk] (h1)--(a) (h3)--(b) (h4)--(c);
  \foreach \v in {h3,h4,h5} \node[hi] at (\v) {};
  \foreach \v in {h1,h2} \node[dom] at (\v) {};
  \foreach \v in {f,d,ee,g} \node[low] at (\v) {};
  \foreach \v in {a,b,c} \node[cand] at (\v) {};
  \node[outp] at (a) {};
  \node[lbl,below=4pt of a] {$a$};
  \node[lbl,below=2pt of b] {$b$};
  \node[lbl,below=2pt of c] {$c$};
  \node[ttl] at (-0.3,2.9) {(b) Step 1: helpers};
  \node[capt,anchor=north] at (2.6,-0.45) {red: marked edges;\ orange: candidates;\\ $S=\{a\}$;\ grey: dominated by $S$};
\end{scope}
\begin{scope}[yshift=-5.4cm]
  \coords
  \begin{scope}[on background layer]
    \node[actbox,fit=(h3)(h4)(h5)] (AA) {};
  \end{scope}
  \draw[faint] (h1)--(h2)--(h3);
  \draw[faint] (h1)--(f) (h1)--(a) (h2)--(a) (h2)--(d) (h3)--(b) (h4)--(c) (h5)--(c) (h5)--(ee);
  \draw[faint] (b)--(c) (f)--(g) (d)--(g);
  \draw[ae] (h3)--(h4)--(h5);
  \foreach \v in {h3,h4,h5} \node[hi] at (\v) {};
  \foreach \v in {h1,h2} \node[dom] at (\v) {};
  \foreach \v in {f,a,d,b,c,ee,g} \node[low] at (\v) {};
  \foreach \v in {a,h3,h5} \node[outp] at (\v) {};
  \node[lbl,anchor=west] at (AA.east) {$\AAA$};
  \node[ttl] at (-0.3,2.9) {(c) Step 2: gather};
  \node[capt,anchor=north] at (2.6,-0.45) {dashed box: active set $\AAA$ (thick: $\G[\AAA]$);\\ $J=\{h_3,h_5\}$, an MIS of $\G[\AAA]$};
  \node[lbl,below left=3pt and 1pt of h3] {$h_3$};
  \node[lbl,below right=3pt and 1pt of h5] {$h_5$};
\end{scope}

\begin{scope}[xshift=6.9cm,yshift=-5.4cm]
  \coords
  \draw[faint] (h1)--(h2)--(h3)--(h4)--(h5);
  \draw[faint] (h1)--(f) (h1)--(a) (h2)--(a) (h2)--(d) (h3)--(b) (h4)--(c) (h5)--(c) (h5)--(ee) (b)--(c);
  \draw[ae] (f)--(g) (d)--(g);
  \foreach \v in {h3,h4,h5} \node[hi] at (\v) {};
  \foreach \v in {h1,h2} \node[dom] at (\v) {};
  \foreach \v in {f,a,d,b,c,ee,g} \node[low] at (\v) {};
  \foreach \v in {a,h3,h5,f,d} \node[outp] at (\v) {};
  \node[lbl,left=5pt of f] {$f$};
  \node[lbl,right=5pt of d] {$d$};
  \node[ttl] at (-0.3,2.9) {(d) Step 3: complete};
  \node[capt,anchor=north] at (2.6,-0.45) {thick: $\G[V_{\mathrm{rest}}]$;\ $K=\{f,d\}$;\\ double circles: $I=S\cup J\cup K$};
\end{scope}
\end{tikzpicture}

\caption{\Cref{alg:main} on a small example. Filled vertices are high and hollow vertices are low; vertices of the output $I$ are drawn with a double circle. (a)~The set-up splits the vertices into $V_{\mathrm{hi}}$ and $V_{\mathrm{lo}}$. (b)~Step~1: the high vertices mark some of their edges to \textcolor{red}{low neighbors (red)}, and the low endpoints $a$, $b$, $c$ become \textcolor{orange}{candidates (orange)}. The candidate $a$ has no candidate neighbor, so $S=\set{a}$; the candidates $b$ and $c$ are adjacent, so neither joins $S$. The two high vertices adjacent to $a$ are dominated by $S$ and drawn grey from now on; they are not active. (c)~Step~2: the other three high vertices form the active set $\AAA$ (dashed box), and $\G[\AAA]$ is the path drawn thick. Every active vertex gathers this component, and the vertices compute the same MIS $J=\set{h_3,h_5}$ of it. (d)~Step~3: the low vertices with no neighbor in $S\cup J$ form $V_{\mathrm{rest}}$, and $\G[V_{\mathrm{rest}}]$ (thick) has maximum degree at most $B$. \Cref{prop_Ghaffari} computes an MIS $K=\set{f,d}$ of it, and $I=S\cup J\cup K$. }
\label{fig:steps}
\end{figure}

%% file: fig_case.tex
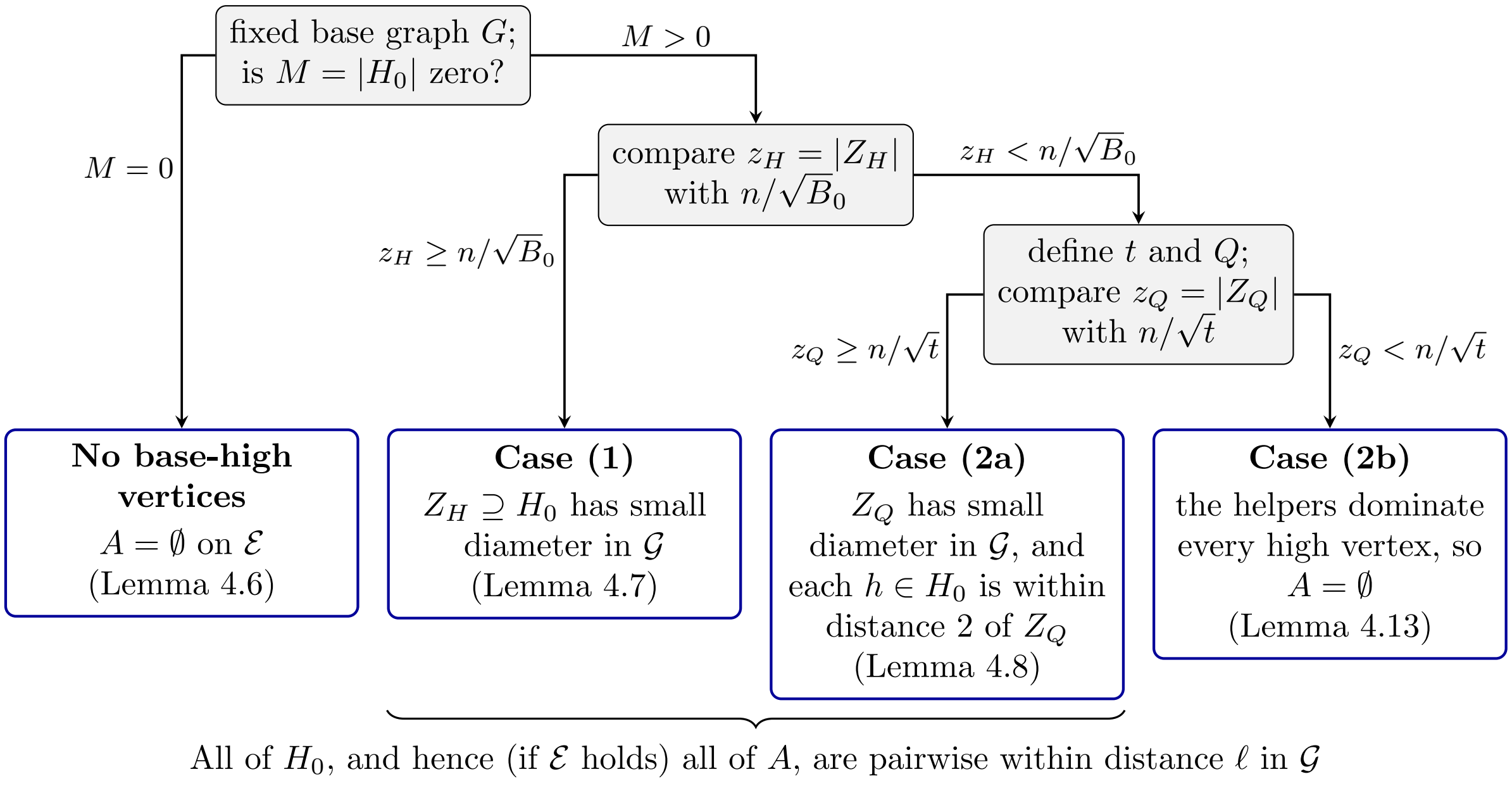
\begin{figure}[htbp]
\centering
\begin{tikzpicture}[
  test/.style={draw=black,rounded corners=3pt,fill=gray!10,font=\small,align=center,inner sep=4pt},
  leaf/.style={draw=blue!60!black,line width=0.8pt,rounded corners=3pt,font=\small,align=flush center,text width=3.4cm,inner sep=4pt,anchor=north},
  arr/.style={->,>=stealth,line width=0.7pt},
  cond/.style={font=\footnotesize,above,inner sep=2pt}
]

\node[test] (root) at (-4.0,0) {fixed base graph $G$;\\ is $M=\abs{H_0}$ zero?};
\node[test] (zh) at (0,-1.25) {compare $z_H=\abs{Z_H}$\\ with $n/\sqrt \D$};
\node[test] (zq) at (4.0,-2.5) {define $t$ and $Q$;\\ compare $z_Q=\abs{Z_Q}$\\ with $n/\sqrt t$};
\node[leaf] (m0) at (-6.0,-3.9) {\textbf{No \mbox{base-high} vertices}\\[2pt] $A=\emptyset$ on $\mathcal E$\\ (\Cref{lem:degree-event})};
\node[leaf] (c1) at (-2.0,-3.9) {\textbf{Case (1)}\\[2pt] $Z_H\supseteq H_0$ has small diameter in $\G$\\ (\Cref{lem:case1})};
\node[leaf] (c2a) at (2.0,-3.9) {\textbf{Case (2a)}\\[2pt] $Z_Q$ has small diameter in $\G$, and each $h\in H_0$ is within distance $2$ of $Z_Q$\\ (\Cref{lem:case2a})};
\node[leaf] (c2b) at (6.0,-3.9) {\textbf{Case (2b)}\\[2pt] the helpers dominate every high vertex, so $A=\emptyset$\\ (\Cref{lem:case2b})};
\draw[arr] (root.west) -| node[cond,pos=0.65,left] {$M=0$} (m0.north);
\draw[arr] (root.east) -| node[cond,pos=0.3] {$M>0$} (zh.north);
\draw[arr] (zh.west) -| node[cond,pos=0.65,left] {$z_H\ge n/\sqrt \D$} (c1.north);
\draw[arr] (zh.east) -| node[cond,pos=0.3] {$z_H<n/\sqrt \D$} (zq.north);
\draw[arr] (zq.west) -| node[cond,pos=0.7,left] {$z_Q\ge n/\sqrt t$} (c2a.north);
\draw[arr] (zq.east) -| node[cond,pos=0.7,right] {$z_Q<n/\sqrt t$} (c2b.north);
\draw[decorate,decoration={brace,mirror,amplitude=5pt},line width=0.6pt]
  ([yshift=-3pt]c1.west |- c2a.south) -- ([yshift=-3pt]c2a.south east)
  node[midway,below=6pt,font=\small,align=center] {All of $H_0$, and hence (if $\mathcal E$ holds) all of $A$, are pairwise within distance $\ell$ in $\G$};
\end{tikzpicture}
\caption{The case analysis in the proof of \Cref{lem:gather-ok}. Each test depends only on the base graph $G$, so exactly one of the four outcomes applies, and it is fixed before the random edges are drawn. In each outcome, the lemma named in the box shows that, with probability at least $1-n^{-c-1}$, either $\AAA=\emptyset$ or all active vertices are pairwise within distance $\ell$ in $\G$.}
\label{fig:cases}
\end{figure}

%% file: fig_2b.tex
\begin{figure}[htbp]
\centering
\begin{tikzpicture}[
  hi/.style={circle,fill=black,inner sep=0pt,minimum size=6pt},
  low/.style={circle,draw=black,fill=white,inner sep=0pt,minimum size=5.5pt},
  smp/.style={circle,draw=black,fill=gray!35,inner sep=0pt,minimum size=6pt},
  ring/.style={circle,draw=blue!60!black,line width=1.1pt,inner sep=0pt,minimum size=11pt},
  base/.style={line width=0.6pt},
  fe/.style={line width=0.5pt,gray!60!black},
  mrk/.style={line width=1.0pt,red!70!black,dashed},
  rnd/.style={line width=1.0pt,orange!80!black,densely dotted},
  nbh/.style={draw=blue!50!black,dashed,line width=0.5pt,circle,minimum size=1.75cm},
  blob/.style={draw=gray!70,fill=gray!15,ellipse,minimum width=1.9cm,minimum height=2.1cm},
  lbl/.style={font=\small},
  note/.style={font=\footnotesize,align=center}
]
\draw[rounded corners=6pt,gray!70,line width=0.6pt] (-6.6,-2.75) rectangle (7.0,1.4);
\node[lbl,anchor=south west] at (-6.6,-2.75) {$L_0$};
\node[blob] (Xp) at (-5.3,-0.15) {};
\node[blob] (Qb) at (5.9,-0.15) {};
\node[lbl] at (-5.3,-1.55) {$X_0^+$};
\node[lbl] at (5.9,-1.55) {$Q$};
\node[note] at (0.9,-2.45) {pool $\mathcal U=L_0\setminus(X_0^+\cup Q)$};

\node[hi] (h)  at (-0.6,3.3) {};
\node[hi] (h2) at (4.6,3.3) {};
\node[hi] (h3) at (-5.3,3.3) {};
\node[lbl,above=2pt of h]  {$h$};
\node[lbl,above=2pt of h2] {$h'$};
\node[lbl,left=2pt of h3]  {$h''$};
\node[lbl,anchor=west] at (6.0,3.3) {$H_0$};

\node[low] (x)  at (-5.55,0.35) {};
\node[smp] (s1) at (-5.0,-0.55) {};
\draw[base] (h3)--(x);
\draw[fe] (x)--(s1);
\node[lbl,left=1pt of x] {$x$};

\node[smp] (s2) at (5.9,-0.45) {};
\foreach \a in {20,90,160,220,320} \draw[fe] (s2)--++(\a:0.55);

\node[smp] (s3) at (-3.95,-0.15) {};
\node[smp] (s4) at (-3.35,-0.75) {};
\draw[fe] (s3)--(s4);

\node[smp] (va) at (-1.6,0.1) {};
\node[low] (wa1) at (-2.1,-0.45) {};
\node[low] (wa2) at (-1.1,-0.45) {};
\draw[fe] (va)--(wa1) (va)--(wa2);
\node[nbh] at (-1.6,-0.15) {};

\node[smp] (vb) at (0.9,0.1) {};
\node[low] (wb1) at (0.4,-0.45) {};
\node[low] (wb2) at (1.4,-0.45) {};
\draw[fe] (vb)--(wb1) (vb)--(wb2);
\node[nbh] at (0.9,-0.15) {};
\node[lbl,below=0pt of wb2] {$w$};

\node[smp] (vc) at (3.4,0.1) {};
\node[low] (wc1) at (2.9,-0.45) {};
\node[low] (wc2) at (3.9,-0.45) {};
\draw[fe] (vc)--(wc1) (vc)--(wc2);
\node[nbh] at (3.4,-0.15) {};

\draw[mrk] (h)--(s1) (h)--(s3) (h)--(s4) (h)--(va) (h)--(vb) (h)--(vc) (h)--(s2);

\draw[mrk] (h2)--(wb2);
\draw[rnd] (h2)--(vc);

\node[ring] at (va) {};
\node[lbl,above left=-1pt and 1pt of va] {$v$};

\node[note,anchor=north] at (-1.6,-1.1) {robustly\\ successful:\\ $v\in S$};
\node[note,anchor=north] at (0.9,-1.1) {(S2) fails:\\ $h'w$ marked};
\node[note,anchor=north] at (3.4,-1.1) {(S1) fails:\\ $h'v'\in R$};
\node[lbl,right=0pt of vc] {$v'$};
\node[note,anchor=north] at (-3.8,-1.1) {not isolated\\ in $F_0^2[\mathcal B_h]$};
\end{tikzpicture}
\caption{Case~(2b), seen from a base-high vertex $h$. Grey vertices form the sample $\mathcal B_h$: the base-low vertices outside $X_0$ joined to $h$ by a marked random edge (red, dashed). Sampled vertices in $X_0^+$, such as the one next to the base edge $h''x$, or in $Q$ are not used, and neither are the two sampled vertices within distance two of each other. The remaining sampled vertices are the test vertices $T_h$; their closed neighborhoods in $F_0$ (dashed circles) are pairwise disjoint, so their outcomes are independent. A test vertex fails if another base-high vertex $h'$ has a random edge to it, violating~\ref{it:S1} (orange, dotted), or a marked edge to one of its neighbors $w$, violating~\ref{it:S2}. A robustly successful test vertex $v$ joins $S$ and dominates $h$ (\Cref{lem:robust-dom}).}
\label{fig:case2b}
\end{figure}
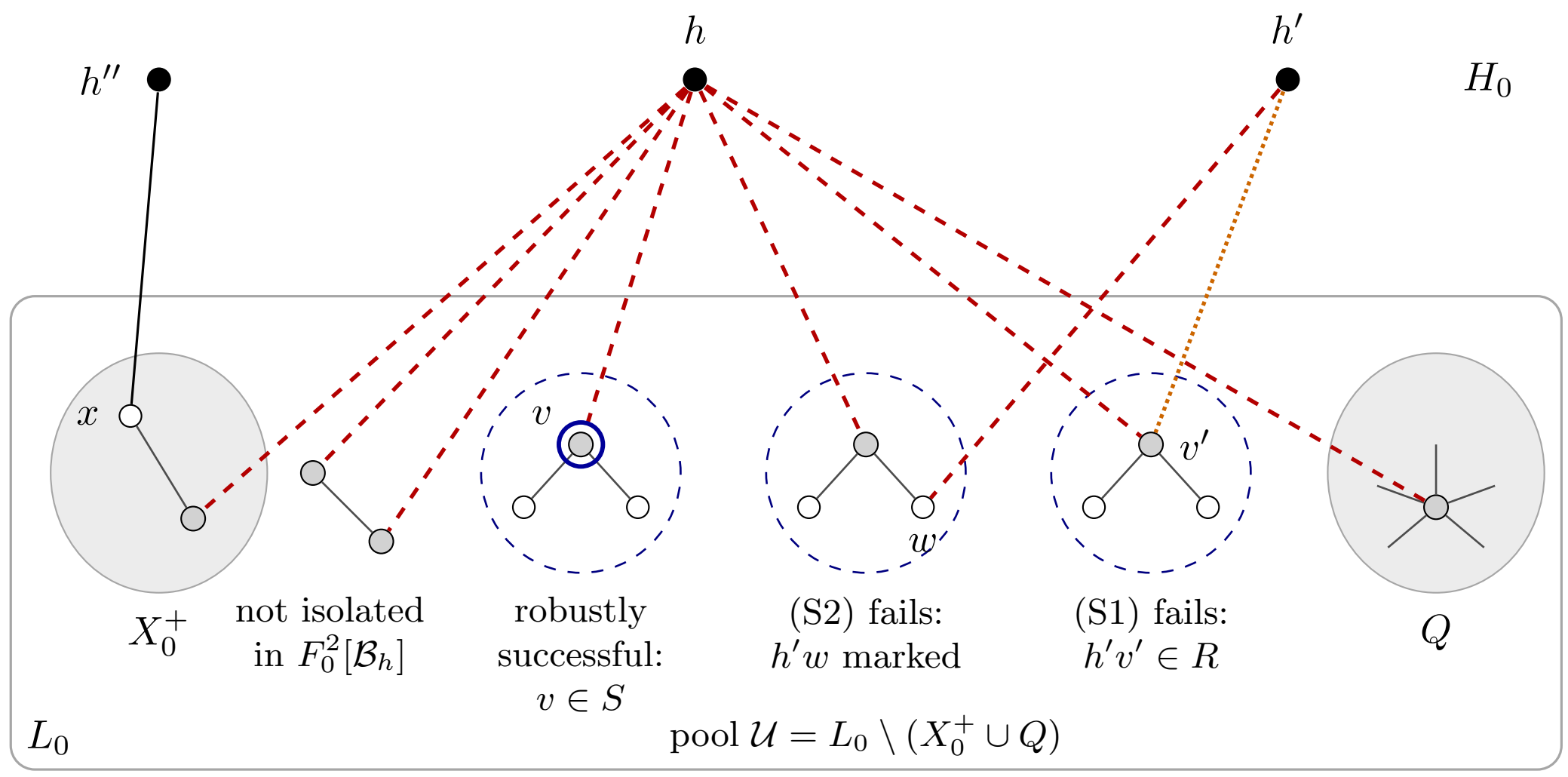

%% file: lower-bound-less-smoothening.tex
The algorithm we provide in Section \ref{sec:upper} shows that when one smooths with parameter $p \geq \beta \ln n/n$ for a sufficiently large
constant $\beta$, MIS can be solved on the resulting graph in $O(\sqrt{\log n})$ rounds with high probability. In this section, we show that this smoothing parameter choice is tight (up to a constant factor), i.e., when the smoothing parameter is $\leq (1/2 - \delta) \ln n / n$, for any constant $\delta \in (0,1/2]$, any algorithm that solves MIS on the smoothed graph takes asymptotically more time. 
More concretely, we show the following theorem.

\begin{theorem}\label{the:lower-bound-less-smoothing}
Fix constants $\delta\in(0,1/2]$ and $\sigma \in (0,1]$, and let $p = p(n)$ satisfy $p \in [0,  (1/2-\delta)\ln n/n]$. Any randomized LOCAL algorithm
that, for every graph $G$ of $n$ nodes computes an MIS of
$\G=G\cup G(n,p)$, with probability at least
$\sigma$, requires
$\Omega\!\left(\frac{\log n}{\log \log n}\right)$ rounds. 

The probability is over the smoothing and the
algorithm's private randomness. 
This bound holds even when nodes know the values of $n$, $p$,
use IDs exactly in $[n]$, and can distinguish between edges of $G$ and edges added by the smoothing.
\end{theorem}

At a high level, we show a reduction from the problem of (non-smoothed) MIS to the problem of MIS on a smoothed graph for smoothing parameter $p \leq (1/2 - \delta) \ln n / n$, for any constant $\delta \in (0,1/2]$. More concretely, we show how to reduce the problem of solving an MIS on a graph $H$ with $h$ nodes to the problem of solving MIS on a smoothed graph $\G$ with $n$ nodes. Thus, any algorithm $\mathcal{A}$ that solves MIS on smoothed graphs in time $T$ can be simulated by nodes in the original instance in $T_{sim}$ rounds, resulting in any instance of MIS being solved in $T_{sim}$. By carefully developing a reduction from any graph $H$ to a smoothed instance $\G$, we ensure that $T_{sim} = T + O(1)$. Note that $\G$ is not the smoothing of $H$ but of a graph constructed from $H$.

This reduction, as we show, holds for any MIS algorithm whose running time satisfies $T<\varepsilon\ln n/\ln\ln n$ (see Lemma~\ref{lb:simulation}), where $\varepsilon$ is a constant given in Section~\ref{subsec:reduce-mis-to-smoothed-mis}.
Since the size of the original graph $h$ is polynomially related to the size of the smoothed graph $n$  (i.e., $\ln h/ \ln \ln h = \Theta(\ln n/ \ln \ln n)$), this condition on the reduction amounts to $T = O(\ln h/ \ln \ln h)$. As a result, when combined with the recent result of Khoury and Schild~\cite{khoury2026} that shows that any randomized algorithm that takes $o(\ln h/ \ln \ln h)$ rounds to solve (non-smoothed) MIS does so with only \textit{sub-constant} probability on some graphs, we see that any algorithm that solves MIS on smoothed graphs with probability at least a constant $\sigma$ must take $\Omega(\ln n/\ln\ln n)$ rounds. 

We now give a proof sketch of Theorem~\ref{the:lower-bound-less-smoothing} in Section~\ref{subsec:lb-proof-sketch}; Section~\ref{subsec:reduce-mis-to-smoothed-mis} proves the reduction, and Section~\ref{subsec:getting-lower-bound} applies the
external lower bound. In Section~\ref{subsec:lower-bound-edge-deletions}, we show how to extend this proof to the smoothing model where both edge insertions and deletions take place.

\subsection{Proof Sketch of Theorem~\ref{the:lower-bound-less-smoothing}}
\label{subsec:lb-proof-sketch}
In Section~\ref{subsec:reduce-mis-to-smoothed-mis}, we show how to reduce the problem of MIS on a graph $H$ to an instance of MIS on a constructed graph $G$ by (i) replacing each node $u$ in $H$ with an appropriately large clique $C_u$, (ii) for every edge $(u,v) \in E(H)$, forming complete bipartite graphs between the two cliques $C_u$ and $C_v$ (i.e., each node of $C_u$ is connected to each node of $C_v$), and (iii) adding several ``padding'' nodes to $G$ that are just isolated nodes that will serve a purpose after smoothing. Once $G$ is constructed, we apply smoothing to it to obtain the smoothed graph $\G = G \cup G(n,p)$. 
We then show that, whenever every replacement clique in $\G$ contains a protected node, any solution to MIS on $\G$, $I$, maps to a solution to MIS on the input graph $H$, $J$, using the following mapping rule $J=\{u\in H:I\cap C_u\ne\varnothing\}$.\footnote{Here, we use the term replacement clique to emphasize that these are cliques in $\G$ formed as a result of replacing nodes in $H$ with a clique in $G$, and explicitly exclude cliques that may form in $\G$ as a result of smoothing.} This is shown by showing that the existence of \textit{protected nodes} in each clique in $\G$ ensures that $J$ is an MIS of $H$ after applying the mapping rule. 
Consider a node $a \in C_u$ in $\G$ (i.e., $a$ is a node in the clique of nodes in $\G$ representing node $u$ in $H$), and let $N_{G}(a)$ and $N_{\G}(a)$ represent $a$'s neighborhoods in $G$ and $\G$ respectively. Node $a$ is a protected node in $\G$ if $N_{G}(a) = N_{\G}(a)$. 
For $J$ to be an MIS of $H$, both independence and maximality conditions must hold. Independence follows from the edges between cliques
corresponding to adjacent nodes of $H$. To establish
maximality, suppose that neither $u$ nor any neighbor
of $u$ belongs to $J$. The protected node in $C_u$
would then neither belong to $I$ nor have a neighbor
in $I$, contradicting maximality of $I$.
We show in Lemma~\ref{lb:protection} that with probability $1 - o(1)$ every replacement clique contains a protected node. 
Then, as we show in Lemma~\ref{lb:projection}, 
any solution to MIS on $\G$, $I$, can be projected to a solution on $H$, $J$, where the mapping is defined as above.

The previous paragraph discussed how to map an MIS on a (virtual) $\G$ to an MIS on $H$. The other main component of the reduction is that if we have a fast MIS algorithm $\mathcal{A}$, by simulating it on $\G$, we can quickly obtain the MIS on $H$. In other words, we need to show that the time taken by nodes in $H$ to simulate $\mathcal{A}$ on $\G$ is not too much. Suppose that node $u \in H$ simulates some node $w \in \G$ and $w$'s neighbor in $\G$ is simulated by node $v \in H$. To simulate a round of $\mathcal{A}$ on $\G$, $u$ must transmit to and receive information from every such node $v$. 
We want to show that one round of the simulation on $\G$ corresponds to one round of actual computation on $H$ with probability $1 - o(1)$. This is shown in steps.

Recall that a node $u \in H$ simulates its clique of nodes $C_u$ in $\G$. We want to show that the adjacency induced by the clique nodes of graph $G$ (which captures the adjacency of the original graph $H$) is preserved in $\G$ with probability $1 - o(1)$. 
We show in Lemma~\ref{lb:core-edges} that for all pairs of nodes $a \in C_u$, $b \in C_v$ such that $u \neq v$ and $C_u,C_v$ are two cliques in $G$, then if $(u,v) \notin E(H)$, it is the case that $(a,b) \notin E(\G)$ with probability $1 - o(1)$. 

\begin{figure}
\centering
\includegraphics[width=.70\linewidth]{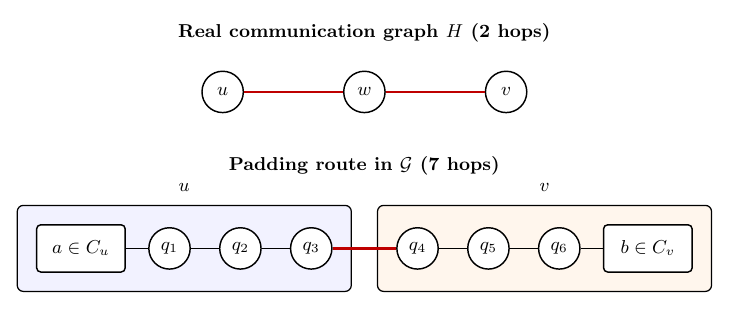}
\caption{An illustrative assignment of virtual nodes
to $u$ and $v$. Nodes $u$ and $v$ are distance $2$ apart in the original graph $H$. However, simulation of a message sent over virtual edge $(q_3,q_4)$ requires communication over the two-hop path between $u$ and $v$ in $H$.}
\label{figure:padding-route-example}
\end{figure}

However, this is not enough to ensure a fast simulation, where one simulated round takes one actual round,  
as paths through padding nodes must also be accounted for. For an illustrative example, suppose $u$ and $v$ locally simulate the nodes in their respective shaded boxes in Figure~\ref{figure:padding-route-example}. Relaying a message across the edge $(q_3, q_4)$ then requires communication between $u$ and $v$ which are two hops apart in $H$.
We show that, with probability $1-o(1)$, paths through padding nodes do not provide additional communication routes between any such $C_u$ and $C_v$ within the $T$ rounds being simulated. More formally, let $F$ be the graph obtained from  $\G$ by removing all inter-clique edges and all intra-clique edges. For each $u \in H$ and $a \in C_u$, take the $T+2$ hop ball centered around $a$ induced in $F$. 
We subsequently construct a virtual forest $\mathcal{F}$ where the root of each tree corresponds to a node $a \in C_u$, for all $u \in H$ and the depth of each tree is at most $T+2$. Lemmas~\ref{lb:population} and~\ref{lb:coupling}, taken together, show that there is a coupling under which, with probability $1 - o(1)$, every virtual tree in $\mathcal{F}$ is identical, including node IDs, to its corresponding radius-$(T+2)$ neighborhood in $F$. 

By combining Lemmas~\ref{lb:core-edges},~\ref{lb:population}, and~\ref{lb:coupling}, we see that with probability $1 - o(1)$, each node $u \in H$ can simulate its clique $C_u$ and the attached virtual trees using one actual round for each round of the simulation.
More generally, we show in Lemma~\ref{lb:simulation} that the simulation of MIS on $\G$ by nodes in the original graph $H$ and projection of the solution on $\G$ back to the original graph $H$ takes $T + O(1)$ rounds, for $T < \frac{\varepsilon \ln n}{\ln \ln n}$, where $\varepsilon$ is a constant that satisfies certain conditions seen in the analysis.

Finally, we complete the lower bound proof in Section~\ref{subsec:getting-lower-bound} by leveraging the lower bound graph of Khoury and Schild~\cite{khoury2026}. We apply our reduction to that lower bound graph, which takes on the role of input graph $H$. If any algorithm $\mathcal{A}$ existed that allowed us to solve MIS on a smoothed graph in $T = o(\ln n/ \ln \ln n)$ time with constant probability, the nodes of the original graph $H$ could simulate its execution on the virtual graph $\G$ in $T+ O(1)$ rounds. The resulting MIS output on $\G$ could then be locally mapped to an MIS on the original graph $H$ (without expending additional rounds). This would solve MIS on $H$ in $o(\ln n/ \ln \ln n) = o(\ln h/ \ln \ln h)$ rounds, which would contradict the lower bound of~\cite{khoury2026}. 
\subsection{Reduction from MIS to MIS on Smoothed Graphs}
\label{subsec:reduce-mis-to-smoothed-mis}

This section proceeds as follows. We first state the various parameter choices and introduce all relevant constants used throughout the section. Subsequently, we formally describe the construction of the smoothed graph $\G$ from the input graph $H$. We then explain how to map the output MIS of the simulation on $\G$ to an MIS for the input graph $H$. Finally, we bound the time taken by nodes in $H$ to perform the simulation.

\paragraph{Parameter choices.}
Throughout the rest of this proof, we use the following 
notation and parameter choices. Fix constants $\delta\in(0,1/2]$ and $\sigma\in(0,1]$. Choose a fixed integer
$d>1/\delta$ and constants $\alpha,\varepsilon$ satisfying
$a_0:=\frac12-\delta+\frac1d<\alpha<\frac12$ and $0<\varepsilon<\alpha-a_0$. 
For sufficiently large $n$, set
$h=\lfloor n^{1/d}\rfloor$, $k=\left\lceil20e^{np}\ln h\right\rceil$, $r=hk$, and
 $s=\lfloor n^\alpha\rfloor$.

Since $e^{np}\le n^{1/2-\delta}$, we have
\[
r=hk
\le n^{1/d}\bigl(20n^{1/2-\delta}\ln n+1\bigr)
\le n^{a_0+o(1)}.
\]
Consequently, $r\le s<n$ for sufficiently large $n$.
Moreover,
\[
ps^2=O(n^{2\alpha-1}\ln n)=o(1).
\]

\begin{figure}
\centering
\includegraphics[width=.90\linewidth]{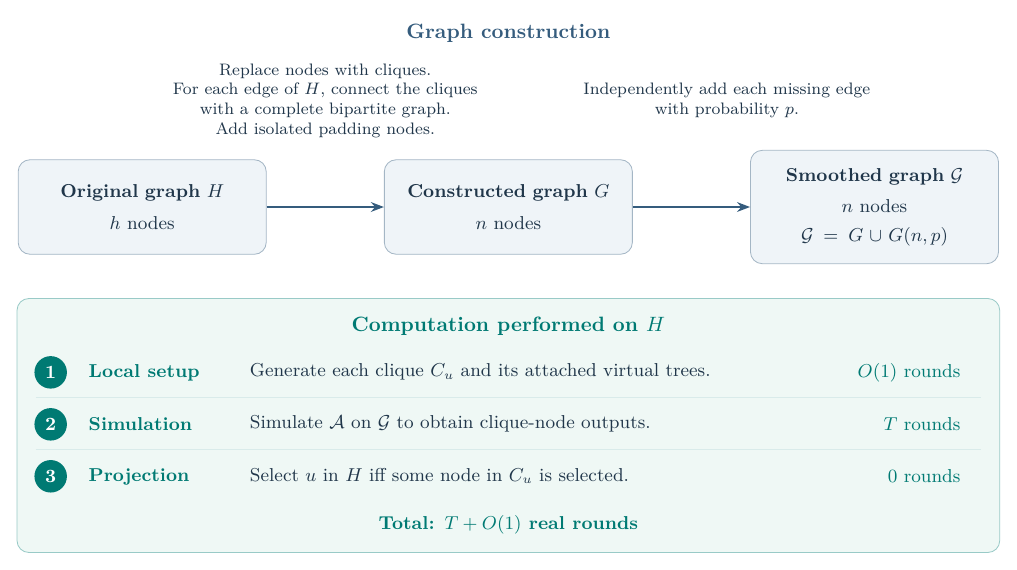}
\caption{Construction and simulation in the reduction.
The input graph $H$ is expanded to $G$ and smoothed
to $\G$. The nodes of $H$ simulate $\mathcal A$ on
$\G$ in $T+O(1)$ rounds and project its clique-node
outputs back to an MIS of $H$.}

\label{figure:01-reduction-overview}
\end{figure}

\paragraph{Constructing $\G$ from input graph $H$.}
The original graph is $H$ and our goal is to construct a virtual graph $G$, apply smoothing to it to obtain $\G$ and subsequently simulate an MIS algorithm $\mathcal{A}$ on $\G$. This is visually represented in Figure~\ref{figure:01-reduction-overview}. 

\begin{figure}
\centering
\includegraphics[width=.90\linewidth]{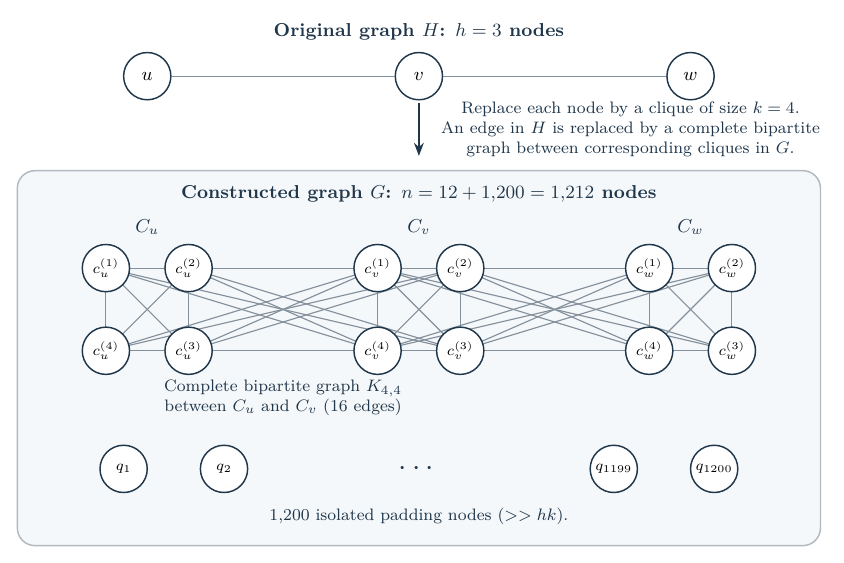}
\caption{An example of how the original graph $H$ of $h$ nodes is converted into the graph $G$ with $n$ nodes. For visual clarity, we choose values of $k$ and $n$ for illustrative purposes rather than correct values. Here, $h=3$, $k=4$, and the number of padding nodes is $1200$, which captures the idea that the number is significantly larger than $hk$.}
\label{figure:02-construction}
\end{figure}

Now, we describe the construction of $G$ from $H$. For each node $u \in H$, replace it with a clique of $k$ nodes $C_u=\{c_u^{(1)},\ldots,c_u^{(k)}\}$. Let $\mathcal{C} = \bigcup_{u \in H} C_u$, and let $r = |\mathcal{C}|$. In total, there are $r = hk$ clique nodes in $G$, which are interconnected as follows: for each edge $(u,v) \in E(H)$, add edges between all nodes in $C_u$ and $C_v$ such that the edges form a complete bipartite graph between the cliques. Finally, we add $n - r$ padding nodes $q_1$, $q_2$, $\ldots$, $q_{n-r}$ to $G$, which are just isolated nodes (i.e., with no neighbors in $G$). An example of this construction is shown in Figure~\ref{figure:02-construction}.
Finally, we discuss nodes' IDs. For the nodes in $H$, we assume that their IDs are unique and in the range $[h]$. 

We assign unique IDs, in the range $[n]$, to nodes in $G$ as follows: assign some clique node $c_u^{(i)}$ the ID $(\operatorname{id}_H(u)-1)k+i$, and assign some padding node $q_j$ the ID $r+j$.

\paragraph{Developing a mapping from MIS on $\G$ to MIS on $H$.}
We now show how to take a potential solution to the MIS on $\G$ and project it to the nodes in the input graph $H$. In particular, we show that if an MIS $I$ is calculated on $\G$, that will translate to an MIS $J=\{u\in H:I\cap C_u\ne\varnothing\}$ on $H$ with probability $1 - o(1)$ by leveraging the existence of protected nodes. Recall that a protected node $a \in \G$ is defined such that its neighborhood in $\G$ is the same as its neighborhood in $G$, i.e., $N_{G}(a) = N_{\G}(a)$. The existence of such a node in each clique allows us to show that the maximality of the MIS is conserved. 
We show that each clique contains such a protected node with probability $1 - o(1)$ in Lemma~\ref{lb:protection}. Subsequently, we show that, when every replacement clique has such a protected node, the way $J$ is defined results in a valid MIS on $H$ in Lemma~\ref{lb:projection}.

\begin{lemma}\label{lb:protection}
With probability at least $1-h^{-9}$, every replacement clique contains a
protected node.
\end{lemma}

\begin{proof}
A clique node has at most $n$ potential edges (edges not in $G$) to padding and other clique nodes. The probability of a node being a protected node (not having any such edges added during smoothing) is at least

\begin{align*}
(1-p)^n &= \exp(n \ln (1-p)) &\\
&= \exp(-np-O(np^2)) &\text{Taylor expansion of $\ln (1-p)$; $p\leq 1/2$ for sufficiently large $n$}\\
&= \exp(-np - o(1)) &\text{since $np^2=O((\ln n)^2/n)=o(1)$}\\
&\geq \tfrac{1}{2}e^{-np}&
\end{align*}
for sufficiently large $n$. For distinct nodes in the same clique,
the potential edge sets are disjoint: their only possible shared
edge joins those two nodes and is already an existing intra-clique edge. Thus their
protection events are independent, even though they may share potential
outside neighbors. Consequently,
$\Pr[C_u\text{ has no protected node}]
\le(1-\tfrac12e^{-np})^k
\le\exp(-ke^{-np}/2)\le h^{-10}$.
A union bound over the $h$ cliques results in the claim.
\end{proof}

\begin{lemma}\label{lb:projection}
If every replacement clique contains a protected node, then every MIS $I$ of $\G$
projects to an MIS of $H$ via
$J=\{u\in H:I\cap C_u\ne\varnothing\}$.
\end{lemma}

\begin{proof}
In any MIS $I$ on $\G$, there is at most one selected node in each clique. If $(u,v) \in E(H)$, then the complete bipartite graph between $C_u$ and $C_v$ prevents the two cliques from both containing a selected node. Thus the MIS $J$ on $H$ is independent.

To show the maximality of $J$, suppose that some node $u$ and all its neighbors  in $H$ are not in $J$. This implies that all clique nodes representing $u$ and its neighbors in $H$ are also not in $I$. We show that this leads to a contradiction with the assumption that $I$ represents an MIS. Consider a protected node $c_u^{(i)}$ in $\G$. It must not be in $I$. Its neighborhood consists of nodes in $C_u$ and nodes in the cliques representing neighbors of $u$ in $H$. As mentioned earlier, none of these nodes are in the MIS $I$. The protected node $c_u^{(i)}$ has no added neighbor outside $C_u$. Hence it has no neighbor in $I$, contradicting the maximality of $I$.
Therefore, every node not in $J$ has a neighbor in $J$. As a result, $J$ is an MIS of $H$.
\end{proof}

\paragraph{Simulation time preservation.}
We now show that nodes on the input graph $H$ will take $T + O(1)$ rounds to simulate algorithm $\mathcal{A}$ on $\G$. We do this by first showing that from $G$ to $\G$, no additional inter-clique edges are added with probability $1 - o(1)$ in Lemma~\ref{lb:core-edges}. Conditioning on this event, we subsequently show that each node $u$ in $H$ can locally simulate 
the radius-$(T+2)$ neighborhoods rooted at the nodes
of $C_u$, after removing all inter-clique and intra-clique edges. 
This is done by first upper bounding the number of such nodes that need to be simulated in Lemma~\ref{lb:population}  
and subsequently showing that such subgraphs are disjoint with probability $1 - o(1)$ for every pair of nodes $u,v \in H$ in Lemma~\ref{lb:coupling}.

\begin{lemma}\label{lb:core-edges}
$\G$ contains an edge not present in $G$ between two clique nodes with probability at most $o(1)$. 
\end{lemma}

\begin{proof}
Recall that there are $r = hk$ clique nodes in $G$. There are at most $\binom r2\le r^2$ possible edges between clique nodes in $G$ that can be added during smoothing to get $\G$. Each edge is added with probability $p$ during smoothing. A union bound gives that an edge not present in $G$ between two cliques is contained in $\G$  with probability at most $pr^2$. Since $p = O(\ln n/ n)$ and $r \leq n^{a_0+o(1)}$ (and recall that $a_0 < 1/2$), we have $pr^2 \leq n^{2a_0-1+o(1)} =o(1)$.
\end{proof}

The smoothing of $G$ to $\G$ is a random process. We focus on the radius $T+2$ neighborhoods of clique nodes in $F$, defined below, 
and show a coupling between the smoothing used to generate this graph and a random process to construct a virtual forest $\mathcal{F}$ of trees of depth at most $T+2$, where each tree is rooted at a clique node $a \in \mathcal{C}$. We show that the two constructions can be coupled, 
while preserving the correct distribution of each construction, so that, with probability $1 - o(1)$, all virtual trees in $\mathcal{F}$ with root nodes $a \in \mathcal{C}$ match the corresponding neighborhoods $F[P_{T+2}(a)]$, including nodes IDs.

Together with Lemma~\ref{lb:core-edges}, this allows each node $u\in H$ to simulate $\mathcal{A}$ on its clique $C_u$ and its attached virtual trees locally, while simulating communication between adjacent cliques over the corresponding edges of $H$.

\begin{figure}
\centering
\includegraphics[width=.90\linewidth]{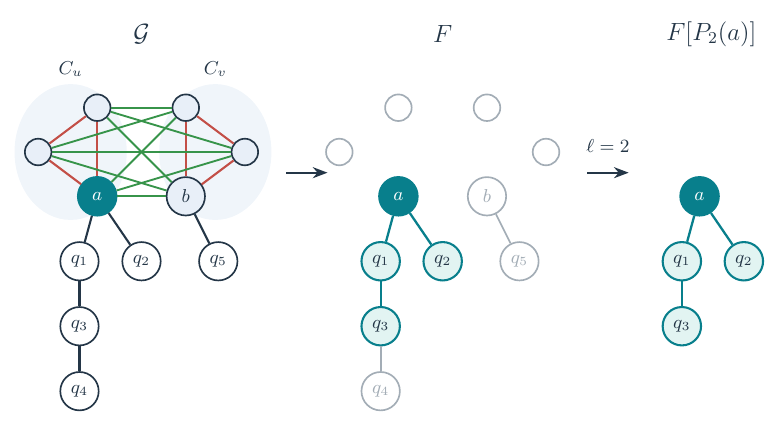}
\caption{An example of how $F[P_2(a)]$ is generated from graph $\G$. Both inter-clique edges (green) and intra-clique edges (red) are removed from $\G$ when creating $F$.}
\label{figure:padding-neighborhood-example}
\end{figure}

 We first define $F[P_{T+2}(a)]$ formally. Define a graph $F=\left(V(\G),\ E(\G)\setminus\binom{\mathcal{C}}{2}\right)$. Observe that \emph{all} inter-clique and intra-clique edges of $\G$ are removed. For each clique node $a\in\mathcal{C}$, define $P_{T+2}(a)=\{z:\operatorname{dist}_F(a,z)\le T+2\}$, where $\operatorname{dist}_F(a,z)$ represents the shortest distance between nodes $a$ and $z$ in $F$. Then, $F[P_{T+2}(a)]$ is the subgraph of $F$ induced by $P_{T+2}(a)$. Lemmas~\ref{lb:population} and~\ref{lb:coupling} imply that, with probability $1-o(1)$, all the neighborhoods $F[P_{T+2}(a)]$, for all $a\in\mathcal{C}$, simultaneously are identical to 
 their corresponding virtual trees, and hence are trees. 
See Figure~\ref{figure:padding-neighborhood-example} for an example illustration of one such $F[P_{T+2}(a)]$.

We now describe the construction of the virtual forest $\mathcal{F}$. At each of the $r = hk$ clique nodes, generate a virtual tree to depth $T+2$ as follows. Start with a given clique node as the root at level $0$ and recursively perform the following process. Independently for each node at level $< T+2$, draw the number of children of the node, $W$,
from a binomial distribution $\operatorname{Bin}(n-r,p)$. Notice that $\E[W]=(n-r)p$. Choose $W$ distinct IDs for these children uniformly at random from $[r+1,n]$. Note that this may result in the same ID appearing multiple times in a given tree or in multiple trees. We show that this happens with probability $o(1)$ by Lemma~\ref{lb:population} and Lemma~\ref{lb:coupling}. Recall that $s=\lfloor n^\alpha\rfloor$. Subsequently, we use $X$ to denote the total number of nodes in the virtual forest $\mathcal{F}$.

\begin{lemma}\label{lb:population}
 If $T<\varepsilon\ln n/\ln\ln n$, then $\Pr[X>s] = o(1)$.
\end{lemma}

\begin{proof}
Each node generated below level $T+2$ independently has expected $(n-r)p$ children. By induction, the expected number of nodes at level $j$, across all $r$ trees is $r((n-r)p)^j$. Summing up over all levels gives $\E[X]\le r(T+3)\max\{1,np\}^{T+2}$.

Recall that the first term $r\leq n^{a_0+o(1)}$. Notice that for sufficiently large $n$,  $\max\{1,np\} \leq \ln n$, and also notice that $T+2< \varepsilon\ln n/\ln\ln n + 2$. From this we see that the second term $T + 3 = n^{o(1)}$. Finally, we can bound the third term too: $\max\{1,np\}^{T+2} \leq (\ln n)^{\varepsilon\ln n/\ln\ln n + 2}  \leq n^{\varepsilon+o(1)}$. Thus we get $\mathbb E[X] \leq  n^{a_0+\varepsilon+o(1)}$. Applying Markov's inequality  gives us $\Pr[X>s] \leq \mathbb E[X]/s =  o(1)$ because $a_0+\varepsilon<\alpha$.
\end{proof}

\begin{lemma}\label{lb:coupling}
For $r\leq s=o(n)$, there is a coupling of the construction of the virtual forest $\mathcal{F}$
with a smoothing leading to $\G$ such that, with probability
at least
$1-\Pr[X>s]-O(ps^2)$,
the following is true for all $a\in\mathcal{C}$: the virtual tree rooted at $a$ is identical to
$F[P_{T+2}(a)]$, including node IDs. 
These neighborhoods are pairwise disjoint, and no edge
of $F$ joins two different neighborhoods. 
In particular, all nodes in the virtual forest have
distinct IDs.

\end{lemma}

\begin{proof}
We use the following equivalent description of the preceding virtual forest construction: 
at each node $i$ at a level $<T+2$, independently for each padding ID $j \in [r+1,n]$, create a child of $i$ with ID $j$ with probability $p$. We use deferred decisions to couple the virtual forest
construction with the smoothing process and bound the
probability of any disagreement between a virtual tree
and its corresponding neighborhood in $F$. Generate $\mathcal F$ and reveal $F$
together, exploring in breadth-first order from all $r$
clique nodes. 
Initially, the set of discovered nodes consists of all
$r$ clique roots. The goal is to discover all nodes of the virtual forest
and, without failure, verify that $F$ contains no edges
between different trees and no edges within a tree
other than its parent-child edges.
We stop the comparison and declare failure if either
(i) a mismatch occurs, as specified below, or
(ii) adding another virtual node would make the total
number of virtual nodes, including the roots, exceed $s$.

When expanding a node at a level $<T+2$, consider each
padding node ID in turn and perform a Bernoulli trial with probability $p$. If the ID is undiscovered at the start of the trial (i.e., has not been assigned to a node in $\mathcal{F}$ before), use the
same coin to decide whether it becomes a virtual child
and whether the corresponding edge is present in $F$. 
(It was not previously decided if this edge was in $F$ or not.) If the coin selects this ID, create the corresponding
virtual child, mark its ID as discovered, and add the corresponding edge to $F$.
If the ID was already discovered at the start of the trial,
then it is considered a mismatch if the trial selects it.

Before the size cutoff, at most $s$ nodes are expanded,
and at most $s$ padding IDs have been discovered.
At each trial involving an already discovered ID,
conditional on everything revealed so far, the probability of the ID being chosen  is $p$. Taking a union bound over the at most $s^2$ such trials, we see that the
probability of a mismatch in the form of an ID repetition before the cutoff is
$\leq ps^2$.

If the exploration finishes without declaring a failure, i.e., $X \leq s$ and no mismatch as described above occurs, we proceed with the checks on inter-tree and intra-tree edges. In particular, for every undecided edge in $F$ between a pair of distinct discovered nodes having at least one padding endpoint, flip a coin with probability $p$ to determine whether the edge exists.\footnote{We mention that at least one endpoint is a padding node to avoid checks between two clique nodes.} If any such edge exists, it is considered a mismatch. These checks
include pairs within a tree, between different trees,
and between nodes at level $T+2$. There are at most $s^2$ such pairs to check. Taking a union bound over all such pairs, the probability of a mismatch is $\leq ps^2$. 

Every virtual coin is fresh, and every shared coin
reveals a previously unexposed edge of $F$. After any
stop, complete the virtual construction according to
its original sampling rule and complete all remaining
smoothing trials independently. This includes trials
between clique nodes, whose edges do not enter $F$.
Thus both constructions retain their correct
distributions.

If $X\le s$ and neither type of mismatch occurs, all
virtual IDs are distinct and the explored edges are
exactly the tree edges. Consequently, the virtual trees
agree simultaneously with $F[P_{T+2}(a)]$ for every
$a\in\mathcal C$. These neighborhoods are pairwise
disjoint, with no edges of $F$ between them. Edges from
level-$(T+2)$ nodes to undiscovered padding nodes are
allowed, since they lead outside the specified radius.

The probability that any claimed property fails is
therefore at most $\Pr[X>s]+2ps^2$, which proves the lemma.
\end{proof}

\begin{lemma}\label{lb:simulation}
Let $\mathcal{A}$ be a randomized LOCAL algorithm that,
for every $n$ node base graph $G$, computes an MIS of
$\G=G\cup G(n,p)$ with probability at least $\sigma$,
over the smoothing and its private randomness,
using at most $T<\varepsilon\ln n/\ln\ln n$ rounds.

The nodes of $H$ can simulate $T$ rounds of the virtual execution on $\G$ and
project the solution back to the original graph $H$ in $T+O(1)$ rounds. The projected solution is an MIS of $H$ with probability at least $\sigma - o(1)$. 
\end{lemma}

\begin{proof}
There are two things to argue here. First, that a solution of MIS on $\G$ can be mapped to a valid solution of MIS on $H$. 
By
Lemmas~\ref{lb:core-edges}, \ref{lb:population},
and~\ref{lb:coupling}, the virtual trees and the
edges between clique nodes are identical to 
those of a
genuine $\G$ with probability $1-o(1)$.
The depth $T+2$ ensures that omitted neighbors of
boundary padding nodes cannot affect clique nodes'
outputs within $T$ rounds. Thus, with probability $1-o(1)$, the local simulation
reproduces the clique nodes' outputs of a genuine
execution of $\mathcal A$ on $\G$. Algorithm $\mathcal{A}$, when executed on $\G$, outputs an MIS with probability
at least $\sigma$. By Lemmas~\ref{lb:protection} and~\ref{lb:projection}, we see that the solution on an MIS for $\G$ can be mapped to an MIS on $H$ with probability $1- o(1)$.
A union bound over the failure of the simulation,
the failure of $\mathcal A$, and the failure of protection
gives overall success probability at least
$\sigma-o(1)-h^{-9}=\sigma-o(1)$.

Second, we need to argue that the total time to simulate $\mathcal{A}$ on $\G$ by nodes in $H$ is $T + O(1)$ rounds. Each node $u\in H$ locally generates and
simulates its clique $C_u$ and the attached virtual
trees. After $O(1)$ rounds of initialization,
communication within a clique and its trees is
simulated locally. For every $(u,v)\in E(H)$, all
messages between $C_u$ and $C_v$ are combined into
one message over $(u,v)$. Since LOCAL messages have
unbounded size, one virtual round takes one real
round. Thus the simulation takes $T+O(1)$ rounds. Furthermore, the mapping of the MIS obtained for $\G$ to one for $H$ can be computed by all nodes in $H$ locally. Thus the total time for the simulation is $T + O(1)$ rounds.
\end{proof}

\subsection{Obtaining the $\Omega(\ln n/ \ln \ln n)$ Lower Bound}
\label{subsec:getting-lower-bound}

In this section, we prove that Theorem~\ref{the:lower-bound-less-smoothing} holds true. 
We now combine Lemma~\ref{lb:simulation} with the
lower bound of Khoury and Schild~\cite{khoury2026}.
Their result implies that there exists a constant
$c_{\mathrm{KS}}>0$, depending on $\sigma$, such that, for all sufficiently
large $h$, any randomized LOCAL algorithm that computes
an MIS with probability at least $\sigma/2$ on every graph
of $h$ nodes requires at least $c_{\mathrm{KS}}\frac{\ln h}{\ln\ln h}$ rounds. This holds even when the nodes have IDs exactly
in $[h]$ and know $h$.

\begin{proof}[Proof of
Theorem~\ref{the:lower-bound-less-smoothing}]
Fix $\delta$, $\sigma$, and $p=p(n)$ as in the theorem. Use the parameters chosen in
Section~\ref{subsec:reduce-mis-to-smoothed-mis}, and
choose a constant $\eta \in (0,\min\left\{\varepsilon,\frac{c_{\mathrm{KS}}}{2d}\right\})$. 
Suppose that, for some sufficiently large $n$, an
algorithm $\mathcal A$ satisfying the theorem's
success guarantee has round complexity $T<\eta\frac{\ln n}{\ln\ln n}$.

Recall that $h=\lfloor n^{1/d}\rfloor$. Since
$\eta<\varepsilon$, Lemma~\ref{lb:simulation} applies:
for every input graph $H$ of $h$ nodes, its nodes
can simulate $\mathcal{A}$ on the constructed smoothed
graph and project the outputs to an MIS of $H$ in
$T+O(1)$ rounds, with probability at least $\sigma-o(1)$.
For sufficiently large $n$, this probability is at
least $\sigma/2$. The construction parameters are fixed
independently of $H$, so this gives one randomized
LOCAL algorithm that succeeds on every graph of
$h$ nodes.

Since $d$ is fixed and $h=\lfloor n^{1/d}\rfloor$,
\[
\frac{\ln h}{\ln\ln h}
=
\left(\frac1d+o(1)\right)
\frac{\ln n}{\ln\ln n}.
\]
Our choice of $\eta$ therefore gives, for sufficiently
large $n$,
\[
T+O(1)
<
\frac{c_{\mathrm{KS}}}{2d}
\frac{\ln n}{\ln\ln n}
<
c_{\mathrm{KS}}\frac{\ln h}{\ln\ln h}.
\]
Thus the derived algorithm solves MIS on every graph
of $h$ nodes with probability at least $\sigma/2$ in fewer
rounds than permitted by the Khoury--Schild lower
bound, which is a contradiction.

Consequently, any algorithm satisfying the theorem's
success guarantee requires at least
$\eta\ln n/\ln\ln n$ rounds for sufficiently large $n$.
Since $\eta$ is a positive constant, this proves the
claimed $\Omega(\ln n/\ln\ln n)$ lower bound.

The construction assigns IDs exactly in $[n]$, and
the simulation can provide the values of $n$ and $p$
and distinguish edges of $G$ from edges added by
smoothing. Hence the argument also establishes
the final assertion of the theorem.
\end{proof}

\subsection{Allowing Both Edge Additions and Deletions during Smoothing}
\label{subsec:lower-bound-edge-deletions}

The same lower bound holds when smoothing independently
adds each missing edge with probability $p$ and deletes
each existing edge with probability $p$. That is, $\G_s = G \triangle G(n,p)$ instead of $\G = G \cup G(n,p)$.

\begin{theorem}\label{cor:lower-bound-edge-deletions}
Fix constants $\delta\in(0,1/2]$ and $\sigma \in (0,1]$, and let $p = p(n)$ satisfy $p \in [0,  (1/2-\delta)\ln n/n]$. Any randomized LOCAL algorithm
that, for every graph $G$ of $n$ nodes computes an MIS of
$\G_s=G\triangle G(n,p)$, with probability at least
$\sigma$, requires
$\Omega\!\left(\frac{\log n}{\log \log n}\right)$ rounds. 

The probability is over the smoothing and the
algorithm's private randomness. 
This bound holds even when nodes know the values of $n$, $p$,
use IDs exactly in $[n]$, and can distinguish between surviving edges of $G$ and edges added by the smoothing.
\end{theorem}

\begin{proof}
Use the same construction and parameter choices as in
Section~\ref{subsec:reduce-mis-to-smoothed-mis}.
Since the padding nodes are isolated in $G$, every edge
of $G$ has both endpoints among the $r$ clique nodes.
Consequently, $|E(G)|\leq \binom r2$. 
A union bound therefore gives
\[
\Pr[\text{some edge of $G$ is deleted}]
\leq p|E(G)|
\leq p\binom r2
\leq n^{2a_0-1+o(1)}
=o(1),
\]
where we use $p=O(\ln n/n)$,
$r\leq n^{a_0+o(1)}$, and $a_0<1/2$. 

Thus, an algorithm that succeeds with probability
at least $\sigma$ under additions and deletions
succeeds with probability at least $\sigma-o(1)$
on the addition-only instances produced by our
construction. The simulation and projection argument
in Section~\ref{subsec:reduce-mis-to-smoothed-mis}
therefore still gives an algorithm on $H$ taking
$T+O(1)$ rounds and succeeding with probability
at least $\sigma-o(1)$.
The argument in Section~\ref{subsec:getting-lower-bound}
then establishes the claimed lower bound.
\end{proof}

\section*{AI Disclosure}
\addcontentsline{toc}{section}{AI Disclosure}

We investigated the potential of applying smoothed analysis to distributed MIS, and in particular, the possibility of obtaining an $O(\sqrt{\log n})$-round MIS algorithm via smoothing, in March 2026. This was inspired by the work that uses smoothed analysis for graph isomorphism, that appeared in STOC 2025~\cite{AKM25}. More recently, we then asked
ChatGPT (GPT-6.1 Sol Ultra) to explore this possibility, and it produced most of the technical results presented in this paper. We significantly reworked the proofs and exposition to make them understandable to humans. The authors take full responsibility for all the results.

%% file: appendix.tex
\appendix
\section{Proof of \texorpdfstring{\Cref{lem:er-diam}}{the random-graph diameter lemma}}\label{app:er-diam}
We prove \Cref{lem:er-diam} with $c'=1024$. Let $X\sim G(N,q)$ with $0<q\le1/2$ and $\nu=Nq\ge1024(\ln N+1)$. In particular $\nu\ge1024$ and $\ln N+1\le\nu/1024$. Put
\[
  b_0=\floor{1/q},\qquad r=\ceil{\frac{\ln N}{\ln(\nu/8)}}+1.
\]
For a vertex set $U$, its \emph{external neighbors} are the vertices of $N_X(U)\setminus U$. The proof uses the following two properties of $X$.
\begin{enumerate}[label=(P\arabic*)]
  \item \emph{Small sets expand:} every vertex set $U$ with $1\le\abs U\le b_0$ has at least $\nu\abs U/8$ external neighbors.
  \item \emph{Large sets are adjacent:} any two disjoint vertex sets $U,W$ with $\abs U,\abs W\ge\ceil{N/16}$ are joined by an edge of $X$.
\end{enumerate}
\Cref{lem:app-p1,lem:app-p2} show that each property fails with small probability, and \Cref{lem:app-diam} shows that together they bound the diameter.

\begin{lemma}\label{lem:app-p1}
$\Prob[\text{(P1) fails}]\le2e^{-\nu/64}$.
\end{lemma}

\begin{proof}
\emph{One fixed set.} Fix $U$ with $\abs U=b$, where $1\le b\le b_0$. Since $b_0\le1/q=N/\nu\le N/2$, there are $N-b\ge N/2$ vertices outside $U$. Each of them is an external neighbor of $U$ independently, with probability
\[
  1-(1-q)^b\ \ge\ 1-e^{-qb}\ \ge\ \frac{qb}2.
\]
The first inequality uses $1-q\le e^{-q}$. For the second, put $y=qb$; then $0<y\le1$ because $b\le b_0\le1/q$, and $e^{-y}\le1-y+y^2/2\le1-y/2$ for $0\le y\le1$. Hence the number of external neighbors is a sum of independent indicators with mean at least $(N/2)(qb/2)=\nu b/4$. By \Cref{fact:tails}(F2) with $\delta=1/2$,
\[
  \Prob\Bigl[\abs{N_X(U)\setminus U}<\frac{\nu b}8\Bigr]\ \le\ e^{-\nu b/32}.
\]

\emph{All sets of one size.} There are $\binom Nb\le(eN/b)^b\le e^{b(\ln N+1)}\le e^{\nu b/1024}$ sets of size $b$. By a union bound, some set of size $b$ fails to expand with probability at most
\[
  e^{\nu b/1024}e^{-\nu b/32}\ \le\ e^{-\nu b/64},
\]
since $1/32-1/1024\ge1/64$.

\emph{All sizes.} Property (P1) fails exactly when some set of some size $b\in\set{1,\dots,b_0}$ fails to expand. Put $y=e^{-\nu/64}$. Since $\nu\ge1024$, we have $y\le e^{-16}\le1/2$. Summing over $b\ge1$,
\[
  \Prob[\text{(P1) fails}]\ \le\ \sum_{b\ge1}y^b\ =\ \frac y{1-y}\ \le\ 2y\ =\ 2e^{-\nu/64}.
  \qedhere
\]
\end{proof}

\begin{lemma}\label{lem:app-p2}
$\Prob[\text{(P2) fails}]\le e^{-\nu N/512}$.
\end{lemma}

\begin{proof}
\emph{One fixed pair.} Fix disjoint sets $U,W$ with $\abs U,\abs W\ge\ceil{N/16}$. There are $\abs U\cdot\abs W\ge N^2/256$ vertex pairs with one end in $U$ and the other in $W$, and each is an edge of $X$ independently with probability $q$. So no edge joins $U$ and $W$ with probability at most
\[
  (1-q)^{N^2/256}\ \le\ e^{-qN^2/256}\ =\ e^{-\nu N/256}.
\]

\emph{Counting the pairs.} An ordered pair $(U,W)$ of disjoint sets is determined by deciding, for each of the $N$ vertices, whether it lies in $U$, in $W$, or in neither. So there are at most $3^N$ such pairs.

\emph{Union bound.} Hence
\[
  \Prob[\text{(P2) fails}]\ \le\ 3^Ne^{-\nu N/256}\ =\ e^{-N(\nu/256-\ln3)}\ \le\ e^{-\nu N/512},
\]
since $\nu/512\ge2>\ln3$ gives $\nu/256-\ln3\ge\nu/256-\nu/512=\nu/512$.
\end{proof}

\begin{lemma}\label{lem:app-diam}
If $X$ has properties (P1) and (P2), then $\diam(X)\le2r+1$.
\end{lemma}

\begin{proof}
For a vertex $v$ and an integer $i\ge0$, let $\mathrm{Ball}_i(v)$ be the set of vertices at distance at most $i$ from $v$ in $X$. A vertex is at distance at most $i+1$ from $v$ exactly when it lies in $\mathrm{Ball}_i(v)$ or is an external neighbor of $\mathrm{Ball}_i(v)$. Fix a vertex $v$.

\emph{Small balls grow fast.} If $\abs{\mathrm{Ball}_i(v)}<b_0$, then (P1) applies to $\mathrm{Ball}_i(v)$, and
\[
  \abs{\mathrm{Ball}_{i+1}(v)}\ \ge\ \Bigl(1+\frac\nu8\Bigr)\abs{\mathrm{Ball}_i(v)}\ \ge\ \frac\nu8\abs{\mathrm{Ball}_i(v)}.
\]
In particular, a ball with fewer than $b_0$ vertices gains a vertex in each step. Since $X$ has $N\ge b_0$ vertices, there is a first index $i^*$ with $\abs{\mathrm{Ball}_{i^*}(v)}\ge b_0$. Since $q\le1/2$, we have $b_0\ge2$, so $i^*\ge1$.

\emph{The ball reaches $b_0$ vertices by radius $r-1$.} By induction, $\abs{\mathrm{Ball}_i(v)}\ge(\nu/8)^i$ for all $i\le i^*$. Applying this at $i=i^*-1$, where the ball still has fewer than $b_0$ vertices,
\[
  \Bigl(\frac\nu8\Bigr)^{i^*-1}\ \le\ \abs{\mathrm{Ball}_{i^*-1}(v)}\ <\ b_0\ \le\ N.
\]
Taking logarithms, $i^*-1<\ln N/\ln(\nu/8)$, so $i^*\le\ceil{\ln N/\ln(\nu/8)}=r-1$.

\emph{One more step reaches $N/16$ vertices.} Choose $U\subseteq\mathrm{Ball}_{i^*}(v)$ with exactly $b_0$ vertices. By (P1), $U$ has at least $\nu b_0/8$ external neighbors, and all of them lie in $\mathrm{Ball}_{i^*+1}(v)$. Since $1/q\ge2$ and $\floor y\ge y/2$ for $y\ge2$, we have $b_0\ge1/(2q)$, so
\[
  \abs{\mathrm{Ball}_r(v)}\ \ge\ \abs{\mathrm{Ball}_{i^*+1}(v)}\ \ge\ \frac{\nu b_0}8\ \ge\ \frac\nu{16q}\ =\ \frac N{16},
\]
using $i^*+1\le r$. As the left-hand side is an integer, $\abs{\mathrm{Ball}_r(v)}\ge\ceil{N/16}$.

\emph{Any two vertices are close.} Let $u,v$ be vertices. If $\mathrm{Ball}_r(u)$ and $\mathrm{Ball}_r(v)$ share a vertex, then $\dist_X(u,v)\le2r$. Otherwise they are disjoint, each with at least $\ceil{N/16}$ vertices, so by (P2) some edge $xy$ has $x\in\mathrm{Ball}_r(u)$ and $y\in\mathrm{Ball}_r(v)$, and $\dist_X(u,v)\le r+1+r$. In both cases $\dist_X(u,v)\le2r+1$.
\end{proof}

\begin{proof}[Proof of \Cref{lem:er-diam}]
By \Cref{lem:app-diam}, the diameter exceeds $2r+1$ only if (P1) or (P2) fails. Since $\nu=Nq\ge1024$ and $q\le1/2$, we have $N\ge2048$, so $\nu N/512\ge\nu/64$. By \Cref{lem:app-p1,lem:app-p2},
\[
  \Prob\bigl[\diam(X)>2r+1\bigr]\ \le\ 2e^{-\nu/64}+e^{-\nu N/512}\ \le\ 3e^{-\nu/64}\ \le\ 3e^{-\nu/1024}=3e^{-\nu/c'}.
  \qedhere
\]
\end{proof}